\documentclass[%
 reprint,
 amsmath,amssymb,
 aps,
showpacs]{revtex4}
\usepackage[utf8]{inputenc}
\usepackage{amsmath, mathtools, tensor}
\usepackage{graphicx}
\usepackage{color}
\usepackage{dcolumn}
\usepackage{bm}

\newcommand{\dd}{\ensuremath{\textrm{d}}}

\newcommand{\const}{\textrm{const}}
\newcommand{\UD}[2]{\ensuremath{^{#1}_{\phantom{#1} #2}}}

\newcommand{\beq}{\begin{equation}}
\newcommand{\eeq}{\end{equation}}
\newcommand{\bea}{\begin{eqnarray}}
\newcommand{\eea}{\end{eqnarray}}
\newcommand{\bean}{\begin{eqnarray*}}
\newcommand{\eean}{\end{eqnarray*}}
\newcommand{\bit}{\begin{itemize}}
\newcommand{\eit}{\end{itemize}}
\newcommand{\bfi}{\begin{figure}}
\newcommand{\efi}{\end{figure}}
\newcommand{\bfic}{\begin{figure*}}
\newcommand{\efic}{\end{figure*}}
\newcommand{\bce}{\begin{center}}
\newcommand{\ece}{\end{center}}
\newcommand{\bt}{\begin{table}}
\newcommand{\et}{\end{table}}
\newcommand{\btb}{\begin{tabular}}
\newcommand{\etb}{\end{tabular}}

\newcommand{\calP}{\ensuremath{\mathcal{P}}}
\newcommand{\calD}{\ensuremath{\mathcal{D}}}
\newcommand{\calM}{\ensuremath{\mathcal{M}}}

\newcommand{\calE}{\ensuremath{\mathcal{E}}}
\newcommand{\calO}{\ensuremath{\mathcal{O}}}

\newcommand{\calQ}{\ensuremath{\mathcal{Q}}}
\newcommand{\calT}{\ensuremath{\mathcal{T}}}
\newcommand{\calW}{\ensuremath{\mathcal{W}}}
\newcommand{\ang}{{ang}}

\newcommand{\qed}{\nobreak \ifvmode \relax \else
      \ifdim\lastskip<1.5em \hskip-\lastskip
      \hskip1.5em plus0em minus0.5em \fi \nobreak
      \vrule height0.75em width0.5em depth0.25em\fi}

\begin{document}

\preprint{APS/123-QED}

\title{Geometric optics in general relativity using bilocal operators}
\author{Michele Grasso}
\email{grasso@cft.edu.pl}
\author{Miko\l{}aj Korzy\'nski}%
 \email{korzynski@cft.edu.pl}
 \author{Julius Serbenta}%
 \email{julius@cft.edu.pl}
\affiliation{%
 Center for Theoretical Physics, Polish Academy of Sciences,
Al. Lotnik\'ow 32/46, 02-668 Warsaw, Poland
}%

\pacs{04.20.-q, 98.80.-k, 95.10.Jk, 97.10.Vm, 97.19.Wn}

\begin{abstract}
We consider the standard problem of observational astronomy, i.e. the observations of light emission from a distant region of spacetime in general relativity. The goal is to describe the changes between the measurements of the light performed by a sample of observers slightly displaced with respect to each other and moving with different 4-velocities and 4-accelerations. In our approach, all results of observations can be expressed as functions of the kinematic variables, describing the motions of the observers and the emitting bodies with respect to their local inertial frames, and four linear bilocal geodesic operators, describing the influence of the spacetime geometry on light propagation. The operators are functionals of the curvature tensor along the line of sight.  The results are based on the assumption that the regions of emissions and observations are sufficiently small so that the spacetime curvature effects are negligible within each of them, although they are significant for the light propagation between them. The new formulation provides a uniform approach to optical phenomena in curved spacetimes and, as an application, we discuss the problem of a fully relativistic definition of the parallax and position drifts (or proper motions).  We then use the results to construct combinations of observables which are completely insensitive to the motion of both the observer and the emitter. These combinations by construction probe the spacetime geometry between the observation and emission regions and in our formalism we may express them as functionals of the Riemann tensor along the line of sight. For short distances one of these combinations depends only on the matter content along the line of sight. This opens up the possibility to measure the matter content of a spacetime in a tomographylike manner irrespective of the motions of the emitter and the observer.
\end{abstract}

\pacs{04.20.-q, 04.25.D; 98.80.-k, 95.10.Jk, 97.10.Vm, 97.10.Wn}

\maketitle


\section{Introduction}

In the previous paper \cite{Korzynski:2018} the expressions for the position, redshift, Jacobi matrix, luminosity and angular distance drift, valid in any spacetime, have been derived assuming only that light propagation can be described using the geometric optics approximation. In general relativity, this simply means that light or gravitational radiation follows the null geodesics and does not influence the spacetime geometry \cite{PhysRev.166.1263, PhysRev.166.1272}. 
This paper extends the results of the previous one by including the GR corrections to the parallax and reformulates all derivations in a simple and convenient, geometric way. We approach here in a general way the typical situation in observational astronomy: in a region of spacetime, whose size is small with respect to the spacetime curvature scale, we have one or more bodies emitting electromagnetic radiation, called emitters or sources. The region will act like a ``stage'', where
various physical processes lead to the emission of radiation from certain points in the spacetime. Depending on the details of the situation the sources can be considered infinitesimally small points or extended luminous objects. They
may move through the spacetime with arbitrary, not necessary geodesic motion and their emission may be continuous or pulsed. 
Far away, in another small region (the ``auditorium'') this radiation is received by a number of observers, again moving in an arbitrary way. Both regions can be considered effectively flat due to their size, but the spacetime between them cannot. The observers register the moment of receiving a signal, measured by their proper time, and the direction from which they have seen it coming (the position on the sky). They are
also able to compare these positions between each other and measure the rate of change of that position across the sky in their proper time (the position drifts, or the proper motions in astronomical terminology).

All observables in question obviously depend on the motions of both the sources and the observers, giving rise to the well-known effects
of the Bradley (or stellar) aberration, parallax, relativistic time dilation etc. In a flat spacetime, they are easy to understand within the framework of special relativity. However, in a general, curved spacetime the results of observations will certainly depend also on the geometry of the spacetime in a nontrivial way.
Recall that gravity affects the light rays by the gravitational ray bending, which results in gravitational lensing, i.e. the distortion and (de)magnification of the images in the background, as well as the delays in the electromagnetic signal arrival times.
We may, therefore, expect it to affect the position drift effects (by the position- and time-dependent light bending and delays) as well as the parallax. Thus all results of observations registered by the observers will inevitably contain contributions from the momentary motions of both the observer and the emitter as well as various effects of light propagation in a curved spacetime. The goal of this paper is to understand the dependence of these results on the spacetime geometry and the motions of the observers and sources in the most general situation.

The key observation from the theoretical point of view is that if the two regions in question are sufficiently small, then the influence of the spacetime geometry on light propagation can be understood using
the first order geodesic deviation equation around a null geodesic. Geometrically the problem becomes the question of the behavior of null geodesics contained in a narrow, 4-dimensional tube built around an arbitrary chosen, fiducial null geodesic connecting both regions. 
Irrespective of the global spacetime geometry,  the geometry within the narrow connecting tube is close to the flat Minkowski geometry plus small corrections due to the curvature \cite{Blau:2006ar}.
These corrections affect the null geodesics and produce the well-known optical effects of linear shape distortions, magnification etc. \cite{Wambsganss1998, schneider-kochanek-wambsganss}. As we will show in this paper, this way of conceptualizing the geometric optics in GR has many advantages: mathematically all resulting effects of light propagation can be encoded in four linear mappings, or bitensors,
called the bilocal geodesic operators, that map the tangent space on one side of the connecting geodesic to the tangent space on the other side. Thanks to the geodesic deviation equation these bilocal mappings can be expressed as functionals of the Riemann curvature tensor along the null geodesic. They can be defined covariantly, without invoking any particular coordinate system or frame. They relate the behavior of null geodesics near the observer's and emitter's end of the narrow tube. Once these mappings are known, including the effects of motions of the observer and the sources is then fairly straightforward, since both the stage and the auditorium are effectively flat regions: we simply need to apply the machinery of special relativity formulated in a geometric manner to calculate the observables. If we additionally assume that we may use the distant observer approximation, in which the perspective distortions are absent and the light cone structure of the spacetime is simplified, we find out that only a part of the four bilocal geodesic operator matters. The relevant information turns out to be
included in two optical operators, constructed in a simple way from the bilocal geodesic operators. The first of them is the well-known Jacobi operator, relating a small
deviation of the direction between two close null geodesics passing through the observation point to the transverse spatial distance between the same geodesics further away. The second one, the emitter-observer asymmetry operator, is related to the parallax effects and the position drift. It has a more complicated geometric interpretation which we shall elucidate in this paper. The natural domains of both operators are appropriate quotient spaces whose elements represent the null geodesics in a reparametrization-invariant way.

The advantage of this approach is that the whole problem can be formulated in a completely covariant and frame-independent way. Therefore in the resulting
expressions for observables, we can clearly separate the dependence on the spacetime geometry (via the curvature along the line of sight) and the dependence
on the momentary motions (i.e. the exact positions, the 4-velocities and the 4-accelerations of the emitter and the observer with respect to their local inertial frames). In the derivations we do not refer to any external structures like a 3+1 splitting
of the spacetime, preferred frames such as the statistical isotropy frames, a background (conformal)  metric, or (conformal) Killing vectors. In general relativity motions are not absolute, so it is possible to derive expressions for observables without invoking any fixed, external reference frames provided this way. Moreover, since we consider geodesics displaced in all 4 dimensions, including time, the formalism includes also the time dependence of the observables, i.e. the drift effects. 

\paragraph{Applications.} The most straightforward application of the mathematical machinery developed in this paper is the study of the parallax and the position drift effects in general relativity and cosmology. Formulas derived here express all possible observables in terms
of the optical operators. These operators, in turn, are given by solutions of a matrix ordinary differential equation (ODE) with the curvature tensor along the line of sight playing the role of the input data. 
Given a single null geodesic in a spacetime, obtained exactly, perturbatively or by numerical integration, we may then integrate
the appropriate matrix ODE's. The solution defines then two bilocal optical operators, which encode all the relativistic light bending effects near the null geodesic. The parallax and the position drift (proper motion)
can be then calculated easily given the details of the motions of the source and the observer, i.e. their momentary 4-velocities and the observer's 4-acceleration. 

This method of calculating the drift and the parallax is particularly useful in numerical relativity since the problem of position drift in a numerically evolved spacetime becomes a question of solving a number of linear ODE's using geometric
data collected along a null geodesic. The problem of the position drift and parallax in the context of cosmological distances  has recently been considered by many authors 
in the context of so-called ``real-time cosmology'' \cite{Quercellini:2008ty, quercellini, rasanen}: it has been shown that the position and redshift drifts provide an additional set of observable data we may use to probe large-scale matter flows, inhomogeneities in the matter distributions and constraint this way cosmological models \cite{quercellini,Quercellini:2008ty,Fontanini:2009qq,Quercellini:2009ni,Krasinski:2012ty, Krasinski:2011iw, Krasinski:2012nw, Krasinski:2010rc, KrasiNSki:2015nta}.
 Moreover, numerical relativistic simulations are currently becoming one of the most important tools of theoretical cosmology 
 \cite{Bentivegna:2012ei, Yoo:2013yea, Bentivegna:2013jta, Yoo:2014boa, Bentivegna:2015flc, Giblin:2015vwq, Mertens:2015ttp, Adamek:2015eda, Adamek:2016zes, Macpherson:2016ict, Macpherson:2018akp}.  
 While the null geodesic tracking is a fairly standard problem in numerical relativity \cite{Vincent:2012kn, Bentivegna:2016fls}, extracting the position drift effects is not. It is, of course, possible to do it using ray tracing together with the shooting method: we begin with one null geodesic connecting the source and the observer and then search by trial and error for another one, connecting observer's and emitter's worldline at a slightly later moment. This procedure is, however, rather cumbersome, while the results of this paper and \cite{Korzynski:2018} offer a relatively simple method to do it.

We also note that the formulation of geometric optics presented here offers a uniform theoretical approach to various optical phenomena connected with light propagation in
a curved spacetime. In particular, since the gravitational lensing, the parallax effects and the position drift effects are all within reach of the framework described above, it offers the possibility to study general and nonperturbative relations between them, valid independently of the details of the spacetime geometry.
As an example, in Sec. \ref{sec:physical} we use the optical operator-based approach to compare and contrast various definitions of parallax in the context of general relativity appearing in the literature and point out their relation to the position drift. We also apply it to study the behavior of the gravitational lensing, the parallax and the drift effects in the vicinity of a caustic,
noting their simultaneous blowup in the general case. 

Finally, a less obvious application is connected with the longstanding problem of determining the spacetime geometry from the optical observations \cite{ELLIS1985315, Fanizza:2015swa}. 
Assume that a known physical process takes place on the stage, such as a type Ia supernova explosion, emission of gravitational waves by a binary system, or even less
energetic ones like a simple motion of one or more luminous bodies. An observer in the auditorium region identifies it and performs a
number of observations using various auxiliary observers contained within the auditorium region. It is assumed that the details of the positions and motions of those auxiliary observers with respect to a local inertial frame are known. The observations do not have to be momentary, in general, they will also involve the time variations of the standard observables. The observer then compares the results
of all these observations with the (inferred) course of events in the stage region, as described in the emitter's own frame. 
The results will obviously depend on the relative motions of the observers and the emitter and their motion with respect to the gravitational field. In standard astronomical or cosmological measurements, such as determining the redshift or luminosity distance, this dependence must be taken into account when interpreting the observational data. Nevertheless, as we will see, it is possible to craft certain observational strategies and define specific combinations of observables in which the dependence on momentary motions on both sides cancels out completely. These combinations depend only on the geometry of the spacetime between the two regions. Since in our formalism the dependence of 
the observables on the momentary motions is explicit, finding these combinations is rather straightforward.
It is also easy to show that they can be expressed as functions of the two optical operators, which in turn constitute fairly simple functionals of the Riemann tensor.
Therefore these combinations effectively probe the value of the curvature along the line of sight, allowing this way 
to discriminate between various models of the spacetime geometry or providing direct information about the tidal forces or the mass distribution within the connecting tube.
Measurements of this kind may be called \emph{direct optical measurements of curvature}. An example of such measurement, based on the notions of parallax distance and 
angular diameter distance, is described in Sec. \ref{sec:peculiar}.

\paragraph{Limits of applicability. } We assume that signal propagation can be treated within the geometric optics approximation. This means that we require the radiation wavelength to be much smaller than the size of the regions considered, and therefore much smaller than the curvature radius scale of the whole spacetime, and that the intensity of the radiation is small enough that its contribution to the stress-energy tensor does not disturb the underlying spacetime geometry \cite{PhysRev.166.1263, PhysRev.166.1272}. Wave effects can be then added as small, frequency-dependent corrections if necessary  \cite{Harte:2018wni}.
On top of that, we assume that the width of the connecting tube is small with respect to the curvature radius, which means that the first order geodesic equation approximation is valid for those geodesics connecting the two regions which are contained within the tube.  This assumption holds if the curvature is small with respect to the width of the tube and roughly constant across each of its cross sections, although
it may vary strongly and rapidly along the tube. The tube itself may be arbitrary long and it may also pass through
strongly curved regions of spacetime.

We will focus in this paper on the case of regions positioned sufficiently far away that we may apply the distant observer approximation. 
Within this approximation, we assume that the relation between null tangent vector and the observed position of an object on the observer's sky can be linearized
around the fiducial geodesic and that the condition for geodesics to be null can be linearized as well. In a flat spacetime, these assumptions work very well when
the emitter is positioned sufficiently far away from the observers as measured in the observer's frame. In a nonflat spacetime, with strong lensing between the two regions considered, the applicability of this approximation is a more complicated issue and we discuss their
limits of validity in more detail in Sec. \ref{sec:doa}. 
In most astrophysical applications of GR, we expect the distant observer approximation to work fairly well. Note also that the position drift formula (\ref{eq:Positiondrift}) as a formula for the momentary proper motion holds independently of the distant observer approximation, as explained in Sec. \ref{sec:positiondriftformula}.

Note that we only linearize the equation for neighboring geodesics and the expressions for observables in the three transverse directions, but not along the connecting tube. 
This means that all the curvature effects (lensing, signal delays etc.) along the connecting tube are considered exactly, without any approximations. In particular, the formalism captures
the inherently nonlinear way the curvature corrections accumulate as light passes through various regions of the spacetime. 
 
 Throughout this work, we assume that the signals are of electromagnetic nature. However, the results should also apply to the gravitational radiation as long
as other assumptions listed above hold as well. Most results should also hold in any other, modified theory of gravity as long as the signal propagation follows the null geodesics
of a Lorentzian metric, with the only exception of the results of Sec. \ref{sec:peculiar}, in which we make use of the Einstein equations.

\paragraph{Structure of the paper. } The paper is structured as follows. In the next section, we formulate the physical problem and introduce the necessary mathematical machinery, including the displacement vectors, direction deviation vectors and the first order geodesic deviation equation which
connects those vectors in both regions. We define the bilocal geodesic operators obtained from the general solution of the geodesic deviation equation,
explain their relation to the curvature tensor and prove a number of algebraic relations satisfied by them. 
Finally, we discuss the relation of the geometric objects introduced above to the results of measurements by arbitrary observers, defining also the notion
of a seminull frame connected to an observer.
The main technical results of Sec. \ref{sec:geometricsetup}, used later throughout the paper, consist of the linear relations (\ref{eq:positiondeviation1})-(\ref{eq:directiondeviation1}) linking the displacement and direction deviation vectors in the stage and the auditorium regions, the direct relations between the bilocal geodesic operator and the curvature (\ref{eq:AODE1})-(\ref{eq:WLLODE}) and the equation for the apparent position on the observer's sky (\ref{eq:position1}).

In Sec. \ref{sec:infinitesimaldisplacement} we introduce first the distant observer approximation and note that under its assumptions we may pass to the appropriate quotient spaces when discussing the displacement and direction deviation vectors, getting rid this way of the gauge degrees of freedom. 
We then prove the two most important technical results of the paper, the time lapse formula (\ref{eq:timelapse1}) and the direction deviation equation (\ref{eq:directiondeviationabstract}), linking the behavior of a perturbed null geodesic in the stage and the auditorium regions.
We introduce for that purpose the two optical operators $\calD$ and $m$ and derive their general relation to the Riemann curvature tensor.

Sec. \ref{sec:physical} contains most of the physical results of the paper, including the results about the relativistic corrections to the parallax. We begin by a clarification of various definitions of parallax in general relativity. 
We then discuss the relation of the Jacobi operator $\calD$ to the magnification matrix given by Eq. (\ref{eq:maginficationmatrix}) (a standard result) and the relation
of the emitter-observer asymmetry operator $m$ to the parallax given by Eq. (\ref{eq:parallaxmatrix}). We then rederive the Eq. (\ref{eq:Positiondrift}) for the position drift using the optical operators and
finally consider the most physically relevant case of an observer and an emitting body contained both in gravitationally bound systems undergoing a geodesic motion
as a whole. We show that the time-dependent position drift with respect to the center of mass frame consists of the proper motion term given by the position drift formula plus
the expected parallax terms, as in Eq. (\ref{eq:driftplusparallax}).  Finally, we discuss the behavior of the parallax, the image magnification and the position drift in the vicinity of a caustic.

In Sec. \ref{sec:peculiar} we show how one can combine the data about the parallax and about the image magnifications and distortions into observables which
do not depend on the momentary motions on both the emitter's and the observer's end of the fiducial geodesic. The combinations
are defined by Eqs. (\ref{eq:wperpdef}) and (\ref{eq:mudef}). We show how these observables can be related to the
curvature and the matter content along the line of sight and propose also a simple measurement of the amount of gravitating matter along the line of sight. It is based on Eq. (\ref{eq:muintegral3}), relating one of those observables to a weighted integral of the stress-energy tensor along the line of sight for short distances.

We conclude with a ummary.

\paragraph{Notation. } In this paper we will use both coordinate bases and bases related to special frames. Within these frames, we will need to separate out sets of 3 and 2 components. 
We, therefore, need to distinguish 4 types of indices in the paper. 
 Greek letters $\mu,\nu,\dots$ run from 0 to 3 and will denote components expressed in coordinate bases in the spacetime $\calM$. Boldface greek indices $\bm\mu, \bm\nu, \dots$ also run from 0 to 3, but will be used for geometric objects expanded in frames.  
The boldface Latin indices $\bm i, \bm j,\dots$ run from 0 to 2 and denote the first three components in a frame. Finally the boldface capital latin indices
$\bm A, \bm B, \dots$ denote the 1 and 2 components expressed in a frame. $\delta\UD{\mu}{\nu}$ will denote the standard Kronecker delta with any set of indices.

Boldface letters will also be reserved for geometric objects in quotient spaces $\calQ_p$ and $\calP_p$ defined in Sec. \ref{sec:infinitesimaldisplacement}, while objects defined in tangent spaces will be denoted by standard letters.

Throughout the text, we assume the speed of light $c=1$. 

\section{Geometric setup} \label{sec:geometricsetup}

\paragraph{Description of the physical problem. }Consider two regions in the spacetime, the stage and the auditorium, separated by a large distance, but at the same time causally connected. Both regions
by definition extend in both space and time, but we assume they are small enough that their local geometry can be treated as flat.
On the other hand, the region connecting them is curved, although the curvature is small enough so that light propagation can be treated within the first order geodesic deviation equation. We assume that from every point in the stage region it is possible to find a null geodesic which
crosses the auditorium in the future.
In this setup the communication between the two regions is possible only one way: light signals emitted from one of them, the stage, can reach later the
other one, the auditorium, provided that they have been
sent in the right direction. 

We assume that each of the regions hosts one or more participants who travel through the spacetime along timelike worldlines (not necessary geodesic). The emitters
in the stage region send time-dependent electromagnetic radiation at all directions, while the observers in the auditorium perform the
standard astronomical observations, registering the energy and the direction from which they perceive the light coming at a given moment of their worldline, as well
as their variations in their own proper time.
The problem we will be concerned with in this paper can be phrased in the following way: knowing the trajectories of the emitters and the observers in their locally flat regions, for example with respect to a local inertial frame, describe the results of the observations made by the observers in the geometric optics approximation. 
 
 As we have already noted in the Introduction, the results will depend on the motions of both the emitters and the observers and also on the properties of the spacetime between them. In this paper, in the spirit of \cite{Korzynski:2018}, we would like to separate these dependencies.  Our aim is to obtain expressions for the observables which contain the quantities describing the
momentary motions on both sides (i.e. the 4-velocities, 4-accelerations and momentary positions with respect to the local inertial frames) as well as geometric objects describing the effects of gravitational fields on the light propagation and defined in a frame- and coordinate-independent way.
The problem may seem at first sight impossible to tackle in full generality, without any assumptions about the spacetime geometry. 
Recall that in general relativity the light propagation is affected by gravity in many ways: variations of the gravitational field induce the gravitational light bending,
causing also the Shapiro and the geometric delays in the signal arrivals. On top of that, the gravitational fields may also increase or decrease the energy of the photon. In a
time-dependent gravitational field, all effects may also be time- and position-dependent,
influencing this way the time variations of the observables recorded by the observers.
In the most general approach, without any simplifying assumptions, the light propagation in the geometric optics approximation is governed by the null geodesic equation which contains the Christoffel symbols \cite{PhysRev.166.1263}.
The Christoffel symbols, in turn, depend on the values of the metric tensor and its first derivatives. It seems then that without further assumptions about the metric little can be
said about the relations between the motions of the participants in both regions and the results of observations.

Fortunately, the problem simplifies very much when we realize that all the information about the gravitational fields we need is actually the geometry
inside the narrow tube connecting the two regions. This can be seen as follows:
let us single out for the sake of convenience a pair of points, $\calE$ in the stage region and $\calO$ in the auditorium,  connected by a null geodesic 
 $\gamma_0$ representing a single ray of light. The two points will serve as a reference for the positions of objects in the corresponding regions, while the null geodesic $\gamma_0$ will play the role of the fiducial null geodesic for all other light rays. We then build a long, thin 4-dimensional tube around it, extending in both the 2 spatial dimensions as well as time. Its with should be comparable to the size of both regions. Now, one can show that as long as the connecting tube is narrow with respect to the typical curvature radius of the manifold, the geometry of the spacetime inside it is quite simple: it must be close to the flat space up to the leading order corrections proportional to the value of the Riemann tensor along $\gamma_0$. This may seem slightly surprising at first, but we can see that clearly if we introduce the equivalent of the 
 Fermi coordinates in the neighborhood of $\gamma_0$. 
 
 The Fermi coordinates around a timelike geodesic are well known and described in many textbooks \cite{Poisson2011},
 but in the case of a null geodesic they are less known \cite{Blau:2006ar}. In short, given the null geodesic $\gamma_0$ and an appropriate parallel propagated frame along it one constructs coordinates $(\xi^i, \lambda)$, consisting of a 
 coordinate $\lambda$ agreeing with the affine parametrization of $\gamma_0$ and three transverse coordinates $\xi^i$, such that
 $\xi^i =0$ corresponds to $\gamma_0$.  Unlike the timelike case, all three transverse coordinates cannot be made orthogonal to $\gamma_0$. Nevertheless, the metric tensor near $\gamma_0$ can be expanded as a Taylor series in $\xi^i$ just like in the ``standard'' Fermi coordinates \cite{Blau:2006ar}:
 \bea
 g_{\mu\nu} = C_{\mu\nu} + D(\mu,\nu)\,R_{\mu i j \nu}(\lambda)\, \xi^i\,\xi^j + O(|\xi|^3), \label{eq:nullFermi}
 \eea
 where $C_{\mu\nu}$ is a constant matrix representing the flat space in non standard coordinates, $R_{\mu i j \nu}(\lambda)$ denotes
 the components of the Riemann tensor along $\gamma_0$ in the parallel propagated frame and $D(\mu,\nu)$ are irrelevant constant coefficients depending on the indices $\mu$ and $\nu$.  Thus the leading order term in the expansion turns out to be always constant, corresponding to a flat tube cut out of the Minkowski spacetime, plus subleading curvature corrections quadratic in the transverse coordinates. We have shown this way that while in a general coordinate system the fiducial geodesic may have a very complicated form, resulting in a complicated form of the metric in its neighborhood, 
the Fermi coordinates can ``unwind'' the geodesic together with the nearby spacetime geometry, revealing that is has a fairly simple form of a  
flat tube with slight deformations seen in its outer parts.

 In this setting \emph{all} the light propagation effects (gravitational light bending, propagation delays etc.), taking place between the stage and the auditorium, can be understood in terms of the geometry of the connecting tube,  given by (\ref{eq:nullFermi}), and its influence on null geodesics. 
 Consequently, all the relativistic corrections to the light propagation effects should be \emph{expressible as functionals of the only relevant ingredient of the spacetime geometry, i.e. the Riemann tensor along the line of sight}. The geometry outside the connecting tube on the other hand, possibly very complicated and not expressible easily in terms of data defined along $\gamma_0$, is irrelevant from the point of view of the
problem. Let us stress again that the statements above hold for \emph{arbitrary} geodesics and along their \emph{whole} length, even if they are long enough to feel the global spacetime curvature or they pass through
strongly curved regions. The global decomposition (\ref{eq:nullFermi}) of the narrow tube geometry works for those geodesics equally well as for short ones.

The main mathematical tool of this paper, applicable under the assumptions defined above to the problem of null geodesics connecting both regions, is the first order geodesic deviation equation (GDE).
Recall that it is a linear ODE relating directly the behavior of neighboring geodesics to the Riemann tensor along a fiducial geodesic we use as a reference \cite{Bazanski1, Bazanski2, Vines:2014oba, Uzun:2018yes}.  One may obtain it from the geodesic equation in metric (\ref{eq:nullFermi}) by neglecting the quadratic terms in the velocity deviations. 

The geodesic deviation equation of the first and higher orders (and their generalizations) around a timelike geodesic, representing the relative motion of free falling bodies,
has been discussed by many authors (see for example \cite{pirani, szekeres, Bazanski1, Bazanski2, Aleksandrov1979, CiufoliniDemianski, Vines:2014oba, Puetzfeld:2015uxi, Flanagan:2018yzh}), but the null case has attracted less attention \cite{Bartelmann_2010, Clarkson:2016zzi, Clarkson:2016ccm, Korzynski:2018, Uzun:2018yes}.
The general solution of the first order GDE around a null geodesics  allows for studying the behavior of the rays of light in a completely covariant  manner, i.e. without invoking explicitly a
coordinate system or a frame.
This is consistent with the fully relativistic and geometric point of view we adopt in this paper: we delay the introduction of coordinate systems, frames or other auxiliary structures as much as possible in order
to understand the ``pure'' geometric relations, valid independently of them.

\paragraph{Mathematical formulation of the problem.} Let $\calM$ be the spacetime with a Lorentzian metric $g$, of signature $(-,+,+,+)$. We consider two regions of spacetime: the stage $N_\calO$ and the auditorium $N_\calE$.  Both are 4-dimensional domains in the vicinity of two points $\calO \in N_\calO$ and $\calE \in N_\calE$ respectively. We assume
that $\calO$ and $\calE$ are connected by a null geodesic $\gamma_0$, called the \emph{fiducial geodesic} (or the \emph{optical axis} in nonrelativistic optics literature), i.e. $\gamma_0(\lambda_\calO) = \calO$ and $\gamma_0(\lambda_\calE) = \calE$ for
the values $\lambda_\calO$ and $\lambda_\calE$ of the affine parameter $\lambda$  of $\gamma_0$. For convenience we also assume that we have parametrized our geodesic backwards in time, i.e.
$\lambda_\calO < \lambda_\calE$. As null geodesics do not have a preferred, normalized parametrization, we may always reparametrize $\gamma_0$
by an affine transformation
without violating the assumptions above: 
\bea
 \lambda \to \lambda' = E\cdot \lambda + F \label{eq:affinereparametrization},
 \eea
 with $E > 0$ and arbitrary $F$. Then the tangent vector to $\gamma_0$ transforms according to
 \bea
  l^\mu \to l'^\mu = \frac{1}{E}\,l^\mu. \label{eq:laffine}
 \eea
In geometric optics we are interested only in the incidence relations between null geodesics and points, i.e. in the question whether or not a null geodesic passes through
a given event and what null direction it follows at that moment. These relations are invariant with respect to reparametrization, because the value of the affine parameter 
at which the null geodesic intersects a given point carries no physical meaning. Therefore we may identify all null geodesics which share the same path and consider
affine reparametrizations (\ref{eq:affinereparametrization})  gauge transformations from the point of view of geometric optics. They should therefore  leave all physical observables invariant.

The regions $N_\calO$ and $N_\calE$ are assumed to be of the size of $L$ much smaller than the characteristic curvature scale of the spacetime $R_c$, i.e. $L \ll R_c$.
In this case, we may introduce in both regions locally flat coordinate systems, centered at $\calO$ and $\calE$ respectively, in which the metric tensor is the flat Minkowski metric up to quadratic terms in $x^\mu$:
\bean
g_{\mu\nu} = \eta_{\mu\nu} + O(x^2).
\eean
The additional terms are due to the local spacetime curvature and therefore scale like $(L/R_c)^2$. We may consider them negligibly small because the size of both regions $L$ is too small for any curvature effects to be directly detected by experiments performed within each region. Therefore we will effectively treat 
both $N_\calO$ and $N_\calE$ as flat. Stating the same in a more coordinate-invariant manner: with the curvature effects being negligible in regions of size $L$,  we may 
simply identify the points in $N_\calO$ and $N_\calE$ with points in the tangent spaces $T_\calO \calM$ and $T_\calE \calM$ in the vicinity of $0$, using, for example, the exponential map. Under this identification, the physical spacetime metric agrees with the flat metric on the appropriate tangent space up to quadratic terms in the distance from 0.

\bfi
\includegraphics[width=0.9\textwidth]{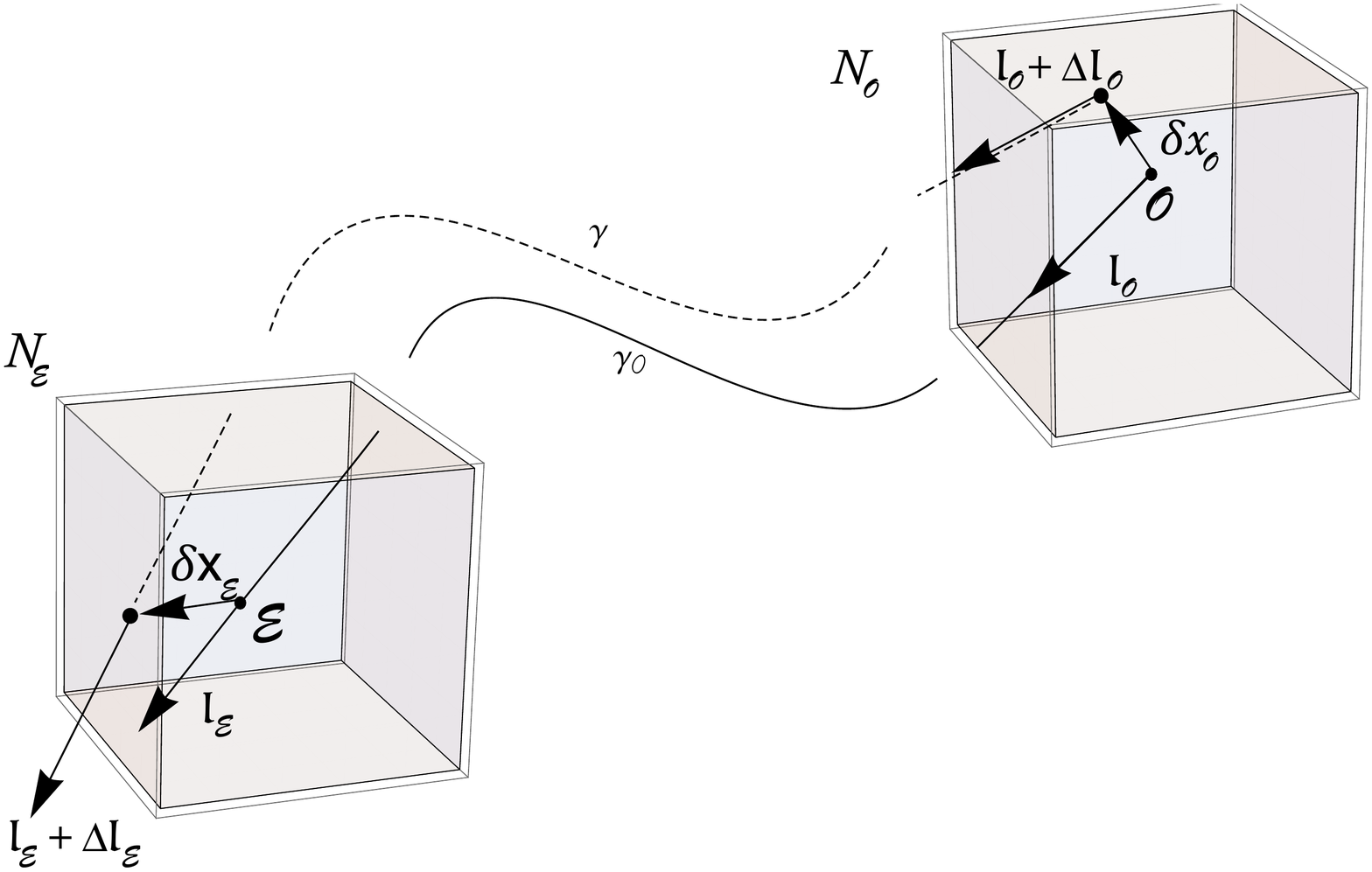}
\caption{Flat, distant regions $N_\calE$ and $N_\calO$ connected by a family of null geodesics. The fiducial null geodesic $\gamma_0$ connects events $\calO$ and $\calE$. We consider
other geodesics lying entirely within a narrow tube around $\gamma_0$. The geodesics may be parametrized by a pair of vectors in the tangent space 
$T_\calO \calM$ giving the initial displacement ($\delta x_\calO$) and the initial direction deviation ($\Delta l_\calO$) with respect to $\gamma_0$. Similar parametrization
$(\delta x_\calE, \Delta l_\calE)$ can also be used in $N_\calE$, at point $\lambda = \lambda_\calE$ along the displaced geodesic. 
}
\label{fig:geometricsetup}
\efi

Consider now all geodesics connecting points from $N_\calO$ with $N_\calE$ contained in a 4-dimensional tube around $\gamma_0$, sufficiently narrow so that we can apply the metric expansion
(\ref{eq:nullFermi}) for its geometry. This means we can use the first order geodesic deviation equation for those geodesics.
 The geodesics are uniquely specified by giving an initial point $i_\calO$ in $N_\calO$
and the initial tangent vector $v_\calO$ in $N_\calO$. In locally flat coordinates they look like straight lines,
although their propagation through the spacetime in between may more complicated, with details depending on the coordinate system, see Fig. \ref{fig:geometricsetup}. They form an 8-parameter family of curves. 

Now, let $l_\calO^\mu$ be the tangent vector to $\gamma_0$ at $\calO$ and $l_\calE^\mu$ at $\calE$. We would like to parametrize the geodesics from the family considered by their deviation from the fiducial null geodesic $\gamma_0$. Obviously we can parametrize every perturbed geodesic $\tilde\gamma$ by
its initial position $p$ at $\lambda = \lambda_\calO$ and the tangent vector $v^\mu$ at $p$. With the point-vector the identification described above we may use simply the 
\emph{initial displacement vector} $\delta x_\calO \in T_\calO \calM$
as the parametrization of the initial position. In a locally flat coordinate system $(y^\mu)$ in $N_\calO$ it is defined as
\bea
\delta x_\calO^\mu =y^\mu\left(\tilde\gamma(\lambda_\calO)\right) - y^\mu\left(\calO\right). \label{eq:deltaxOdef}
\eea
Note that $\delta x_\calO$ is a fully 4-dimensional vector, because we consider here displacements in all 3 spatial directions and in time.

We now move on to $v^\mu$. Fixing an arbitrary coordinate system in $N_\calO$ we may use the difference between the components of $v^\mu$
and $l_\calO^\mu$ to parametrize the tangent vector:
\bea
\delta l_\calO^\mu = v^\mu - l_\calO^\mu \label{eq:deltalOdef}
\eea
in analogy with (\ref{eq:deltaxOdef}). However, $\delta l_\calO^\mu$ defined this way is not a proper vector, because it does not transform like a vector under general coordinate system transformations. In other words, expression (\ref{eq:deltalOdef}) defines, in general, different elements of the tangent space $T_\calO \calM$ when applied in different coordinate systems. As we know from elementary differential geometry, this is because we have subtracted here the components of vectors defined at two close but distinct
 points, $p$ and $\calO$. Of course, this can be fixed by adding an appropriate term involving the Christoffel symbols term: we define
 \bea
\Delta l_\calO^\mu = \delta l_\calO^\mu + \Gamma\UD{\mu}{\nu\sigma}(\calO)\,l_\calO^\nu \,\delta x_\calO^\sigma, \label{eq:DeltalOdef}
\eea
where $\Gamma\UD{\mu}{\nu\sigma}(\calO)$ are the Christoffel symbols at $\calO$. 
The resulting expression $\Delta l_\calO^\mu$ parametrizes the initial tangent vector equally well, but unlike the ``bare'' $\delta l_\calO^\mu$ it is a proper vector. 
From the geometric point of view, Eq. (\ref{eq:DeltalOdef}) yields the \emph{same} vector in $T_\calO \calM$ independently of the coordinate system chosen.
It is obvious, though, that in locally flat coordinate systems like $(y^\mu)$ - and only in those - we have $\Delta l_\calO^\mu = \delta l_\calO^\mu$ and both definitions coincide.
The vector $\Delta l_\calO^\mu$, which we will call the \emph{initial direction deviation vector}, corresponds to the difference between $v^\mu$ parallel transported
from $p$ to $\calO$ and $l_\calO^\mu$. 
The  pair $(\delta x_\calO^\mu, \Delta l_\calO^\mu)$ defines a unique geodesic considered as a perturbation of $\gamma_0$ and will be referred to
as \emph{the displacement vectors}, see again Fig. \ref{fig:geometricsetup}.

\paragraph{Remarks. }Since the geodesics are supposed to be confined within the narrow tube all along $\gamma_0$ both the initial displacement and the initial direction deviation cannot be too large. Additionally, the condition for applicability of the first order GDE means that the gravitational lensing produces only a linear distortion of the image of all objects in $N_\calE$, which excludes the possibility of multiple imaging for light rays contained within the tube. 

The choice of the particular value $\lambda_\calO$ of the affine parameter $\lambda$ at which we parametrize the initial displacement and direction deviation is arbitrary: in principle we could choose any point and any value for that purpose. This choice is nevertheless consistent with the assumption that the geodesics are only linearly perturbed with respect to the fiducial one: we may expect that for slightly perturbed geodesics the endpoint given by $\lambda = \lambda_\calO$ will lie very close to the corresponding endpoint of $\gamma_0$, i.e. $\calO$, and therefore within $N_\calO$. Thus we can parametrize the position by a small displacement vector. The same reasoning applies to the other endpoint.

\subsection{Geodesic deviation equation and bilocal geodesic operators} \label{sec:bilocal}
We will now consider the relation between the displacement vectors around $\calO$ and around $\calE$ for geodesics passing through $N_\calO$ and $N_\calE$. 
Assume that for $\lambda = \lambda_\calE$ the deviated geodesic passes through $N_\calE$ and let $\delta x_\calE$ and $\Delta l_\calE$ denote the displacement vector
and the direction deviation vector respectively, see Fig. \ref{fig:geometricsetup}. Assuming the geodesics deviate from $\gamma_0$ only by distances small with respect $R_c$, we may obtain these vectors 
by solving the ODE for the first order perturbation of the geodesic equation around $\gamma_0$, i.e. the first order geodesic deviation equation (GDE):
\bea
\nabla_l \nabla_l \xi^\mu - R\UD{\mu}{\alpha\beta\nu}\,l^\alpha\,l^\beta\,\xi^\nu = 0, \label{eq:GDE}
\eea
with the initial data
\bea
\xi^\mu(\lambda_\calO) &=& \delta x_\calO^\mu  \label{eq:GDEID1}\\
\nabla_l \xi^\mu(\lambda_\calO) &=& \Delta l_\calO^\mu. \label{eq:GDEID2}
\eea
With this setup we obtain the displacements at the other end from the values of the solution for $\lambda=\lambda_\calE$: $\delta x_\calE^\mu = \xi^\mu(\lambda_\calE)$ and $\Delta l_\calE^\mu = \nabla_l \xi^\mu(\lambda_\calE)$. The combination $R\UD{\mu}{\alpha\beta\nu}\,l^\alpha\,l^\beta$ is often referred to as the \emph{optical tidal matrix}.

Since the GDE is linear, the solution at $\lambda_\calE$ must be a linear function of the initial data at $\calO$:
\bea
\delta x_\calE^\mu = { W_{XX} }\UD{\mu}{\nu}\,\delta x_\calO^\nu + { W_{XL} }\UD{\mu}{\nu}\,\Delta l_\calO^\nu \label{eq:positiondeviation1} \\
\Delta l_\calE^\mu = { W_{LX} }\UD{\mu}{\nu}\,\delta x_\calO^\nu + { W_{LL} }\UD{\mu}{\nu}\,\Delta l_\calO^\nu \label{eq:directiondeviation1},
\eea
with $W_{XX}$, $W_{XL}$, $W_{LX}$, $W_{LL}$ being bilocal operators  (also known as 2-point tesors \cite{SyngeBook} or bitensors \cite{Poisson2011, Vines:2014oba}), acting from $T_\calO \calM$ to $T_\calE \calM$. 
Together they form the resolvent operator, or the Wroński matrix \cite{Fleury:2014gha} for the GDE, $W = W(\lambda_\calE,\lambda_\calO)$. It is 
a linear mapping between vector sums of two copies of the tangent space, i.e.
\bean
W: T_\calO \calM \oplus T_\calO \calM \to T_\calE \calM \oplus T_\calE \calM,
\eean
defined by the relation $W(\delta x_\calO,\Delta l_\calO) = (\delta x_\calE,\Delta l_\calE)$. As noticed by
Uzun \cite{Uzun:2018yes}, it is also a symplectic mapping, since in GR the ODE's for null geodesics can be formulated as a Hamiltonian system,
in general as well as in the first order perturbation theory \cite{Fleury:2014gha} \footnote{Note that unlike \cite{Fleury:2014gha} we consider displacements not only in the two perpendicular, spatial directions, but in all 4 possible directions, including the time. The resulting system of ODE's has, therefore, a twice larger dimension. However, the Wroński matrix formalism extends to the fully 4-dimensional GDE without any problems. One can also check that the extended Wroński matrix is symplectic too, but this is beyond the scope of this article.}.

In this paper the four operators $W_{XX}$, $W_{XL}$, $W_{LL}$ and $W_{LX}$ will be referred to as the \emph{bilocal geodesic operators} (BGO). The notation we introduced for them highlights the fact that they constitute four parts of a larger geometric object. In the context of timelike geodesics
the first two have already been introduced by DeWitt and Brehme \cite{DeWittBrehme} and Dixon \cite{Dixon2, Vines:2014oba}, under the name of \emph{Jacobi propagators},  denoted by $K$ and $H$ (the definition of the latter involves often an additional prefactor). They can also be obtained by
differentiating the Synge's world function \cite{SyngeBook, Dixon2, Vines:2014oba}, but here we will not make any use of the world function formalism.
Recently the BGO's defined along a timelike geodesic have been used as a tool to study of the gravitational waves memory effect \cite{Flanagan:2018yzh}. In the rest of the paper, we will focus exclusively on the  null case and explore
the relation of the BGO's to the spacetime geometry and to the optical observations made by observers in $N_\calO$.

\paragraph{Bilocal geodesic operators and the Riemann curvature tensor.}
It follows easily from the geodesic deviation equation (\ref{eq:GDE}) and from Eqs. (\ref{eq:positiondeviation1})-(\ref{eq:directiondeviation1}) that the BGO's can be expressed as solutions to ODE's defined along $\gamma_0$ and involving the curvature tensor. Assume we fix a
parallel-propagated frame $e_{\bm \mu}$ along the null geodesic. Then we solve the following ODE for a 4-by-4 matrix-valued function $A\UD{\bm\mu}{\bm\nu}(\lambda)$ with initial data at $\lambda_\calO$: 
\bea
\ddot A\UD{\bm \mu}{\bm \nu} - R\UD{\bm \mu}{\bm \alpha\bm\beta\bm\sigma}\,l^{\bm\alpha}\,l^{\bm\beta}\,A\UD{\bm\sigma}{\bm\nu} = 0 \label{eq:AODE1}\\
A\UD{\bm\mu}{\bm\nu}(\lambda_\calO) = \delta\UD{\bm\mu}{\bm\nu} \label{eq:AODE2}\\ 
\dot A\UD{\bm\mu}{\bm\nu}(\lambda_\calO) =0, \label{eq:AODE3}
\eea
where dot denotes the derivative with respect to the affine parameter $\lambda$ and $R\UD{\bm \mu}{\bm \alpha\bm\beta\bm\sigma}\,l^{\bm\alpha}\,l^{\bm\beta}$ denotes
the components of the optical tidal matrix in the parallel propagated frame.
From the definition (\ref{eq:positiondeviation1})-(\ref{eq:directiondeviation1}) we can prove that the two BGO's are given in terms of $A\UD{\bm\mu}{\bm\nu}(\lambda)$ and
its derivative at the emission point:
\bea
{W_{XX}}\UD{\bm\mu}{\bm\nu} = A\UD{\bm\mu}{\bm\nu}(\lambda_\calE)  \label{eq:WXXODE} \\
{W_{LX}}\UD{\bm\mu}{\bm\nu}=\dot A\UD{\bm\mu}{\bm\nu}(\lambda_\calE). \label{eq:WXLODE}
\eea
The other two operators can be obtained from the solution of the same equation with different initial data, namely we take the matrix-valued function $B\UD{\bm\mu}{\bm\nu}(\lambda)$ satisfying
\bea
\ddot B\UD{\bm\mu}{\bm\nu} - R\UD{\bm\mu}{\bm\alpha\bm\beta\bm\sigma}\,l^{\bm\alpha}\,l^{\bm\beta}\,B\UD{\bm\sigma}{\bm\nu} = 0 \label{eq:BODE1}\\
B\UD{\bm\mu}{\bm\nu}(\lambda_\calO) = 0 \label{eq:BODE2}\\
\dot B\UD{\bm\mu}{\bm\nu}(\lambda_\calO) = \delta\UD{\bm\mu}{\bm\nu} . \label{eq:BODE3}
\eea
In that case we have
\bea
{W_{XL}}\UD{\bm\mu}{\bm\nu} = B\UD{\bm\mu}{\bm\nu}(\lambda_\calE)  \label{eq:WLXODE} \\
{W_{LL}}\UD{\bm\mu}{\bm\nu}=\dot B\UD{\bm\mu}{\bm\nu}(\lambda_\calE). \label{eq:WLLODE} 
\eea

Relations presented above clarify the dependence of the BGO's on the spacetime geometry. They show that they can be expressed as nonlocal functionals of the Riemann tensor along the line of sight.  
We stress here that even though the GDE and the matrix equations
(\ref{eq:AODE1})-(\ref{eq:BODE3}) are linear, the BGO's are \emph{nonlinear} functionals of the curvature tensor along $\gamma_0$. 
The reader may check that if we take two solutions of (\ref{eq:AODE1})-(\ref{eq:AODE3}) or (\ref{eq:BODE1})-(\ref{eq:BODE3}) corresponding to two different optical tidal tensor functions
$R\UD{\bm \mu}{\bm \alpha\bm\beta\bm\sigma}\,l^{\bm\alpha}\,l^{\bm\beta}(\lambda)$, then a linear combination of these solutions \emph{does not} satisfy the same Eqs.
 (\ref{eq:AODE1})-(\ref{eq:AODE3}) or (\ref{eq:BODE1})-(\ref{eq:BODE3}) for the same linear 
combination of the optical tidal tensor functions.  The situation is  analogous to the dependence of the evolution operator $U(t)$ on the Hamiltonian $H(t)$ in quantum mechanics: $U(t)$ is defined by
a first order ODE and initial condition
\bea
i \hbar \dot U(t) &=& H(t)\,U(t) \\
U(0) &=& {\bm 1}.
\eea
Now, adding a perturbation term to the Hamiltonian results a nonlinear change of the evolution operator, which can be expressed using the well-known path-ordered exponential formula \footnote{Another example for this nonlinearity is the second order vector differential equation $\displaystyle \frac{d^2 \mathbf x}{d\mathbf \lambda^2} + A \mathbf x = 0$, where $A$ is a constant diagonalisable matrix with positive eigenvalues. The general solution has the following form: $\mathbf x = \cos{[\sqrt{A} (\lambda - \lambda_0)]}{\mathbf x}_0 + \frac{1}{\sqrt{A}}\sin{[\sqrt{A}(\lambda - \lambda_0)]} \dot{\mathbf x}_0$. Even if $A$ enters the differential equation linearly, the general solution contains a nonlinear dependence.}.

The nonlinearity of the BGO's as functionals of curvature reflects the fact that our formalism captures all nonlinear effects of combined light bending at different points along the fiducial geodesic.
Note also that although all four BGO's are effectively functionals of the \emph{same} optical tidal tensor $R\UD{\bm \mu}{\bm \alpha\bm\beta\bm\sigma}\,l^{\bm\alpha}\,l^{\bm\beta}(\lambda)$,  they are in fact \emph{different functionals}  and therefore without any further assumptions regarding the curvature their values
for any pair of points $\calO$ and $\calE$ should be treated as completely independent from each other.

\paragraph{Algebraic properties of the BGO's.} In \cite{Vines:2014oba, Korzynski:2018} two general properties of the solutions of the GDE have been proved. We will show here that they immediately translate to
corresponding two properties of the bilocal geodesic operators, which hold irrespective of the spacetime geometry or whether $\gamma_0$ is null or not.

Let $\xi^\mu$ denote a solution of the GDE (\ref{eq:GDE}). Then we have 
\bean
\xi^\mu \,l_\mu = A + B\,\lambda,
\eean
where $A,B = \const$. This way we have defined 2 constants of motion for the GDE, namely $B = \nabla_l\xi^\mu\,l_\mu$ and $A = \xi^\mu \,l_\mu - B\,\lambda$.

The second property is that the expression
\bea
\xi^\mu = (C + D\,\lambda)\,l^\mu \label{eq:secprop}
\eea
with $C,D = \const$ is always a solution and, since the first order GDE is linear, it can be added to any other solution without
affecting equation (\ref{eq:GDE}). Both properties are easy to verify using Eq. (\ref{eq:GDE}).

Consider the solution (\ref{eq:secprop}) at $\calO$ and $\calE$: we have $\delta x_\calO^\mu = (C + \lambda_\calO\,D)\,l_\calO^\mu$, 
$\Delta l_\calO^\mu = D\,l_\calO^\mu$ and $\delta x_\calE^\mu = (C + \lambda_\calE\,D)\,l_\calE^\mu$, 
$\Delta l_\calE^\mu = D\,l_\calE^\mu$. We substitute these equations to (\ref{eq:positiondeviation1})-(\ref{eq:directiondeviation1}) and 
assuming the resulting relations must hold for all $C$ and $D$ we get
\bea
{ W_{XX} }\UD{\mu}{\nu}\,l_\calO^\nu &=& l_\calE^\mu \label{eq:Wprop1}\\
{ W_{LX} }\UD{\mu}{\nu}\,l_\calO^\nu &=& 0 \label{eq:Wprop2} \\
{ W_{XL} }\UD{\mu}{\nu}\,l_\calO^\nu &=& (\lambda_\calE - \lambda_\calO)\,l_\calE^\mu \label{eq:Wprop3} \\
{ W_{LL} }\UD{\mu}{\nu}\,l_\calO^\nu &=& l_\calE^\mu \label{eq:Wprop4}.
\eea
The second set of relations can be obtained from the conservation of the two constants of motion defined above. Namely, for any initial data $\delta x_\calO^\mu$ and 
$\Delta l_\calO^\mu$ the values of $A$ and $B$ need to remain equal in $\calO$ and $\calE$. This means that
\bean
l_{\calO\,\mu}\,\Delta l_\calO^\mu &=& l_{\calE\,\mu}\,\Delta l_\calE^\mu \\
l_{\calO\,\mu}\,\delta x_\calO^\mu - \lambda_\calO\,l_{\calO\,\mu}\,\Delta l_\calO^\mu &=& l_{\calE\,\mu}\,\delta x_\calE^\mu - \lambda_\calE\,l_{\calE\,\mu}\,\Delta l_\calE^\mu.
\eean
We again make use of (\ref{eq:positiondeviation1})-(\ref{eq:directiondeviation1}) in order to express $\delta x_\calE^\mu$ and $\Delta l_\calE^\mu$ by
$\delta x_\calO^\mu$ and $\Delta l_\calO^\mu$. The resulting equations turn out to be equivalent to the following 4 relations:
\bea
l_{\calE\,\mu}\,{ W_{XX} }\UD{\mu}{\nu} 
 &=& l_{\calO\,\nu} \label{eq:Wprop5} \\
l_{\calE\,\mu}\,{ W_{LX} }\UD{\mu}{\nu} &=& 0 \label{eq:Wprop6} \\
l_{\calE\,\mu}\,{ W_{XL} }\UD{\mu}{\nu} &=& (\lambda_\calE - \lambda_\calO)\,l_{\calO\,\nu} \label{eq:Wprop7} \\
l_{\calE\,\mu}\,{ W_{LL} }\UD{\mu}{\nu} &=& l_{\calO\,\mu}.  \label{eq:Wprop8}
\eea 

Finally, let us note that two of the BGO's undergo rescalings under the affine reparametrizations of the fiducial null geodesic $\gamma_0$. Namely, under
the reparametrization
 (\ref{eq:affinereparametrization}) we have the following transformation law:
\bea
 W_{XX} &\to & W'_{XX} = W_{XX} \\
 W_{XL} &\to & W'_{XL} = E\cdot W_{XL} \\ 
 W_{LX} &\to & W'_{LX} = \frac{1}{E} \cdot W_{LX} \\
 W_{LL} &\to & W'_{LL} = W_{LL}.
\eea

\subsection{Observed position on the sky and seminull frames}

Let $u_\calO^\mu$ be the 4-velocity of an observer and $l^\mu$ the past-pointing, null tangent vector to a null geodesic passing through $\calO$. The direction from which the observer sees the light coming is defined as a normalized, spatial vector, orthogonal to $u_\calO^\mu$, pointing in the same direction as $l^\mu$:
\bea
r^\mu = \frac{1}{l_\sigma\,u_\calO^\sigma}\,l^\mu + u_\calO^\mu. \label{eq:directionformula}
\eea
Formula (\ref{eq:directionformula}) defines an observer-dependent mapping
\bean
V(u_\calO,\cdot): {\cal N}^-_\calO \to \textrm{Dir}(u_\calO)
\eean
from the set of past-oriented null vectors ${\cal N}^-_\calO = \left\{ X \in T_\calO \calM | X^\mu\,X_\mu = 0, X^0 < 0 \right\}$ to the
observer's sphere of directions, i.e. the set of normalized, purely spatial vectors for the observer, i.e. $\textrm{Dir}(u_\calO) = \left\{ X\in T_\calO \calM | X^\mu\,X_\mu = 1, u_\calO^\mu\,X_\mu = 0 \right\}$ \cite{perlick, low, Korzynski:2018}.
The space ${\cal N}^-_\calO$ does not have a well-defined metric and, therefore, by using it, it is not possible to calculate the angular distance between points on the sky: one really needs to pass to the observer-dependent space $\textrm{Dir}(u_\calO)$ in order to do that. It turns out that this introduces the dependence of the angle
measurements between apparent positions on the sky from the observer's 4-velocity. This is commonly referred to as the \emph{light aberration effect}, or \emph{stellar aberration} and may be explained by a relative tilt of the spheres of directions (and the whole simultaneity planes) for observers with different 4-velocities, see \cite{liebscher1998} for a more detailed discussion
involving the historical background.
 
In the context of relativistic geometric optics it is customary to introduce a frame, or vierbein, connected with the observer. The standard approach is to use the Sachs orthonormal frame, consisting of $u_\calO^\mu$, the
direction vector $r^\mu$ and two perpendicular, spatial vectors $e_{\bm A}^\mu$ spanning the Sachs screen space \cite{perlick-lrr, sachs, ehlers-jordan-sachs}. We have found out that it is more convenient to use a related, but slightly modified frame, which we will call
the \emph{seminull frame} (SNF). The frame consists of  $u_\calO^\mu$, the same two perpendicular, spatial vectors $e_{\bm A}^\mu$ and the null vector $l^\mu$ instead of
$r^\mu$. It is not orthonormal and we can check that
the products of the basis vectors read
\bean
l^\mu\,l_\mu &=& 0 \\
e_{\bm A}^\mu\,l_\mu &=& 0 \\
u_\calO^\mu\,e_{\bm A\,\mu} &=& 0 \\
e_{\bm A}^\mu\,e_{\bm B\,\mu} &=& \delta_{\bm A\bm B} \\
u_\calO^\mu\,u_{\calO\,\mu} &=& -1 \\
l^\mu\,u_{\calO\,\mu} &=& Q > 0,
\eean
where we have introduced the notation $Q$ for the product of $l^\mu$ and $u_\calO^\mu$, a constant, but so far undetermined.

We remind the reader that in the subsequent calculations the frame indices will be denoted by boldface letters, with capital latin indices $\bm A$, $\bm B$, ... running over the spatial components $\bm 1$ and $\bm 2$,
small latin indices $\bm i$, $\bm j$, ... running over $\bm 0$, $\bm 1$ and $\bm 3$ and the boldface Greek indices $\bm \mu$, $\bm \nu$, ... running over all 4 dimensions 
from $\bm 0$ to $\bm 3$.

The position on the observer's sky $r_0^\mu$, determined by the fiducial geodesic $\gamma_0$, will serve as the point of reference for all other points on the sky considered here. Namely, let $k^\mu$ be another past-oriented
null vector, corresponding to another source of light, and let $\tilde r^\mu = \frac{1}{k_\sigma\,u_\calO^\sigma}\,k^\mu + u_\calO^\mu$ be the corresponding
direction vector. If the position of the second source
lies on the same hemisphere as $r_0^\mu$ (we will assume that throughout the paper) then it can be uniquely determined from the transversal components of $\tilde r^\mu$ in the seminull frame of the observer, denoted by $\tilde r^{\bm A}$, $\bm A = 1,2$. We will, therefore, use these components as the main variables describing of positions on the observer's sky.
Note that for a source which lies close to $r^\mu$ we have a simple, direct relation between $\tilde r^{\bm A}$ and the angular coordinates on the sky $\delta\theta^{\bm A}$ centered at $r^\mu$, directed along 
$e_{\bm 1}^\mu$ and $e_{\bm 2}^\mu$ and expressed in radians. Namely, we have
\bea
 \delta \theta^{\bm A} \approx \tilde r^{\bm A} \label{eq:positionapprox}
\eea
for $\delta\theta^{\bm A} \ll 1$. The relation for larger angles requires the use of the standard  trigonometric formulas.

In the next sections, we will need to compare the positions on the sky of various objects as registered by observers with different 4-velocities and at different points
in $N_\calO$. In a general spacetime, this is not a trivial task, because, as we mentioned, the position on the sky is a vector in the observer-dependent space of directions. Two problems arise here:
\begin{enumerate}
  \item how do we compare position vectors at different points, 
  \item how do we compare directions on the sky registered by
observers boosted with respect to each other since the notion of a spatial vector is different for each of them. 
\end{enumerate}
It turns out that, owing to the flatness of 
$N_\calO$, this is possible using the 
reference direction given by the null vector $l_\calO$.

Since the region $N_\calO$ is effectively flat in our approximation, we may
simply identify the tangent spaces at all points with $T_\calO \calM$ using the parallel propagation. This is possible independently of the paths connecting points with $\calO$ we may choose for that purpose. Thus the problem of comparing vectors at different points is solved.
We also introduce, consistently with the approach above, 
a parallel propagated SNF $(u_\calO^\mu, e_{\bm A}^\mu, l_\calO^\mu)$ from $\calO$ throughout the whole region $N_\calO$. This way any vector or tensor can be directly compared componentwise with a corresponding
object at another point.

From now on we assume that all equations are written using this type of parallel frame at $N_\calO$ and a similar one at $N_\calE$.
Following Sec. \ref{sec:bilocal}, we will write $l^\mu = l_\calO^\mu$ and
from (\ref{eq:deltalOdef}) and (\ref{eq:DeltalOdef}) we have simply $k^\mu = l^\mu_\calO + \Delta l_\calO^\mu$ for the other null geodesic. Then it is easy to see that
for the observer $u_\calO^\mu$ we have
\bea
\tilde r^{\bm A} = \frac{\Delta l^{\bm A}_\calO}{u_{\calO\,\sigma}\,\left(l_\calO^\sigma + \Delta l_\calO^\sigma\right)} = \frac{\Delta l^{\bm A}_\calO}{u_{\calO\,\sigma}\,l_\calO^\sigma}\,\left(1 - \frac{\Delta l_\calO^{\bm 0}}{u_{\calO\,\sigma}\,l_\calO^\sigma} + \Delta l_\calO^{\bm 3}\right)^{-1} \label{eq:position1}
\eea
in the SNF. 

 Consider now an observer with a different 4-velocity $U^\mu$. Obviously he or she uses a different screen space, but we may introduce a SNF $(U^\mu, f_{\bm A}^\mu, l_\calO^\mu)$ such that
the spatial vectors are aligned along $e_{\bm A}^\mu$, i.e. $f_{\bm A}^\mu = e_{\bm A}^\mu + C_{\bm A}\,l_\calO^\mu$, with appropriate $C_{\bm 1}$ and $C_{\bm 2}$ \cite{Korzynski:2018}. This way we can compare also the positions
on the sky of observers boosted with respect to each other: both $u_\calO^\mu$ and $U^\mu$ may use the fiducial null vector $l_\calO^\mu$ as providing the
 reference direction on their skies and the screen vectors
$e_{\bm A}^\mu$ and $f_{\bm A}^\mu$ as the two perpendicular directions on the celestial sphere used for defining the two-dimensional angular distance from the fiducial direction. The two spatial components of $\tilde r$, i.e. $\tilde r^{\bm A}$, evaluated for both observers and expressed in the corresponding SNF's, may now be used to compare the registered directions on the sky among them. 

The construction can be repeated at the emitter's end of the null geodesic. Given the emitter's momentary 4-velocity $u_\calE^\mu$ and 
the tangent vector $\tilde k^\mu = l_\calE^
\mu + \Delta l_\calE^\mu$ we can define the \emph{viewing direction vector} $s^\mu$,
pointing in the spatial direction (in emitter's frame) from which a given observer watches the events in $N_\calE$:
\bea
 s^\mu = -\left(\frac{1}{\tilde k_\sigma\,u_\calE^\sigma}\,\tilde k^\mu + u_\calE^\mu\right).
\eea

\section{Infinitesimal displacement formulas} \label{sec:infinitesimaldisplacement}

BGO's are fundamental objects describing the properties of all geodesics in the vicinity of $\gamma_0$. However, from the point of view of geometric optics, they contain too much information. Apart from null geodesics they also describe the spatial and timelike geodesics in the vicinity of $\gamma_0$. On top of that, they
distinguish differently parametrized null geodesics sharing the same path, which is indistinguishable in geometric optics.  
The affine parametrization carries no physical information and therefore the formalism effectively contains two gauge degrees of freedom corresponding to the affine reparametrizations of the null geodesics. We would like to isolate them in our formalism and focus on the remaining, physical degrees of freedom.

The condition for the displaced geodesic to remain null reads
\bea
g_{\mu\nu}\,(l_\calO^\mu + \Delta l_\calO^\mu)(l_\calO^\nu + \Delta l_\calO^\nu)= 0. \label{eq:nullcond1}
\eea
It needs to be imposed only once along the geodesic, in this case at $\lambda = \lambda_\calO$.
We would also like to identify null geodesics sharing the same path. This means that we identify the initial data for which the initial tangent vector is proportional, while the
initial points differ only by a vector proportional to the tangent. We identify therefore the initial data pairs of the form
\bea
\left(\begin{array}{l}
\delta x_\calO^\mu\\
\Delta l_\calO^\mu
\end{array} \right) \sim \left(\begin{array}{l}
\delta x_\calO^\mu +  C_1\left(l_\calO^\mu + \Delta l_\calO^\mu\right)\\
\Delta l_\calO^\mu +  C_2\left(l_\calO^\mu + \Delta l_\calO^\mu\right)
\end{array} \right) \label{eq:identification1} \label{eq:equiv1}
\eea
for some nonvanishing constants $C_1$ and $C_2$.

\subsection{Distant observer approximation} \label{sec:doa}

In the rest of this paper, we will assume that we work within the validity regime of the relativistic distant observer approximation (DOA) or the relativistic counterpart of the paraxial approximation.
Relativistic DOA is straightforward to explain mathematically, but it is more difficult to understand its physical meaning and its limits of applicability. 
Mathematically it boils down to linearizing all equations and relations involved in the displacement variables $\delta x_\calO^\mu$ and $\Delta l_\calO$, thereby neglecting all quadratic and higher terms, i.e. 
$\delta x_\calO^\alpha\,\delta x_\calO^\beta$, $\Delta l_\calO^\alpha\,\Delta l_\calO^\beta$ as well as the cross terms $\delta x_\calO^\alpha\,\Delta l_\calO^\beta$. 
This is consistent with the use of the first order GDE for the linearized relations between the deviation vectors $\delta x$ and direction deviations $\Delta l$ at $N_\calO$ and $N_\calE$,
ignoring this way higher order effects in light propagation.
Physically this is equivalent to
two distinct types of approximation:
\begin{enumerate}
\item \emph{Flat light cones approximation (FLA). } We assume we may neglect higher order terms in $\Delta l_\calO$ in Eq. (\ref{eq:nullcond1}), in this way obtaining the null condition in the linearized form
\bea
\Delta l_\calO^\mu\,l_{\calO\,\mu} = 0 \label{eq:nullcond2}.
\eea 
In Sec. \ref{sec:EOasymmetry} we will show that this approximation is equivalent to assuming that the light cones originating on one end of $\calO$ degenerate
to flat null hypersurfaces on the other end. In other words, the regions $N_\calO$ and $N_\calE$ are small enough that the bending of the surface of the light cone
with apex in the opposite region is negligible. 

 \item  \emph{Parallel rays approximation (PRA). }  We assume we can drop the $\Delta l_\calO^{\bm 0}$ and the $\Delta l_\calO^{\bm 3}$ term in Eq. (\ref{eq:position1}). This way we effectively
linearize the whole expression in the direction vector deviation:
\bea
\tilde r^{\bm A} =  \frac{\Delta l^{\bm A}_\calO}{l_{\calO\,\sigma} \,u_\calO^\sigma}. \label{eq:position2}
\eea
This is equivalent to linearizing the whole mapping $V(u_\calO,\cdot)$ around $l_\calO$.
It is straightforward to verify using equations (\ref{eq:Wprop1})-(\ref{eq:Wprop8}) that this way we neglect the dependence of the position on the sky on the null SNF components $\delta x_\calO^{\bm 3}$ and $\delta x_\calE^{\bm 3}$, leaving just the transversal $\bm 1$ and $\bm 2$ as well as the timelike components $\bm 0$ of the position deviation vectors. This, in turn, means that we treat all null rays considered within $N_\calO$ and $N_\calE$ as effectively parallel to each other (and
in turn also to the original $l_\calO$), neglecting the small change of the transversal position of the light rays between the front and the back of $N_\calO$ due to the direction deviation, see Fig. \ref{fig:PRA}. 
Indeed, for sufficiently small $N_\calE$ and $N_\calO$, we may neglect the convergence or divergence of null light rays when
discussing the relation between the direction deviation and displacement vectors. From the observational point of view, it is important to note that in PRA we neglect all perspective distortions of the images as perceived by the observers in
$N_\calO$.

 \bfi
\centering
\includegraphics[width=70mm]{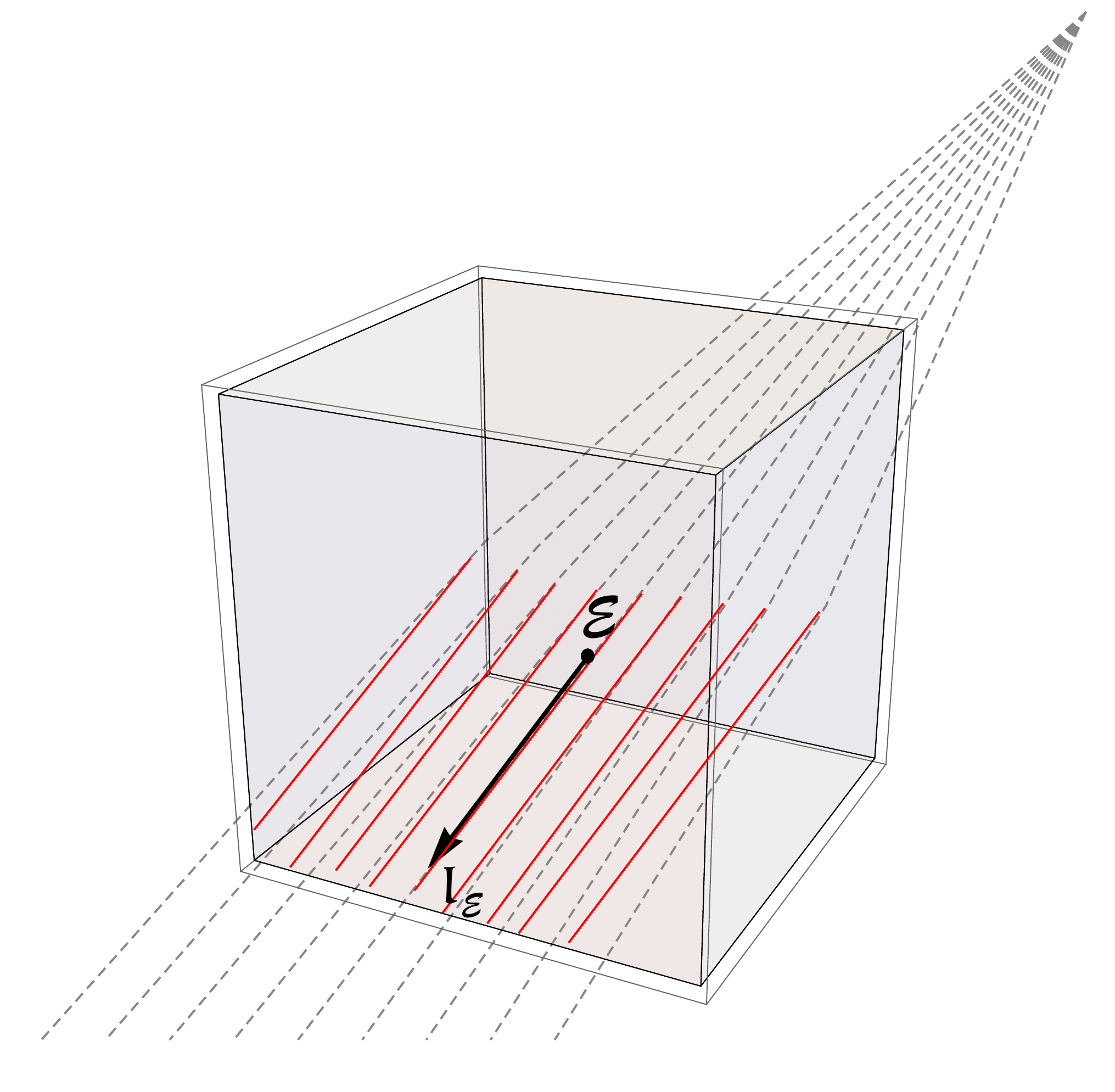}
\qquad\qquad
\includegraphics[width=80mm]{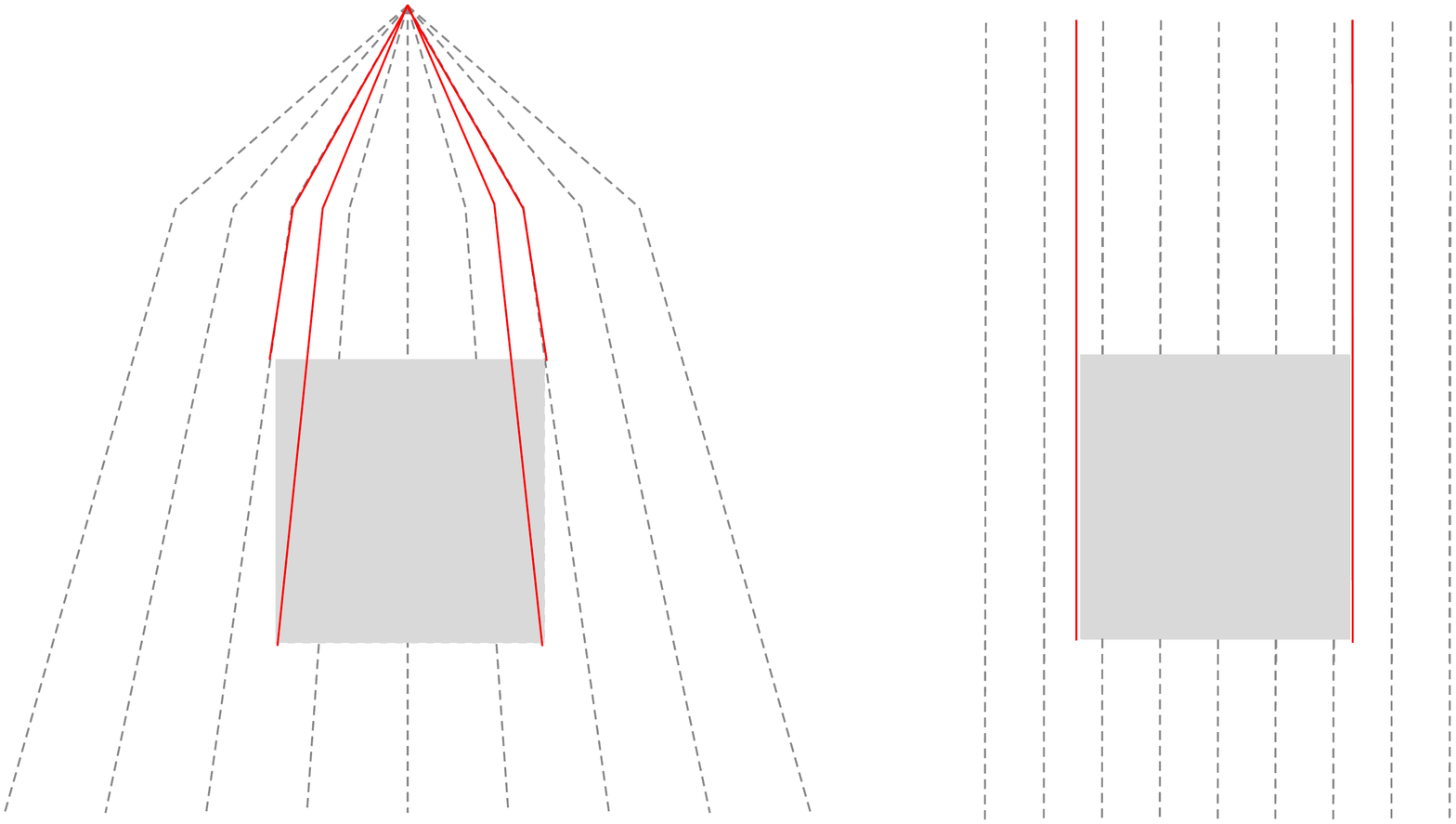}
\caption{In the parallel light rays approximation, we assume that the regions $N_\calO$ and $N_\calE$ are small enough and at
the same time sufficiently separated that we can treat the light rays of null geodesics passing through them as parallel when discussing
the relation of the direction deviation vector at $\calO$ to the displacement vectors at both ends, see the figure on the left-hand side. 
This amounts to neglecting all possible perspective distortions of three-dimensional images and allows to consider only flat, two-dimensional projections to the screen space, see the figure on the right-hand side.
 }
\label{fig:PRA}
\efi  
  
\end{enumerate}

We will now consider the limits of applicability of both approximations in a concrete physical situation. We will begin by the flat case and then move to the case of a
general spacetime with an arbitrary metric.

\paragraph{Limits of applicability - flat case.} In the Minkowski space we have simply ${W_{XX}}\UD{\bm \mu}{\bm \nu} = \delta\UD{\bm\mu}{\bm\nu}$ and
${W_{XL}}\UD{\bm \mu}{\bm \nu} = (\lambda_\calE - \lambda_\calO)\,\delta\UD{\bm\mu}{\bm\nu}$. Let $x_\calE^\mu$ and $x_\calO^\mu$ denote the 
coordinates of $\calE$ and $\calO$ respectively and let $d_\calO$ denote the spatial distance between $\calO$ and $\calE$ in the
observer's frame, given by $d_\calO = (x_\calE^\mu - x_\calO^\mu)\,u_{\calO\,\mu}$. The reader may verify that 
in that case the following expansions for the dimensionless combinations of the $\Delta l_\calO$ components are valid:
\bean
 \frac{\Delta l_\calO^{\bm A}}{l_{\calO\,\sigma}\,u_\calO^\sigma} &=& O\left(\frac{L}{d_\calO}\right) \\
 \Delta l_\calO^{\bm 3} &=& O\left(\frac{L}{d_\calO}\right)  \\
 \frac{\Delta l_\calO^{\bm 0}}{l_{\calO\,\sigma}\,u_\calO^\sigma} &=& O\left(\frac{L^2}{d_\calO^2}\right).
\eean
(Note that the component $\Delta l_\calO^{\bm 3}$ is by definition dimensionless, unlike the other two which require dividing by $l_{\calO\,\mu}\,u_\calO^\mu$). We may substitute these expansions to (\ref{eq:position1}) and (\ref{eq:nullcond1});
it follows then easily that both FLA and PRA are equivalent to neglecting the subleading, quadratic terms in the $\frac{L}{d_\calO}$ expansion, while keeping the linear ones.
Also, the angular size of the region $N_\calE$ on the observer's sky expressed in radians, estimated by applying the expansion above to (\ref{eq:position2}), scales linearly like
$\frac{L}{d_\calO}$. This justifies the use of Eq. (\ref{eq:positionapprox}) for the angular variables $\delta \theta^{\bm A}$ determining the position in the sky.

Obviously, the approximations work well as long as the spatial distance between both regions is much larger than their size. This justifies the name distant observer approximation. Note, however, that the applicability of both PRA and FLA depends not only on the position of the points $\calO$ and $\calE$ but also on the observer's 4-velocity $u_\calO$. 
Even if they work well for a given $u_\calO$ they may fail for another, strongly boosted observer, because in his or her frame the spatial distance $d_\calO$ will be much smaller due to the Lorentz contraction. This, in turn, will lead to a relative increase of the subleading terms in the $\frac{L}{d_\calO}$ expansion and
thus to the increase of the perspective distortions and the light cone bending effects. On the other hand, we point out that this problem can always be cured by shrinking the regions $N_\calO$
and $N_\calE$ appropriately and thus decreasing $L$. This analysis should also apply to almost flat spacetimes and weak lensing.

\paragraph{Limits of applicability - general case.} In a completely general spacetime drawing the precise limits of applicability of the FLA and PRA is much more difficult.
Guided by the results for the flat case we assume that the following dimensionless combinations of components of $\Delta l_\calO$, expressed in the SNF, are small:
\bea
\frac{\Delta l_\calO^{\bm A}}{l_{\calO \bm \sigma}\,u_\calO^{\bm \sigma}} &=& 
\frac{1}{l_{\calO \bm \sigma}\,u_\calO^{\bm \sigma}}\, {{W_{XL}}^{-1}}\UD{\bm A}{\bm \nu}\,\left(\delta x_\calE^{\bm \nu} - {W_{XX}}\UD{\bm \nu}{\bm \sigma}\,\delta x_\calO^{\bm \sigma}\right) \ll 1 \label{eq:DOAvial1}\\
 \Delta l_\calO^{\bm 3} &=& {{W_{XL}}^{-1}}\UD{\bm 3}{\bm \nu}\,\left(\delta x_\calE^{\bm \nu} - {W_{XX}}\UD{\bm \nu}{\bm \sigma}\,\delta x_\calO^{\bm \sigma}\right) \ll 1 .\label{eq:DOAvial2}
\eea
These equations can be viewed as conditions for the BGO's $W_{XX}$ and $W_{XL}$, the observer's 4-velocity $u_\calO$ and the size $L$ of both domains determining the scale of the terms $\delta x_\calO$ and $\delta x_\calE$. 
Substituting these relations to Eq. (\ref{eq:nullcond1}) expressed in the SNF yields an expansion for the remaining component of $\Delta l_\calO$:
\bean
 \frac{\Delta l_\calO^{\bm 0}}{l_{\calO\,\sigma}\,u_\calO^\sigma} = O\left( \left( \frac{\Delta l_\calO^{\bm A}}{l_{\calO \bm \sigma}\,u_\calO^{\bm \sigma}}\right)^2 \right).
\eean
It is quadratic, therefore, negligible in comparison with the other three components. Neglecting $ \Delta l_\calO^{\bm 0} $, on the other hand, is precisely equivalent to the FLA defined by Eq. (\ref{eq:nullcond2}).

As for the PRA, we note that with conditions (\ref{eq:DOAvial1})-(\ref{eq:DOAvial2}) we have
\bean
 \tilde r^{\bm A} =   \frac{\Delta l^{\bm A}_\calO}{l_{\calO\,\bm \sigma} \,u_\calO^{\bm \sigma}} + O\left(\frac{\Delta l^{\bm A}_\calO}{l_{\calO\,\bm \sigma} \,u_\calO^{\bm \sigma}} \cdot \Delta l_\calO^{\bm 3}\right).
\eean
Again keeping the leading, linear order is equivalent to the PRA condition (\ref{eq:position2}). 
Just like in the flat case, for a given spacetime and fixed $\calE$ and $\calO$ the conditions (\ref{eq:DOAvial1})-(\ref{eq:DOAvial2}) are sensitive to the 4-velocity of the observer. 
But in contrast to the flat case, there is no \emph{a priori} relation between
the BGO's and the distance between the two points $\calO$ and $\calE$, because in general both $W_{XX}$ and $W_{XL}$ may take any value
limited only by conditions (\ref{eq:Wprop1}), (\ref{eq:Wprop3}), (\ref{eq:Wprop5}) and (\ref{eq:Wprop7}). In fact, with very strong lensing the applicability of the DOA
for fixed $L$ and observer can even begin to \emph{decrease} further down
$\gamma_0$ as we move away from $\calO$. We may illustrate this with a simple example: with very strong lensing along the way, the beam centered at $\calO$ may become strongly convergent or divergent far down $\gamma_0$, introducing this way
strong perspective distortions observable for regions of size $L$ and violating this way the assumptions of the PRA. However, just like before, the applicability of both approximations can always be restored if we narrow down $N_\calO$ and $N_\calE$ appropriately. 

Summarizing, the applicability of the DOA in a spacetime with an arbitrary metric and for any given observer is a subtle problem and needs to be tested on a case by case basis, for example,
using conditions (\ref{eq:DOAvial1})-(\ref{eq:DOAvial2}). In astrophysics, it is assumed they hold automatically because of the extreme ratio between the distances between $\calO$ and
$\calE$ and their sizes (measured in a typical observer's frame, for example, the CMB rest frame), but if the influence of the curvature is strong or the observer is sufficiently boosted this must be done with care.

In principle it is possible to use the nonlinearized relations (\ref{eq:nullcond1}) and (\ref{eq:position1}) for the observables, taking into account this way the perspective effects and the effects of the bending of the lightcone surfaces, but mathematical consistency requires then to include also the nonlinear terms in the relations between the deviations of the geodesics in $N_\calO$ and $N_\calE$, for example with the help
of the higher order GDE's \cite{Bazanski1, Bazanski2, Vines:2014oba}. This is significantly more cumbersome than the formalism presented here and we leave this for future studies.

\paragraph{Quotient spaces. }Within the regime of validity of the PRA we may simplify the equivalence relation of (\ref{eq:equiv1}) to
\bea
\left(\begin{array}{l}
\delta x_\calO^\mu\\
\Delta l_\calO^\mu
\end{array} \right) \sim \left(\begin{array}{l}
\delta x_\calO^\mu +  C_1\,l_\calO^\mu \\
\Delta l_\calO^\mu +  C_2\,l_\calO^\mu, \label{eq:equiv2}
\end{array} \right) 
\eea
i.e. we identify the initial data for which the position and directional deviations only differ by a multiple of  $l_\calO^\mu$. 
From (\ref{eq:Wprop1})-(\ref{eq:Wprop4}) it is easy to see that adding this type of terms in $N_\calO$ leads to 
a similar change in the final data:
\bea
\left(\begin{array}{l}
\delta x_\calE^\mu\\
\Delta l_\calE^\mu
\end{array} \right) \sim \left(\begin{array}{l}
\delta x_\calE^\mu +  D_1\,l_\calE^\mu \\
\Delta l_\calE^\mu +  D_2\,l_\calE^\mu,
\end{array} \right) \label{eq:equiv3}
\eea
with constants $D_1$ and $D_2$ related to $C_1$ and $C_2$.

This invariance leads to the following idea: instead of considering the optical operators as acting from one tangent space to another one we may consider them
as mappings between the appropriate quotient spaces $\calQ_\calO = T_\calO \calM / l_\calO$ and $\calQ_\calE = T_\calE \calM / l_\calE$, see Fig. \ref{fig:quotientspaces}.
Namely, let $\calQ_\calE$ be the quotient of $T_\calE\calM$ by the equivalence relation $X^\mu \sim Y^\mu$ iff $X^\mu = Y^\mu + c\,l^\mu_\calE$ for any real number $c$. 
$\calQ_\calO$ can be defined in an analogous way, with $l_\calE$ replaced by $l_\calO$ and $T_\calE\calM$ replaced by $ T_\calO \calM$.

 Within these 3-dimensional spaces
we also consider the 2-dimensional subspaces orthogonal to $l_\calO^\mu$ and $l_\calE^\mu$ respectively, i.e.
$\calP_\calO = l_\calO^\perp / l_\calO$ and $\calP_\calE = l_\calE^\perp / l_\calE$, see again Fig. \ref{fig:quotientspaces}. They  will be referred to as the \emph{perpendicular spaces}, see again Fig. 
\ref{fig:quotientspaces}. Unlike $\calQ_\calO$ and $\calQ_\calE$, the perpendicular spaces $\calP_\calO$ and $\calP_\calE$ inherit a positive definite metric $q$ from the Lorentzian spacetime metric $g$ \cite{Korzynski:2018}:
let $\bm X$ and $\bm Y$ be two vectors in $\calP_\calE$ and let $X$ and $Y$ be any two corresponding vectors in $T_\calE \calM$, i.e.
$\bm X = [X]$ and $\bm Y = [Y]$, where $[\cdot]$ denotes the linear operation of taking the equivalence class of a vector in $T_\calE \calM$  with respect to the relation $\sim$  defined above. The reader may check that the formula $q(\bm X,\bm Y) = g(X,Y)$ defines the same value of the scalar product of $\bm X$ and $\bm Y$ irrespective of the choice of
the equivalence class representatives $X$ and $Y$.

The angles and distances calculated using $q$ correspond to the angles and distances measured by any observer on points projected down to the plane perpendicular to the direction of observation (the Sachs's screen space) along the null direction of light propagation \cite{Korzynski:2018}. This is a reformulation of the well-known Sachs shadow theorem \cite{sachs, perlick-lrr}
in terms of the quotient spaces.
This fact gives the geometry of the perpendicular spaces an explicitly observer-invariant meaning and we explore it later in the paper in order to separate the dependence
of observables
on the observer's and emitter's frame and on the spacetime geometry.

We end this subsection by noting that both quotient spaces have a very simple interpretation in terms of the vector components in a SNF. Namely, they correspond to vectors with a ``forgotten'' fourth component corresponding to $\bm \mu = \bm 3$. The mapping $[\cdot]$ has then a simple form of $(X^{\bm 0}, X^{\bm 1}, X^{\bm 2}, X^{\bm 3}) \to (X^{\bm 0}, X^{\bm 1}, X^{\bm 2})$. Vectors in any perpendicular subspace $\calP$ along $\gamma_0$ have additionally vanishing first component, i.e. $(0,X^{\bm 1}, X^{\bm 2})$. 

The reader may check that in the GDE (\ref{eq:GDE}) expressed in the SNF the equations for the first three components decouple from the fourth one and that vectors with vanishing $\bm \mu = \bm 0$ component form a subspace of solutions. This way we see that the quotient spaces defined above are compatible with
the properties of the first order GDE around a null geodesic.

 \bfi
\centering
\includegraphics[width=70mm]{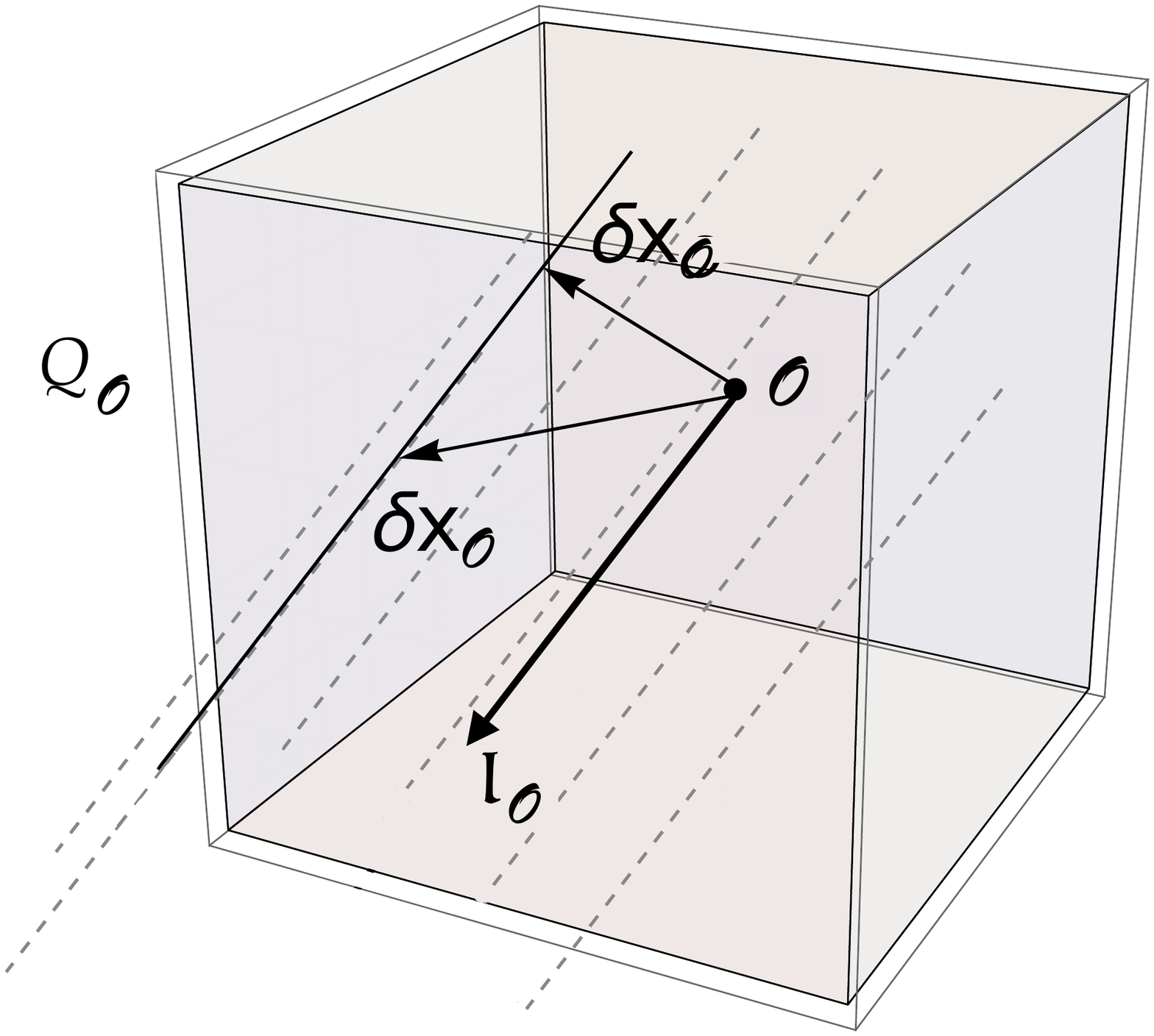}
\qquad\qquad
\includegraphics[width=70mm]{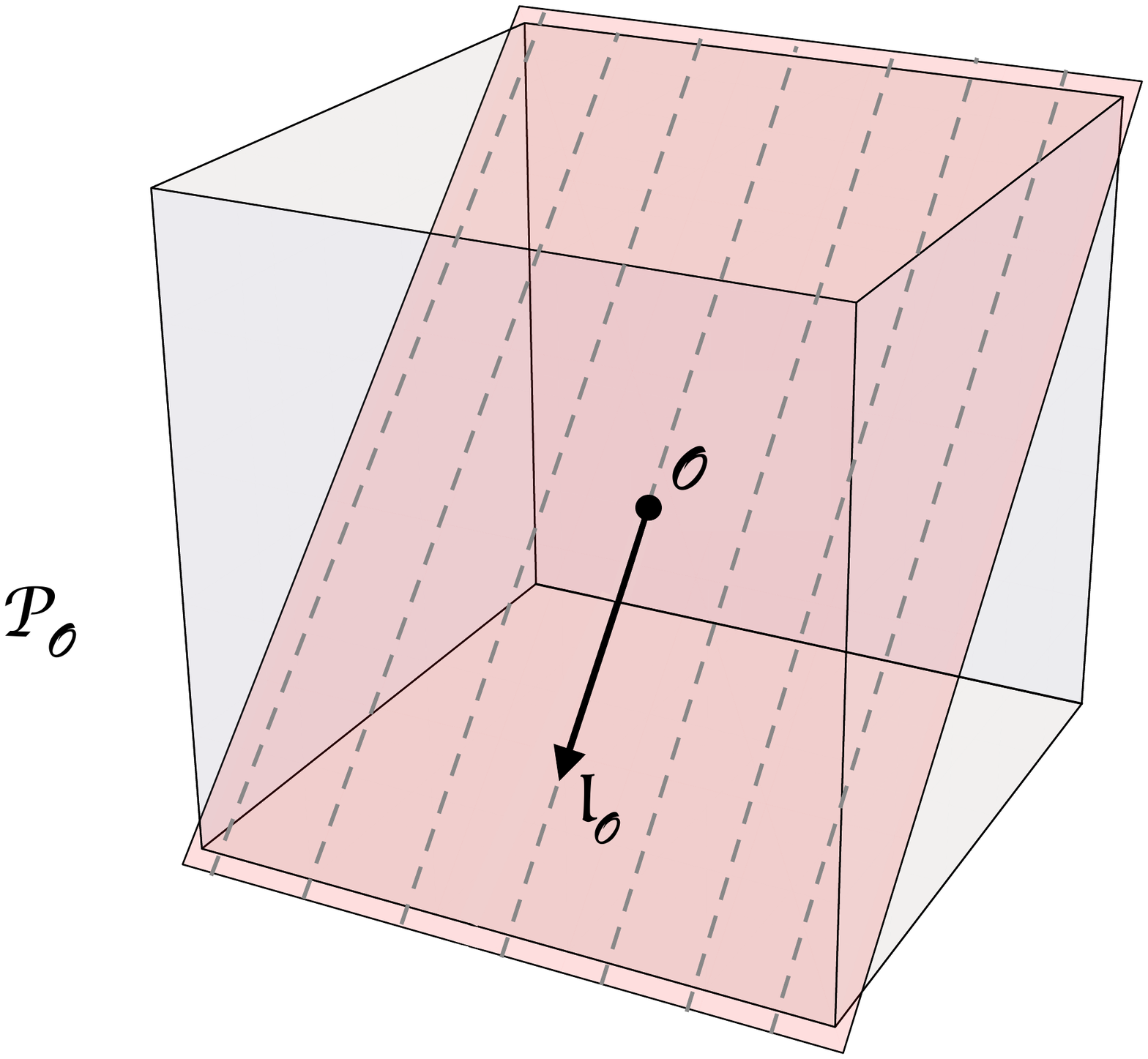}
\caption{Geometry of the quotient spaces $\calQ_\calO$, $\calP_\calO$ and their counterparts on the other side.
Elements of $\calQ_\calO$ correspond to the vectors in $T_\calO \calM$, or points in $N_\calO$, identified if they are separated by a multiple of $l_\calO$.
Geometrically it is the space of null straight lines in $N_\calO$ parallel to $\gamma_0$. The two-dimensional subspace $\calP_\calO$ corresponds to null straight lines
parallel to $\gamma_0$ which additionally lie on the null hyperplane orthogonal to $l_\calO$. }
\label{fig:quotientspaces}
\efi
\subsection{Time lapse formula}

We multiply Eq. (\ref{eq:positiondeviation1}) by $l_{\calE\,\mu}$ from both sides and make use of Eq. (\ref{eq:Wprop5}), (\ref{eq:Wprop7}) to obtain the following relation:
\bean
l_{\calE\,\mu}\,\delta x_{\calE}^\mu = l_{\calO\,\mu}\,\delta x_{\calO}^\mu + \left(\lambda_\calE - \lambda_\calO\right)\,\Delta l_\calO^\mu\,l_{\calO\,\mu}.
\eean
Comparing this formula with (\ref{eq:nullcond2}) we see that the perturbed geodesic is null in the DOA iff 
\bea
l_{\calE\,\mu}\,\delta x_{\calE}^\mu = l_{\calO\,\mu}\,\delta x_{\calO}^\mu. \label{eq:timelapse1}
\eea
We will refer to this equation as the \emph{time lapse formula}. We have just proved that within the FLA any two points in $N_\calO$ and $N_\calE$ respectively can be connected by a null geodesic iff
their deviation vectors satisfy the linear condition (\ref{eq:timelapse1}).
This is the first important result of this paper, because it gives immediately the relation between the time lapse in $N_\calE$ and the time lapse in $N_\calO$, as registered
by the observers, and formulated in a coordinate- and frame-independent way.
Namely, for any observer whose worldline intersects $N_\calO$ we may relate
his or her proper time to the ‘‘null time'' $\chi = l_{\calO\,\mu}\,\delta x^\mu_\calO$ (for an effectively flat space this is a simple special relativity problem). Equation
(\ref{eq:timelapse1}) shows then that
at a given moment the observer can see only the events in $N_\calE$ which lie on the null hypersurface given by $ l_{\calE\,\mu}\,\delta x^\mu_\calE = \chi$. This a simple and elegant way to take into
account the R\o mer delay between the regions $N_\calO$ and $N_\calE$ due to the finite light speed.

\bfi
\centering
\includegraphics[width=0.9\textwidth]{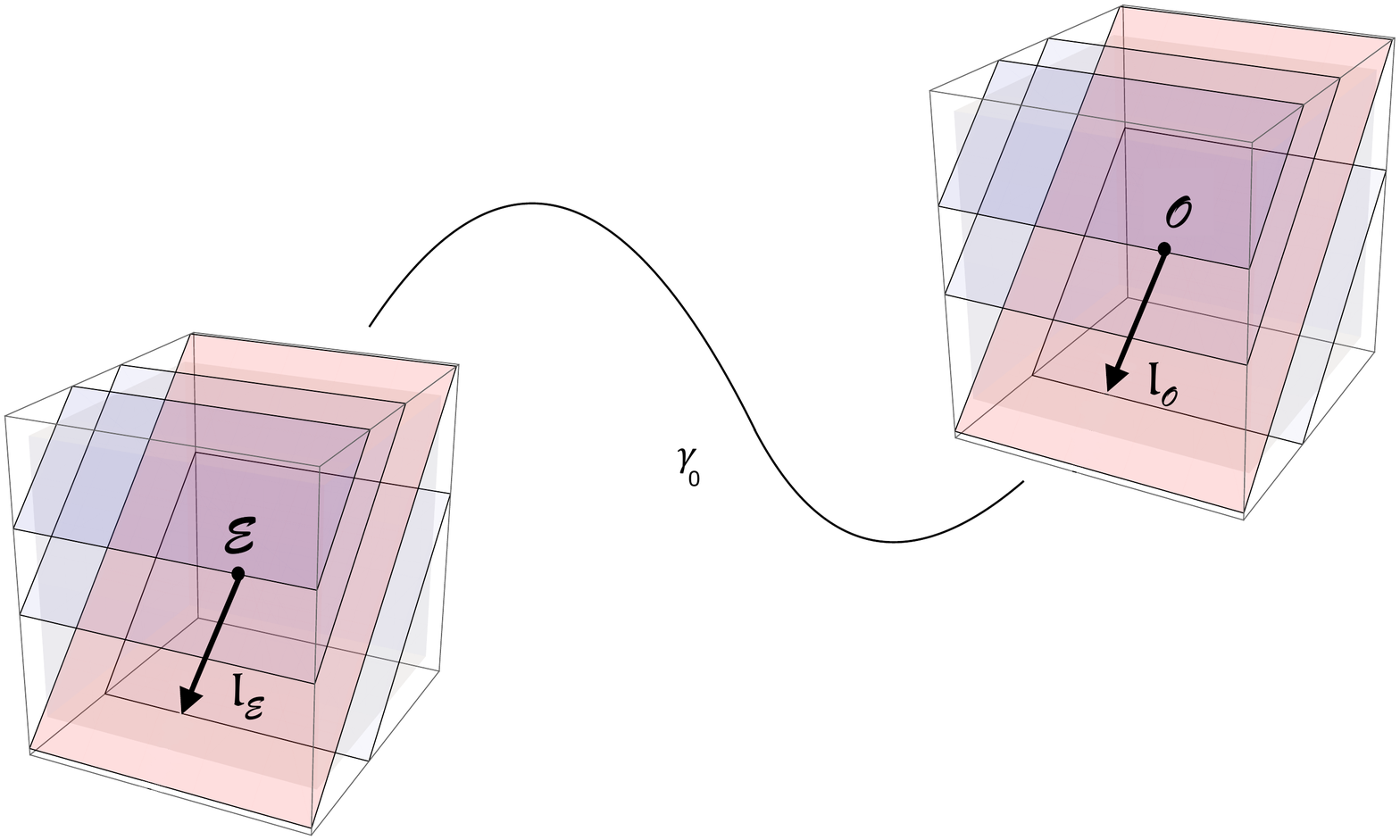}
\caption{Null tangent vectors $l_\calO$ at $\calO$ and $l_\calE$ at $\calE$ define two corresponding foliations by
null hypersurfaces orthogonal to $l_\calO$ and $l_\calE$ respectively. In the flat light cones approximation, only the pairs of points lying on the corresponding leaves can be
connected by a null geodesic. This requirement restricts the events in $N_\calE$ from which an observer in $N_\calO$ can register signals at a given instant of time.
}
\label{fig:foliations}
\efi

Geometrically, the conditions $l_{\calO\,\mu}\,\delta x_{\calO}^\mu = \const$ and $l_{\calE\,\mu}\,\delta x_{\calE}^\mu = \const$ define foliations of $N_\calO$ and $N_\calE$ by
families of null hypersurfaces, see Fig. \ref{fig:foliations}. We can, therefore, rephrase the result of this section as follows: the null vectors $l_\calE$ and $l_\calO$ define
two foliations by null hypersurfaces in the two locally flat regions. For observers located on a leaf in $N_\calO$ only the points lying on the corresponding leaf in $N_\calE$ can be reached by a null geodesic.
It is also clear that while the null tangent vectors $l_\calO^\mu$ and $l_\calE^\mu$ are defined only up to a common rescaling according to 
(\ref{eq:laffine}), this ambiguity does not affect the two foliations or the correspondence between their leaves.

The two null foliations at two ends of $\gamma_0$ are in fact the degenerate families of light cones centered at the opposite ends of $\gamma_0$. If we pick a point $p$ in $N_\calO$ then its past light cone
in $N_\calE$ will degenerate to a flat hypersurface due to the large distance between the two regions and their small size. Moreover, the future light cone of \emph{any} point on that null hypersurface will degenerate in $N_\calO$ to the same null hypersurface containing $p$: in the FLA the light cones with apices contained within a single foliation leaf look exactly the same on the other end of $\gamma_0$. The same argument works with the role of the emission and observation point reversed. These observations explain the name we have coined for flat light cone approximation (\ref{eq:nullcond2}). Physically, the flatness of light cones means that the R\o mer delay is in a good approximation independent of the
perpendicular displacements of the observer and the source because of the large distance between them.

We may pull back formula (\ref{eq:timelapse1}) to the quotient spaces $\calQ_\calO$ and $\calQ_\calE$. Note that the products on both sides are insensitive to
adding multiples of $l_\calO^\mu$ and $l_\calE^\mu$ to the displacement vectors. We may therefore consider $g(\cdot,l_\calO)$ and
$g(\cdot,l_\calE)$  as a mappings defined on the respective quotient spaces
and write
\bea
g([\delta x_\calO],l_\calO) = g([\delta x_\calE],l_\calE) \label{eq:timelapse2}.
\eea

\subsection{Direction variation formula} \label{sec:directionvariationformula}

We will now recast formula (\ref{eq:directiondeviation1}) applied to deviations satisfying (\ref{eq:nullcond2}) in a more convenient form. We begin by rewriting it as follows:
\bea
{W_{XL}}\UD{\mu}{\nu}\,\Delta l_\calO^\nu = \delta x_\calE^\mu - {W_{XX}}\UD{\mu}{\nu}\,\delta x_\calO^\nu \label{eq:directiondeviationfull}
\eea
Equation (\ref{eq:directiondeviationfull}) can be pulled back to the quotient spaces $\calQ_\calO$ and $\calQ_\calE$ respectively.

The operator $W_{XL}$ defines in  an operator acting on the perpendicular space $\calP_\calO$ 
with the image in $\calP_\calE$. Namely, given a vector $\bm Y \in \calP_\calO$ we may pick any vector $Y \in T_\calO \calM$ such that $[Y] = \bm Y$ and define
\bean
 \calD(\bm Y) = \left[ W_{XL}( Y ) \right].
\eean
It is straightforward to verify using (\ref{eq:Wprop1})-(\ref{eq:Wprop8}) that the resulting vector in $\calQ_\calE$ does not depend on the choice of the representative $Y$  of the equivalence class
and that the resulting vector is always perpendicular to $l_\calE^\mu$. Thus we have defined a linear mapping
\bean
  \calD : \calP_\calO \to \calP_\calE.
\eean
This mapping can be represented by a 2-by-2 matrix and is known in the relativistic geometric optics as the Jacobi map \cite{perlick-lrr, Korzynski:2018}.
The Jacobi map as defined here depends on the parametrization we have chosen for the fiducial geodesic $\gamma_0$. Under an affine reparametrization (\ref{eq:affinereparametrization})--(\ref{eq:laffine}) it rescales according to the formula $\calD \to E\cdot \calD$. 

Similar reasoning can be applied to $W_{XX}$: given $\bm Z \in \calQ_\calO$ we may take any $Z \in T_\calO \calM$ such that $[Z] = \bm Z$ and define
\bea
 {\cal W}(\bm Z) = \left[ W_{XX}( Z ) \right]. \label{eq:calWdefinition}
\eea
We check using (\ref{eq:Wprop1})-(\ref{eq:Wprop8}) that this defines a linear mapping between the quotient spaces
\bean
  {\cal W} : \calQ_\calO \to \calQ_\calE.
\eean
$\cal W$ can be used directly, but since we are interested in the curvature effects on the light rays we have found it convenient to separate out the
``nonflat" contribution to the operator, related directly to the curvature.  In a flat space we have ${W_{XX}}\UD{\mu}{\nu} = T\UD{\mu}{\nu}$,
where $T$ is the parallel transport operator from $\calO$ to $\calE$. Let $\calT$ denote the pullback of $T$ to the space $\calQ_\calE$, i.e. mapping $\calT : \calQ_\calO \to \calQ_\calE$ given by the formula
\bean
   \calT(\bm Z) = \left[T(Z)\right],
\eean
where
 $\bm Z \in \calQ_\calO$
and $Z \in T_\calO \calM$ is a vector such that $[Z] = \bm Z$. With this setup we may define
\bean
 m = {\cal W} - {\cal T}.
\eean
The operator $m$ will be referred to as the \emph{emitter-observer asymmetry operator}. While the domain of $m$ consists of the whole space $\calQ_\calO$ its image is automatically perpendicular to $l_\calE^\mu$ -- this follows from (\ref{eq:Wprop5})
and the elementary property of the parallel transport $l_{\calE\,\mu}\,T\UD{\mu}{\nu} = l_{\calO\,\nu}$. Thus the emitter-observer asymmetry operator is a mapping
\bean
  m: \calQ_\calO \to \calP_\calE.
\eean
Since the dimension of $\calQ_\calO$ is 3, while the dimension of $\calP_\calE$ is 2, the operator $m$ is always degenerate.
Unlike the Jacobi map, it is insensitive to affine reparametrizations of the fiducial null geodesic $\gamma_0$, i.e. $m \to m$ if $\lambda \to \lambda' = A\,\lambda + B$.
$m$ has already appeared in \cite{Korzynski:2018}, although without a full discussion of its properties.
The pair of bilocal operators $\calD$ and $m$ will be referred to as the \emph{optical operators}.

We may rewrite the pullback of the whole Eq. (\ref{eq:directiondeviationfull}) using the optical operators:
\bean
\calD\left([\Delta l_\calO]\right) = [\delta x_\calE] - [\delta \hat{x}_\calO] - m\left([\delta x_\calO]\right),
\eean
where $\delta \hat{x}_\calO \equiv \calT(\delta x_\calO)$ is a shorthand notation for the parallel transport from $\calO$ to $\calE$.
In the final step we can use the linearity of the projection $\left[\cdot\right]$ to the quotient space $\calQ_\calE$ to put both deviations vector inside a common square bracket and obtain the
\emph{direction deviation equation}:
\bea
\calD\left([\Delta l_\calO]\right) = [\delta x_\calE - \delta \hat{x}_\calO] - m\left([\delta x_\calO]\right) \label{eq:directiondeviationabstract}
\eea
Note that as long as (\ref{eq:timelapse1}) is satisfied the combination $\delta x_\calE^\mu - \delta \hat{x}_\calO^\mu$ is perpendicular to $l_\calE^\mu$ and the
first term on the right hand side of (\ref{eq:directiondeviationabstract}) is a vector in $\calP_\calE$, even though each position deviation vector individually does not need to
be perpendicular to $l_\calO$ or $l_\calE$. This is of course in full agreement
with the fact that the left-hand side and the second term on the right-hand side are automatically in $\calP_\calE$.
On the other hand, we see that (\ref{eq:directiondeviationabstract}) is impossible to satisfy
when applied to a pair of deviation vectors not satisfying the time lapse formula.

Equations (\ref{eq:timelapse2}) and (\ref{eq:directiondeviationabstract}) describe completely how observers in $N_\calO$ register signals coming from events in $N_\calE$: (\ref{eq:timelapse2}) defines which
events in $N_\calE$ are visible from which points in $N_\calO$ while (\ref{eq:directiondeviationabstract}) yields the direction from which the observers perceive the
light coming. The optical properties of the spacetime between those regions are therefore completely contained in the following geometric structures:
\begin{enumerate}
\item The pair of null vectors $l_\calO$ and  $l_\calE$, given up to a common rescaling. They define three important ingredients of the geometry: 
$l_\calO$ gives the null direction from which an observer in $\calO$ sees the light coming from $\calE$, this way providing the reference direction for all other observations.
 $l_\calE$, on the other hand, gives null direction from which he or she effectively observes the region $N_\calE$. Finally, both vectors together define the two null foliations in $N_\calO$ and $N_\calE$ relating the time lapse in both regions.
\item  The Jacobi map $\calD$ acting
between the perpendicular spaces on the two ends of the null geodesic (also defined up to the same rescaling), encoding the effects of gravitational lensing
between $\calO$ and $\calE$.
\item The parallel transport operator $T$ along $\gamma_0$, preserving the metric.
\item The emitter-observer asymmetry
map $m$ from $\calQ_\calO$ to $\calP_\calE$. It is related to the parallax effects and the apparent position drift. Its function will be explained in detail in Sec. \ref{sec:EOasymmetry}.
\end{enumerate}
All these objects exist independently of any other structures in $M$, in particular independently from
the choice of the coordinate systems, observers and frames in $N_\calO$ and $N_\calE$. 
The last two operators can be in fact be combined into $\calW$ from Eq. (\ref{eq:calWdefinition}), but we have opted to consider them separately because $m$ turns out to have a particularly
elegant geometric interpretation.

Geometrically the pair of equations (\ref{eq:timelapse2}) and (\ref{eq:directiondeviationabstract}) can be seen as a linearized transformation between the two ways in which we may parametrize null geodesics passing
through $N_\calO$ and $N_\calE$. On the one hand, we can parametrize them by giving the initial point $\delta x_\calO^\mu$ and the initial direction, specified by the
direction deviation vector $\Delta l_\calO^\mu$. In the DOA and for geodesics considered without parametrization all we need is the equivalence class $\left[ \delta x_\calO \right]$,
i.e. a vector in the 3-dimensional space $\calQ_\calO$, and the equivalence class of $\Delta l_\calO^\mu$, also in $\calQ_p$. The null condition (\ref{eq:nullcond1}) restricts the
choice of  $[\Delta l_\calO]$ to the 2-dimensional subspace $\calP_\calO$. The space of geodesics we consider is therefore  5-dimensional.

On the other hand, we may also parametrize the null geodesics by the endpoints $\delta x_\calE$ and
$\delta x_\calO$, or --- more precisely --- by their equivalence classes in $\calQ_\calO$ and $\calQ_\calE$ respectively. In the second parametrization, we need to impose
the linear condition given by the time lapse formula (\ref{eq:timelapse2}). The resulting dimension of the null geodesic family is therefore again $3+3-1=5$.
The direction deviation vector at $\calO$ is then given  by (\ref{eq:directiondeviationabstract}) up to an irrelevant multiple of $l_\calO^\mu$.

\subsection{Optical operators from the Riemann curvature tensor}

We will show that the two optical mappings are effectively linear functionals of the spacetime curvature tensor along $\gamma_0$, given by solutions of linear ODE's. 
Recall that $W_{XX}$ and $W_{XL}$ expressed in a parallel propagated SNF can be obtained from the curvature tensor directly via (\ref{eq:AODE1})-(\ref{eq:WXLODE}).

Given any SNF $(u^\mu,e_{\bm A}^\mu, l_\calO^\mu)$ we can construct a corresponding frame $(\bm u,\bm{e}_{\bm A})$ in $\calQ_\calO$ by simply taking
$\bm u = [u]$, $\bm{e}_{\bm A} = [e_{\bm A}]$ and a frame $(\bm{e}_{\bm A})$ in $\calP_\calO$.
By parallel propagating the SNF and repeating this procedure we obtain similar parallel propagated frames 
$(\hat u^\mu,\hat e_{\bm A}^\mu,\hat l_\calO^\mu)$, $(\hat {\bm u},\hat {\bm{e}}_{\bm A})$ and $(\hat{\bm e}_{\bm A})$, in $T_p \calM$, $\calQ_p$ and $\calP_p$ respectively,
defined now at all points $p$
along $\gamma_0$.  Note that the parallel propagation $\hat l_\calO^\mu$ of $l_\calO^\mu$ along $\gamma_0$ is simply the tangent vector $l^\mu$, because the tangent vector along a geodesic is always parallel 
propagated. Moreover, at $\calE$ we have $\hat l_\calO^\mu = l_\calE^\mu$.

We will now prove a direct relation between the optical operators expressed in the frame $(\hat {\bm u},\hat {\bm{e}}_{\bm A})$ and submatrices of $W_{XX}$ and $W_{XL}$ expressed
in the SNF.  Let $\bm \xi_0 \in \calP_\calO$, with the decomposition $\bm \xi_0 = \bm \xi_0^{\bm A}\,\bm e_{\bm A}$. Take the vector $\xi_0$ in $T_\calO \calM$ given by $\xi_0 = \bm \xi_0^{\bm A}\,e_{\bm A}$
as the corresponding vector such that $\bm \xi_0 = [\xi_0]$. Applying the definition we get 
\bean
\calD(\bm \xi_0) = \left[W_{XL}(\xi_0)\right] = \left[W_{XL}(\bm\xi_0^{\bm B} \, e_{\bm B})\right] = \bm\xi_0^{\bm B}\,\left[W_{XL}(e_{\bm B})\right].
\eean
On the other hand we know that
\bean
\left[W_{XL}(e_{\bm B})\right] = \left[{W_{XL}}\UD{\bm 0}{\bm B}\,\hat u + {W_{XL}}\UD{\bm A}{\bm B}\,\hat e_{\bm A} + {W_{XL}}\UD{\bm 3}{\bm B}\,l \right].
\eean
The first term vanishes because of (\ref{eq:Wprop7}): since $e_{\bm B}$ is orthogonal to $l_\calO$, then $W_{XL}(e_{\bm B})$ must be orthogonal to $l_\calE$. It follows that
${W_{XL}}\UD{\bm 0}{\bm B} = 0$ in the seminull frame. The last term $\left[{W_{XL}}\UD{\bm 3}{\bm B}\,l \right]$ on the other hand vanishes because at $\calE$ we have $l = l_\calE$ and the equivalence class of $l_\calE$ in $\calQ_\calE$ is by definition 0.  We are thus left with
\bean
\calD(\bm\xi_0) = \,{W_{XL}}\UD{\bm A}{\bm B}\,\bm\xi_0^B\,\hat {\bm e}_{\bm A}
\eean
or
\bea
\calD\UD{\bm A}{\bm B} = {W_{XL}}\UD{\bm A}{\bm B} \label{eq:Dsubmatrix}
\eea
in the aforementioned frames. Thus the Jacobi operator turns out to be a two-by-two submatrix of $W_{XL}$, independently of the choice of the SNF.
Similar arguments prove  $m$ expressed in $(\hat{\bm u},\hat{\bm e}_A)$ satisfies
\bea
m\UD{\bm A}{\bm i} = {W_{XX}}\UD{\bm A}{\bm i} - T\UD{\bm A}{\bm i}, \label{eq:msubmatrix}
\eea
where $i$ runs through $0,1,2$, $W_{XX}$ is expressed in the SNF and $T$ is the parallel transport operator in that frame:
\bean
  T\UD{\bm A}{\bm i} =\delta\UD{\bm A}{\bm i}= \left(\begin{array}{rrr} 1 & 0 & 0 \\ 0 & 1 & 0 \end{array}\right).
\eean

We will use the results proved above to derive linear ODE's which relate the optical operators directly to the curvature.
We note that in the ODE  (\ref{eq:BODE1}) for   ${W_{XL}}$ the equations for the submatrix (\ref{eq:Dsubmatrix}) decouple from the rest of the equations. Namely,
the relevant components of the ODE's read
\bean
 \ddot B\UD{\bm A}{\bm B} - R\UD{\bm A}{\bm\mu \bm\nu \bm 0}\,l^{\bm \mu}\,l^{\bm \nu} \,B\UD{\bm 0}{\bm B} - R\UD{\bm A}{\bm \mu \bm\nu \bm C}\,
 l^{\bm \mu}\,l^{\bm \nu} \,B\UD{\bm C}{\bm B} - R\UD{\bm A}{\bm\mu \bm\nu \bm 3} \,l^{\bm\mu}\,l^{\bm\nu}\,B\UD{\bm 3}{\bm B} = 0.
 \eean
We know that $R\UD{\bm A}{\bm \mu\bm \nu \bm 3} \,l^{\bm \mu}\,l^{\bm \nu} = R\UD{\bm A}{\bm 3 \bm 3 \bm3} = 0$ because of the symmetries of the Riemann tensor. On the other hand, $ B\UD{\bm 0}{\bm B}$  must vanish,
because $B\UD{\bm \mu}{\bm \nu}$ must satisfy (\ref{eq:Wprop7}) all along $\gamma_0$. We are thus left with the following equation for $\calD\UD{\bm A}{\bm B} = B\UD{\bm A}{\bm B}$ along $\gamma_0$:
\bea
 \ddot \calD\UD{\bm A}{\bm B} - R\UD{\bm A}{\bm\mu\bm\nu \bm C}\,l^{\bm\mu}\,l^{\bm\nu} \,\calD\UD{\bm C}{\bm B} = 0 \label{eq:DODE1}
 \eea
 with the initial data
 \bea
 \calD\UD{\bm A}{\bm B}(\lambda_\calO) &=& 0 \label{eq:DODE2}\\
 \dot \calD\UD{\bm A}{\bm B}(\lambda_\calO) &=& \delta\UD{\bm A}{\bm B}. \label{eq:DODE3}
 \eea
 The value of the solution at the endpoint $\calE$ gives the Jacobi map in the frame $(\hat{\bm e}_A)$ irrespective of the choice of 
 the corresponding SNF.
The derivation of the ODE for $m$ proceeds in a similar way. We begin by noting that in equation (\ref{eq:AODE1}) the ODE's for the components
$A\UD{\bm A}{\bm i}$ decouple from the rest:
\bean
\ddot A\UD{\bm A}{\bm i } - R\UD{\bm A}{\bm\mu\bm\nu \bm 0}\,l^{\bm\mu}\,l^{\bm\nu} \,A\UD{\bm 0}{\bm i} - R\UD{\bm A}{\bm\mu\bm\nu \bm C}\,l^{\bm\mu}\,l^{\bm\nu} \,A\UD{\bm C}{\bm i} - R\UD{\bm A}{\bm\mu\bm\nu \bm 3} \,l^{\bm\mu}\,l^{\bm\nu}\,A\UD{\bm 3}{\bm i} = 0.
\eean
The fourth term vanishes again because $R\UD{\bm A}{\bm\mu\bm\nu \bm 3} \,l^{\bm\mu}\,l^{\bm\nu} = 0$ due to the symmetries of the Riemann. The second one survives, but it can be simplified using the algebraic properties of $W_{XX}$. Namely, $A\UD{\bm 0}{\bm \nu} = \delta\UD{\bm 0}{\bm \nu}$ because of (\ref{eq:WXXODE}) and (\ref{eq:Wprop5}). We get therefore the following ODE for $W_{XX}$ in a SNF:
\bean
{{{\ddot W}_{XX}}}\,\UD{\bm A}{\bm i } - R\UD{\bm A}{\bm\mu\bm\nu \bm C}\,l^{\bm\mu}\,l^{\bm\nu} \,{W_{XX}}\UD{\bm C}{\bm i} &=& R\UD{\bm A}{\bm\mu\bm\nu \bm 0}\,l^{\bm\mu}\,l^{\bm\nu} \,{\delta}\UD{\bm 0}{\bm i} \\
{W_{XX}}\UD{\bm A}{\bm i}(\lambda_\calO) &=& \delta\UD{\bm A}{\bm i} \\
{{\dot W}_{XX}}\,\UD{\bm A}{\bm i}(\lambda_\calO) &=& 0.
\eean
In the final step we rewrite the equations above replacing the components $W_{XX}$ with the
corresponding components of $m$ as variables. The relation between the variables is given by (\ref{eq:msubmatrix}).
The ODE's take the form of
\bea
\ddot {m}\UD{\bm A}{\bm i } - R\UD{\bm A}{\bm\mu\bm\nu \bm C}\,l^{\bm\mu}\,l^{\bm\nu} \,{m}\UD{\bm C}{\bm i} &=&  R\UD{\bm A}{\bm\mu\bm\nu \bm i}\,l^{\bm\mu}\,l^{\bm\nu} \label{eq:mODE1}\\
{m}\UD{\bm A}{\bm i}(\lambda_\calO) &=& 0 \label{eq:mODE2}\\
{\dot m}\UD{\bm A}{\bm i}(\lambda_\calO) &=& 0. \label{eq:mODE3}
\eea 
The value of the solution at $\lambda_\calE$ yields $m$ in the parallel-propagated SNF.

We immediately note that the ODE is an inhomogeneous version of (\ref{eq:DODE1}) with two more components involved. The inhomogeneity is proportional to the Riemann tensor and the initial data is vanishing at $\lambda_\calO$. 
Thus, in a flat space, the resulting operator always vanishes, unlike the Jacobi map. A nonvanishing mapping $m$ is therefore always an effect of the spacetime curvature.

Just like the four BGO's,   $m$ and $\calD$ are two different functionals of the Riemann tensor (or, more precisely, the optical tidal matrix $R\UD{\bm \alpha}{\bm \mu \bm \nu \bm \beta}\,
l^{\bm \mu}\,l^{\bm \nu}$) along the fiducial null geodesic, given by the solutions two different matrix ODE's.  Therefore their values are  in general unrelated to each other.

\subsection{Remarks}

\paragraph{Direction deviation formula in a seminull frame.} The derivation of the direction variation formula (\ref{eq:directiondeviationabstract}) has been presented in an abstract and covariant manner, highlighting this way
the coordinate- and frame-invariance of the formalism. It is nevertheless very instructive to rewrite it in a parallel-propagated SNF. It takes the form of
\bea
\calD\UD{\bm A}{\bm B}\,\Delta l_\calO^{\bm B}  = \left(\delta x_\calE - \delta \hat{x}_\calO\right)^{\bm A} - {m}\UD{\bm A}{\bm i}\,\delta x_\calO^{\bm i}. \label{eq:dirdevsnf}
\eea

\paragraph{The perpendicular part of the emitter-observer asymmetry operator.} We have already seen when discussing the Jacobi operator that the components corresponding to two spatial directions perpendicular to $u$ tend to decouple from the other two in the GDE. This suggests that a similar independence may be present in the spatial submatrix of $m$. In order to show it we define the \emph{perpendicular part of the emitter-observer asymmetry operator} $ m_\perp : \calP_\calO \to \calP_\calE$ as the restriction of $m$ to the
subspace $\calP_\calO$:
\bean
 m_\perp = m\big|_{\calP_\calO}  
\eean
In a SNF $m_\perp$ can be represented by a 2-by-2 submatrix of $m$. The reader may also check that the ODE's (\ref{eq:mODE1})-(\ref{eq:mODE3}) for its components 
${m_\perp}\UD{\bf{A}}{\bf{B}} \equiv m\UD{\bf{A}}{\bf{B}}$ indeed decouple from the
equations for the remaining two components $m\UD{\bf A}{\bf 0}$. We emphasize that while the perpendicular part $m_\perp$ of $m$ is defined independently of any observer, the
splitting of $m$ into the perpendicular part and the timelike components $m\UD{\bf 0}{\bf A}$ requires fixing an observer with a 4-velocity vector $u$.

\paragraph{The geodesic deviation equation as an ODE in the quotient spaces.} The ODE systems  (\ref{eq:DODE1})-(\ref{eq:DODE3}) and (\ref{eq:mODE1})-(\ref{eq:mODE3}) can also be interpreted as equations defined directly in the quotient spaces $\calQ_p$ and 
$\calP_p$ \cite{Korzynski:2018}. Namely, it is possible do define the covariant derivative $\nabla_l$ as a differentiation in the quotient spaces
and pull back the optical tidal matrix to $\calQ_p$. This way we avoid specifying a whole SNF including an observer $u^\mu$. This interpretation highlights
observer-independence of the spatial components of the operators considered.

\paragraph{The emitter-observer asymmetry operator extended to the full tangent space. }It is possible to extend the domain of $m$ to the whole tangent space, defining $m_{ext}(X) = m([X])$, i.e. with identically vanishing $m_{ext}(l)$. This is the way $m$ has been defined in \cite{Korzynski:2018}. The extended operator satisfies a similar ODE of type
\bean
\ddot {m}_{ext}\,\UD{\bm A}{\bm \beta } - R\UD{\bm A}{\bm\mu\bm\nu \bm C}\,l^{\bm\mu}\,l^{\bm\nu} \,{m_{ext}}\,\UD{\bm C}{\bm \beta} &=&  R\UD{\bm A}{\bm\mu\bm\nu \bm \beta}\,l^{\bm\mu}\,l^{\bm\nu} \\
{m_{ext}}\UD{\bm A}{\bm \beta}(\lambda_\calO) &=& 0 \\
{{\dot m}_{ext}}\,\UD{\bm A}{\bm \beta}(\lambda_\calO) &=& 0.
\eean

\section{Parallax, position drift and the physical meaning of the optical operators} \label{sec:physical}

In this section, we will discuss the physical meaning of the optical operators $\calD$ and $m$, and in particular, we will point out their relation to the parallax measurements, but
before that, we need to define what precisely we mean by parallax in the context of general relativity. 

\subsection{Parallax in general relativity} \label{sec:parallaxnotions}

The notion of parallax is straightforward to grasp in a flat space and in nonrelativistic context, but its generalization to general relativity is more ambiguous \cite{rasanen},
with various researchers using different definitions \cite{mccrea, weinberg-letter, kasai, rosquist, rasanen, Marcori:2018cwn}. We will, therefore, begin this section by a short clarification what one can mean by parallax in the context of general relativity.
 
In the broadest possible sense parallax is the difference of the apparent position of an object on the celestial sphere when regarded from at least two different points of view.
The difference may be due to many effects, including the gravitational light bending. 
Note that the definition above requires a method of comparing the celestial spheres defined at a different point of the spacetime, and possibly registered by different observers. 
In GR this already introduces a great deal of ambiguity, as there exist infinitely many ways to identify points in two distinct celestial spheres. The problem is made
even more complicated by the fact that the positions on the sky, defined as vectors in the sphere of direction $\textrm{Dir}(u_\calO)$, depend on the observer's 4-velocity $u_\calO$ via the aberration effect, see Eq. (\ref{eq:directionformula}).
This means that the direction identification is also nontrivial among two observers at the same event but with different 4-velocities $u$ and $v$.
While infinitely many such identifications are possible, it is reasonable to require that they preserve the metric structure of the celestial sphere, i.e. the
angle measured between any two points on the sky should be invariant under the identifications. 

The definitions of parallax can be classified according to the way we compare the directions on the sky at different events and according to how we select the points between
which we make the comparison. In the literature the possibilities have been considered:

\paragraph{Parallax with respect to the local inertial frame.} If the region of spacetime in which we consider the measurements
can be considered locally flat, like $N_\calO$, then we can employ the parallel transport of vectors for the purpose of direction identification.
Physically this assumption means that we use \emph{local physical phenomena} and the \emph{local geometry} to define the notion of parallel directions on the celestial spheres among all nearby observers. Assume we fix an observer with a given 4-velocity $u_\calO$
at one of the observation points, say $\calO$, and a compatible SNF $(u_\calO,e_{\bm A}, l_\calO)$. Its parallel transport $(\hat u_\calO,\hat e_{\bm A}, \hat l_\calO)$
defines corresponding frames at all points in $N_\calO$. This way we have defined a way to compare the directions on the sky for
all nearby observers whose 4-velocities are equal to $\hat u_\calO$: the spatial vectors $\hat e_{\bm A}$ and the projection of the null direction $\hat l_\calO$ define 3
orthogonal directions on the spheres of direction $\textrm{Dir}(\hat u_\calO)$ at each point. These, in turn, allow for angle-preserving identification of all other points among all spheres $\textrm{Dir}(\hat u_\calO)$.

In practice, the observations at different points will most likely be performed by observers with different 4-velocities $v \neq \hat u_\calO$. A simple way around that is to assume we
perform the corresponding boost of the celestial sphere right after the observations, subtracting this way the effects of the Bradley aberration and reconstructing this way the sky looks at a given point for a fictitious observer with 4-velocity $\hat u_\calO$. The change of the apparent position of an object in that reference frame will be referred to as
\emph{the parallax with respect to a local inertial frame}.  

\paragraph{Classic parallax. } In general, the observations of a distant object performed at different events will register light emitted in different moments along the emitter's worldline. This means that
the result of observations will not only depend on the spacetime geometry and the observer's frame, but also on the motion of the emitter. One may, however, devise a measurement in which many observers will deliberately measure the emitter's position at carefully chosen moments so that all of them register signals emitted exactly at the same moment $\calE$. This can be achieved by appropriate timing of the observations. In the DOA this corresponds to performing
observations within a single null hypersurface $\delta x_\calO^\mu\,l_{\calO\,\mu} = \const$. The observers can be displaced in two perpendicular directions $e_{\bm A}$, but note the displacement
in the direction of $l_\calO$ does not involve any parallax in the DOA. In this measurement we are considering the parallax
of a \emph{single spacetime event} $\calE$, therefore this notion of parallax seems to be the closest in spirit to the nonrelativistic understanding of the term.  Following Räsänen \cite{rasanen}
it will be called the \emph{classic parallax} (in \cite{rosquist, weinberg-letter} it is referred to as the \emph{trigonometric parallax}, and in \cite{kasai} it is simply called \emph{parallax} or \emph{triangulation}). Since the position measurement is made
by comparison with parallel-propagated basis it is a special case of the parallax with respect to the local inertial frame. This type of measurement is by definition completely independent of the motion of the emitting body, i.e. its momentary 4-velocity, 4-acceleration etc., what matters is only its exact position at a single instant of time.

\paragraph{Single worldline parallax.} The measurements of the classic parallax must be made at points separated by spacelike vectors. They require therefore using more than one physical observer separated by sufficiently large distances. In astronomy, this is usually not feasible. Instead, we rely on observations performed along the worldline of a \emph{single} observer, positioned for example on the Earth or on a spacecraft \cite{rasanen, klioner2003}. This observation will
certainly happen on different null hypersurfaces, involving, therefore, light emitted at different moments along the emitter's worldline. The results of observations
will therefore certainly contain a contribution from the  motion of the source of light.

The comparison of the position at different moments may nevertheless proceed as before: knowing the details of the motion
of the observer in the local inertial frame (for example the knowing the Earth's orbit  in the barycentric reference system) we may subtract any effects of the Bradley aberration, R\o mer delays and possibly the local effects of light bending due to the Sun or massive planets
and consider the ``pure'' parallax with respect to a local inertial frame \cite{klioner2003, gaia-astrometry}. The resulting effects of parallax will be referred to as the \emph{single worldline parallax}.
Again it is a special case of the parallax with respect to the local inertial frame.

This is obviously a different type of measurement than the classic parallax. Nevertheless, it is possible to infer about the classic parallax from 
a single worldline measurement under certain additional assumptions and we will discuss the exact relation between these two types of measurement later in this section.

\paragraph{Position drift. } For a single observer with a given worldline there is another natural way of identifying directions on the observer's spheres of directions at different times, one that does not involve an artificial boosting of the sky in order to subtract the aberration effects. Instead of a parallel-transported basis, we may simply consider the Fermi-Walker transport of vectors in the observer's the sphere of directions $\textrm{Dir}(u_\calO)$ along the worldline as our ``fixed directions on the sky'' \cite{Hellaby:2017soj, Korzynski:2018}. The Fermi-Walker transport
preserves the metric structure of the sphere of directions, just like the parallel transport does in a fixed orthonormal frame \cite{fermi1922sopra,mtw}.
Fixed directions defined this way by an observer correspond physically to directions given by a system of gyroscopes carried by the observer during his or her noninertial motion \cite{PhysRevD.87.024031, mtw}.
The variation of the apparent position of an object defined this way will be referred to as the \emph{position drift} \cite{Korzynski:2018}. It obviously depends on the details of
the motion of both the observer and the emitter, but unlike the single worldline parallax, it also depends on the observer's 4-acceleration \cite{Korzynski:2018, Hellaby:2017soj, Marcori:2018cwn}.

\paragraph{Relative parallax. }The notions of parallax defined above use the properties of the local inertial or noninertial frame and the spacetime's local geometry in order to define the reference directions for measuring the variations of the apparent positions. Therefore, they require not only measuring the apparent position of an object at two distinct points but also comparing it to fixed directions defined in the local inertial or noninertial frame. This may be rather complicated to do in practice and, therefore, in astronomy, where one considers instead just the relative changes of the apparent positions of images of two or more objects with respect to each other, without any reference to the notion of fixed directions. The observables, in this case, are given by the variations of angles between the images. Measurements of this kind are
by far the simplest to perform since they only require a single telescope, without a system gyroscopes or other devices defining fixed directions across the observers' region $N_\calO$. This type of parallax may be called the \emph{relative parallax}. Among all possible definitions, this one is the closest to the way parallax is normally
measured in astronomy: the positions of sources and their time variations are expressed with respect to the Solar System's barycentric reference frame, in which the nonrotating axes are determined by the apparent positions of a number of selected extragalactic radio or optical sources \cite{klioner2003} rather than any local physical phenomena.
Therefore, in the end, the parallax measurements by Gaia or Hipparcos rely only on the relative positions on the celestial sphere of many sources. In the context of relativistic astrophysics and cosmology, relative parallax has been introduced recently (not under that name) by Räsänen, in his theoretical work about the parallax of distant quasars measured by the Gaia mission \cite{rasanen}.

Note that it is very easy to obtain the relative parallax from the parallax with respect to the local inertial frame or the position drift. Since both methods
of identifying directions on the sky preserve the metric structure of the sphere of directions the rate of change of the angle between any two sources can be easily expressed
via their momentary positions and their variations using simple trigonometry:
let $\vec r_1$ and $\vec r_2$ be normalized, spatial 3-vectors pointing towards the apparent  positions of two sources in $\textrm{Dir}(u_\calO)$. Then the angle
$\alpha$ between their images is given by their scalar product via $\cos \alpha = \vec r_1 \cdot \vec r_2$. Therefore a small variation of $\alpha$ is
expressible in terms of the variations of $\vec r_1$ and $\vec r_2$, defined by the parallax with respect to a local inertial frame or by the position drift:
$-\delta\alpha\,\sin\alpha = \delta \vec r_1 \cdot \vec r_2 + \vec r_1 \cdot \delta \vec r_2$.
However, going in the opposite direction is not possible: the change of relative
angles between images of an arbitrarily large number of sources is not enough to recover the values of the image drifts with respect to the local inertial frame. This is because passing to the relative parallax involves the loss of information about the rigid rotation of the whole celestial sphere with respect to the fixed directions in the local inertial frame.
This would manifest itself by a secular, average rotation the images of the faraway sources like quasars when compared to the Solar System's local nonrotating frame.
In the standard cosmological model it is most likely very small due to negligible vorticity on large scales, but note that it is by definition unobservable using only the relative parallax measurements.
 The (unphysical) Goedel metric provides a simple and elegant example of this effect, as discussed in \cite{rasanen}.

In the considerations above we have assumed the region $N_\calO$ to be strictly flat. It is possible however to extend the applicability of all definitions to a more
physically relevant case when the metric is flat plus small, localized Newtonian or post-Newtonian perturbations due to the presence of local masses. One may then use the fictitious, background
flat metric to define fixed directions in $N_\calO$ along the same lines. This would correspond to using the inertial frame defined far away from the local masses as a reference for the parallax. Note, however, that in order to obtain the ``pure'' parallax from the observations of a source one would need to subtract all the local gravitational effects such as the light bending due to the presence of the local masses \cite{klioner2003}.

\subsection{Jacobi operator and the image distortions} \label{sec:Jopimd}

We now move on to the discussion of the role of the optical operators. We begin by explaining the physical meaning of the Jacobi operator. The material presented here consists standard results (see for example the
review paper \cite{perlick-lrr}), but will serve as an introduction to the results of the
next sections in which explain the physical significance of the other optical operator $m$.

\bfi
\centering
\includegraphics[width=0.9\textwidth]{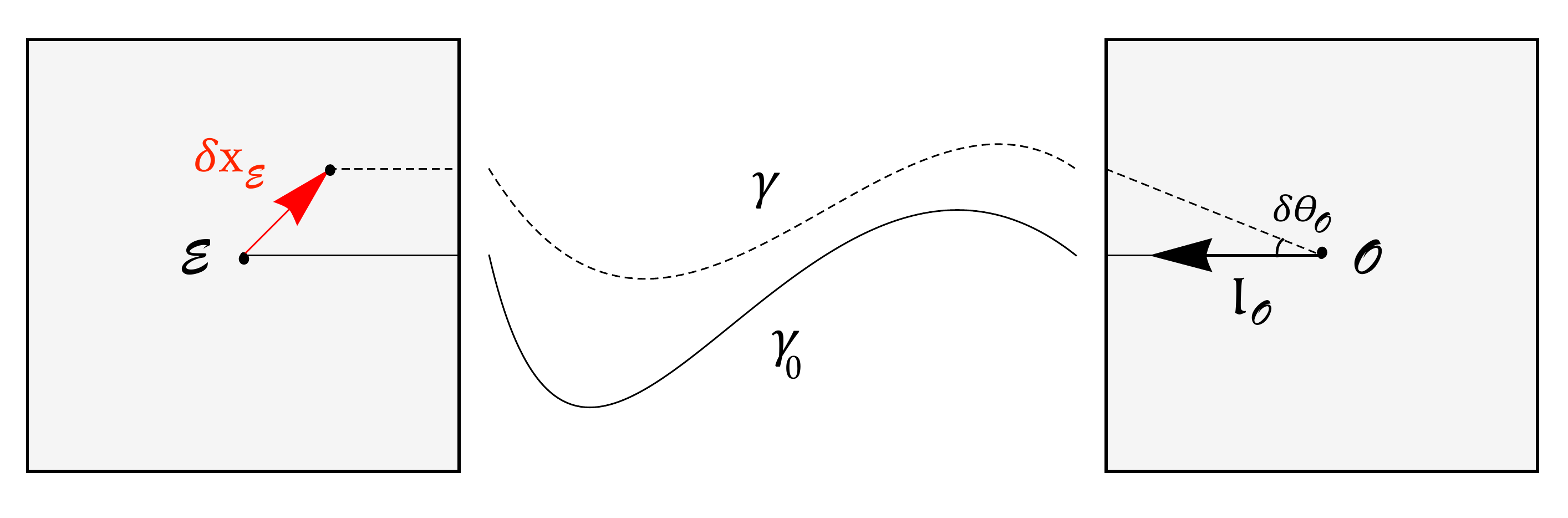}
\caption{A view of the projection of the null hypersurfaces containing $\calE$ and $\calO$. The angle at which the observer sees another point in an extended emitter's
cross section is determined by $\delta x_\calE$, the Jacobi operator and the observer's 4-velocity.}
\label{fig:magnificationmatrix}
\efi

Consider a single observer performing an observation at $\calO$ ($\delta x_\calO = 0$) and a body of finite size passing through $\calE$, see Fig. \ref{fig:magnificationmatrix}. 
Obviously $\delta x_\calO^\mu \,l_{\calE\,\mu} = 0$ and from the time lapse formula (\ref{eq:timelapse1}) we see that the observer registers light from various points of the body emitted
 at the moment given by the condition $\delta x_\calE^\mu \,l_{\calE\,\mu} = 0$.  Thus when we pass to the quotient space we have $\left[\delta x_\calE\right] \in   \calP_\calE$. 
 Therefore Eq. (\ref{eq:dirdevsnf}) takes the form of
 \bea
 \Delta l_\calO^{\bm A} = {\calD^{-1}}\UD{\bm A}{\bm B}\,\delta x_\calE^{\bm B}. \label{eq:directionfromEdisplacement}
\eea
The equation above yields a linear relation between the deviation of the direction of light propagation at $\calO$ and the displacement of a point on the emitting body's cross section 
from $\calE$. This latter can be related to the physical distance from $\calE$ on the emitter's screen space as measured in the body's own frame, regardless of the
emitter's momentary motion: we noted that the distances between points on the cross section, as measured in the body's reference frame, 
 are given by the distances evaluated in $\calP_\calE$ using its internal metric, irrespective of the body's 4-velocity $u_\calE^\mu$. 

\paragraph{Magnification matrix.}
We would like to relate the distances on the emitter's screen space to the angles measured at the observer's sky. This requires taking into account the Bradley aberration effects. According to (\ref{eq:positionapprox}) and (\ref{eq:position2}) we have
\bea
\delta \theta_\calO^{\bm A} = \frac{1}{l_{\calO\,\sigma}\,u_\calO^\sigma}\,{\calD^{-1}}\UD{\bm A}{\bm B}\,\delta x_\calE^{\bm B}   \label{eq:positionfromEdisplacement}
\eea
with $u_\calO^\sigma$ being the observer's 4-velocity. The combination
\bea
M\UD{\bm A}{\bm B} = \frac{1}{l_{\calO\,\sigma}\,u_\calO^\sigma}\,{\calD^{-1}}\UD{\bm A}{\bm B} \label{eq:maginficationmatrix}
\eea
is called the \emph{magnification matrix} in the gravitational lensing theory (the definition here relates the perpendicular displacements on the source plane directly to angles on the sky rather than the position on the image plane, but these two are related by a simple rescaling). It has the dimension of $1/L$, where $L$ denotes length. It depends on the observer's  motion via the aberration effect: observers with different 4-velocities observe the celestial sphere transformed by a conformal mapping with respect to each other. 
This transformation makes 
certain parts of the sky appear larger or smaller depending on the observer's 4-velocity but without any shape distortions of small objects. In (\ref{eq:maginficationmatrix}) this dependence is encoded in the
$\left(l_{\calO\,\sigma}\,u_\calO^\sigma\right)^{-1}$ factor in front of the Jacobi matrix. The magnification matrix for weak lensing in a perturbed Friedmann-Lema\^{\i}tre-Robertson-Walker (FLRW) Universe, together with corrections due to proper motions, has been calculated in \cite{Bonvin:2008ni}.

\paragraph{Angular diameter distance.}
The magnification matrix gives explicitly the relation between perpendicular distances on the emitter's side and in his/her frame and the angles on the observer's sky. 
Its determinant measures the relation of a light emitting body's cross sectional area $A_\calE$  to the stereographic angle $\Omega_\calE$ it takes up on the sky. 
The ratio between these areas is used to define the \emph{angular diameter distance} to the body, also known as the \emph{area distance}:
\bea
 D_{ang} =\sqrt{\frac{A_\calE}{\Omega_\calE}}. \label{eq:Dangdef1}
\eea
 This quantity is related to the magnification matrix via
\bea
D_\ang = \left|\det M\UD{\bm A}{\bm B}\right|^{-1/2}.  \label{eq:Dangdef2}
\eea
Equivalently, we can express the angular diameter distance directly via the Jacobi map:
\bea
D_\ang = \left(l_{\calO \,\sigma}\,u_\calO^\sigma\right)\,\left|\det \calD\UD{\bm A}{\bm B}\right|^{1/2}. \label{eq:Dang}
\eea
Just like $M\UD{\bm A}{\bm B}$, $D_{ang}$ depends on the observer's  motion via the aberration effects, but not on the emitter's motions. It is related
to the luminosity distance $D_{lum}$ via the Etherington's formula
\bean
 D_{lum} = D_\ang (1 + z)^2,
\eean
see \cite{perlick-lrr, etherington}, the latter republished as \cite{etherington2}.
 The off-diagonal part of $M\UD{\bm A}{\bm B}$ contains the information about the image distortions \cite{schneider-kochanek-wambsganss}.

A simple physical interpretation of the angular diameter distance can be explained as in Fig. \ref{fig:triangles}.
\bfi
\centering
 \includegraphics[width=0.7\textwidth]{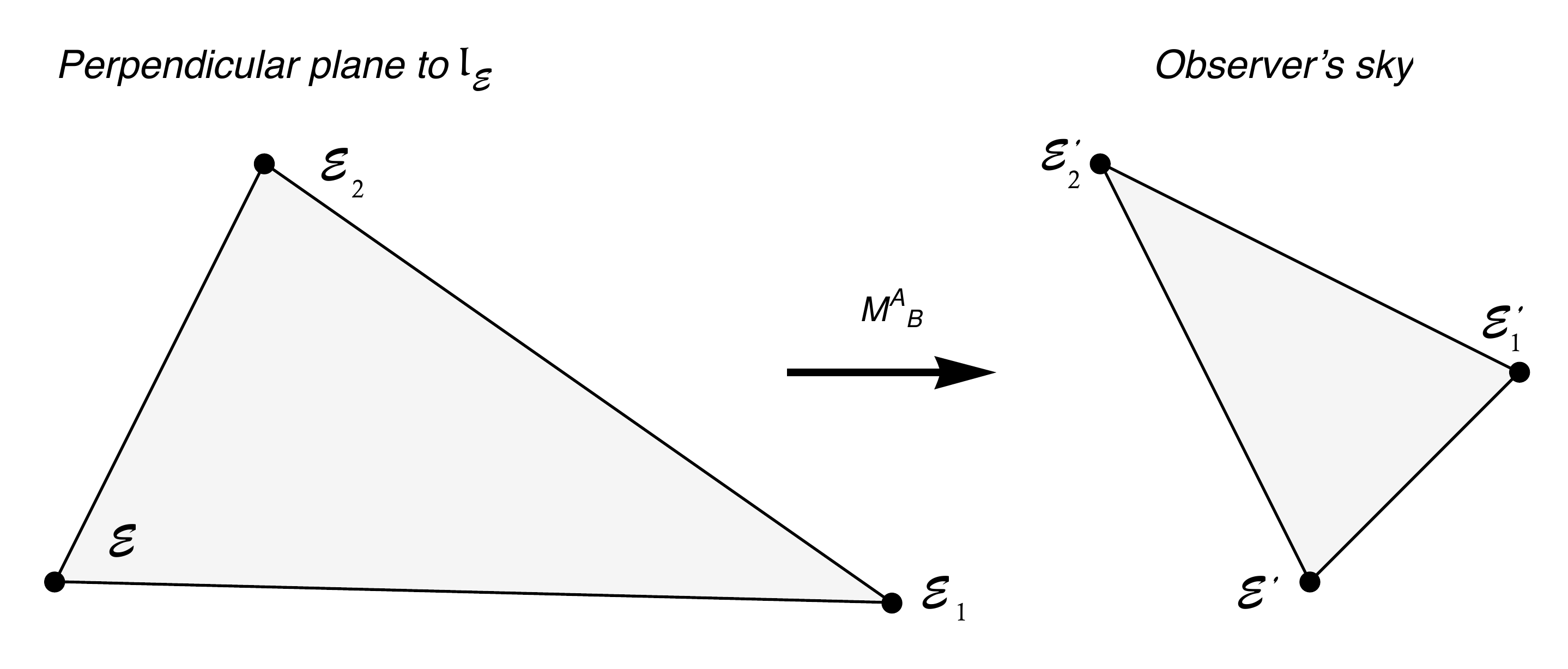}
\caption{A pointlike luminous object at $\calE$ and two other pointlike luminous objects nearby, forming a triangle on the cross section plane
in the direction of the propagation of light. The angular diameter distance between $\calO$ and $\calE$ is obtained as the square root of the ratio between
the area of the triangle on the cross section plane and
the stereographic angle occupied by the triangle with vertices at the apparent positions of the 3 objects at the observer's sky.}
\label{fig:triangles}
\efi
Assume we observe from the point $\calO$ the light emitted by three pointlike objects at the events $\calE$, $\calE_1$ and $\calE_2$, all lying on
the same null hypersurface $l_{\calE\,\mu}\,\delta x_{\calE}^\mu = 0$.  Their projections from 
a triangle of area $A_\calE$ on the cross sectional plane perpendicular to $l_\calE$ (this value is independent of the choice of the observer in $\calE$ due to the Sachs shadow theorem). On the other hand, their images form a triangle on the observer's sky, covering the stereographic angle of $\Omega_\calE$. 
In a spacetime with strong lensing the image may be subject to a strong deformation, including a rescaling, shear, and rotation, but
the angular diameter distance is defined simply by the ratio of $A_\calE$ and $\Omega_\calE$ via (\ref{eq:Dangdef1}).
 
\subsection{Emitter-observer asymmetry operator and the classic parallax} \label{sec:EOasymmetry}

\bfi
\centering
\includegraphics[width=0.8\textwidth]{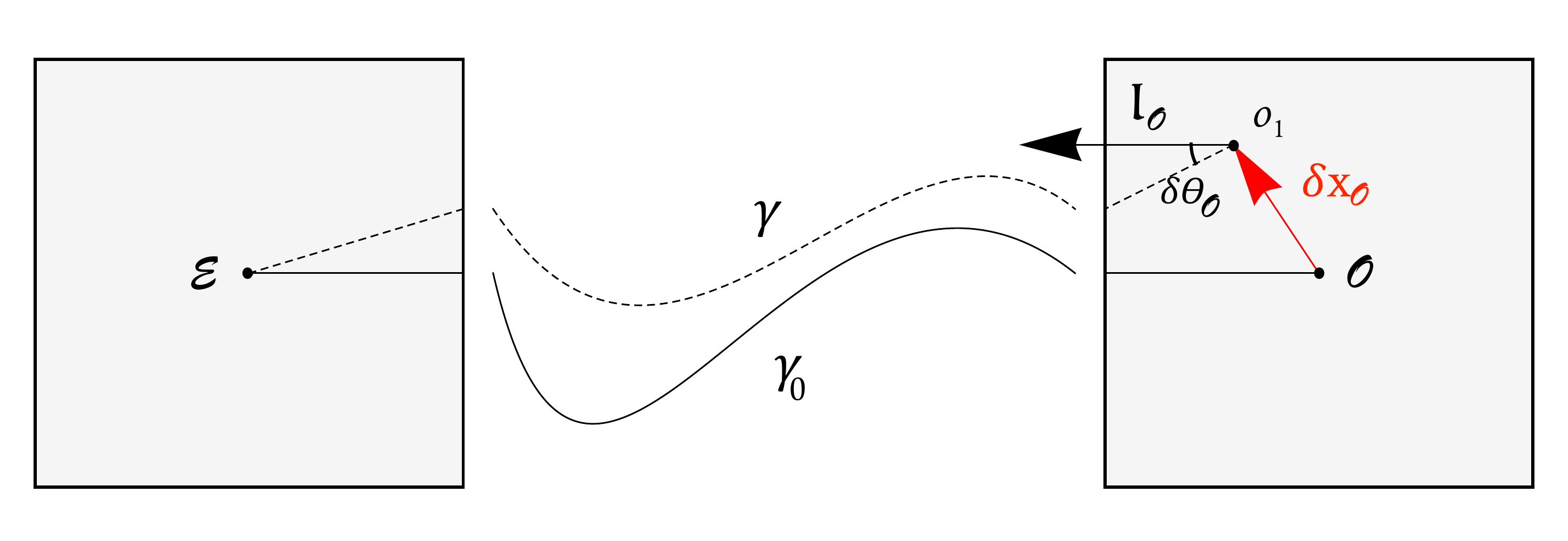}
\caption{A view of the projection of the null hypersurfaces containing $\calE$ and $\calO$. The angle at which the displaced observer sees 
the point source at $\calE$ as compared to the one at $\calO$, i.e. the classical parallax, 
is determined by the Jacobi operator, the emitter-observer asymmetry operator, and the observer's 4-velocity.
}
\label{fig:parallaxmatrix}
\efi

Consider now the opposite situation: we have a single point source emitting light while passing through the point $\calE$ and a number of observers $o_1$, $o_2$, ... , displaced with respect to each other and such that $o_1$ passes through $\calO$. Each observer $o_i$ will register the emitter's apparent position $e_i$ on their celestial spheres.
 We assume that the observers register the emitter's position on the sky using light emitted only at point $\calE$, i.e. we have $\delta x_\calE^\mu = 0$. From (\ref{eq:timelapse1}) this means that they all register the position of the source \emph{at the moment they cross the null 
hypersurface $\delta x_\calO^\mu\,l_{\calO\,\mu} = 0$}.
For simplicity we also assume that they are comoving: their 4-velocities at the moment of measurement are exactly the same, i.e. in locally flat coordinates they are all equal to a fixed $u_\calO^\mu$. This way we do not need to consider the aberration effects when comparing the results of their measurements.

Since $\delta x_\calO^\mu$ must be orthogonal to $l_\calO^\mu$ for all observers at the moment of observation we have
$[\delta x_\calO] \in \calP_\calO$ for the equivalence class of their displacement vectors. From (\ref{eq:dirdevsnf}) we get the following relation
between the observers' displacement vectors and the direction deviation:
\bea
\calD\UD{\bm A}{\bm B}\,\Delta l_\calO^{\bm B} = -\delta \hat x_\calO^{\bm A} -
{m_{\perp}}\UD{\bm A}{\bm B}\,\delta \hat x_\calO^{\bm B} = -(\delta\UD{\bm A}{\bm B} +
{m_\perp}\UD{\bm A}{\bm B})\,\delta \hat x_\calO^{\bm B} \label{eq:directionfromOdisplacement}.
\eea
We see that the Eq. (\ref{eq:directionfromOdisplacement}) differs from (\ref{eq:directionfromEdisplacement}) by the sign at the position
displacement term and the presence of a term involving the perpendicular part $m_\perp$ of the emitter-observer asymmetry operator $m$, see Fig. \ref{fig:parallaxmatrix}.
In order to elucidate its physical meaning, we will now compare the situation when $m_\perp$ vanishes (for example because the spacetime is flat) and when it does not.
 
In a flat space, where $m_\perp$ vanishes between $\calE$ and
$\calO$, (\ref{eq:directionfromOdisplacement}) is formally identical to (\ref{eq:directionfromEdisplacement}) with $\delta x_\calE$ 
replaced by $-\delta \hat x_\calO$. 
In other words, a perpendicular displacement of the observer $\left[\delta
x_\calO\right]$ is
precisely equivalent to a displacement of the emitter in the opposite
direction, i.e. $\left[\delta x_\calE \right] = - \left [ \delta \hat x_\calO
\right]$. The notion of opposite direction, used here for displacement vectors
defined at two different endpoints of the fiducial null geodesic $\gamma_0$,
is defined by the parallel transport along it.

These conclusions follow already from (\ref{eq:dirdevsnf}), where, in the absence of $m$, any value of the left-hand side (LHS)
can be attributed to either $\delta \hat x_\calO$ or $-\delta x_\calE$
at will. This is easy to understand geometrically if we realize that in the flat space the perpendicular
displacement vectors and the null geodesics form in this case a thin but long
parallelogram,
 see Fig. \ref{fig:parallelogram}. The two null geodesics corresponding to the displacements on two ends of $\gamma_0$ 
 are initially parallel, so the angles between them and the direction of the fiducial geodesic must be the same.

\bfi
\centering
 \includegraphics[width=0.8\textwidth]{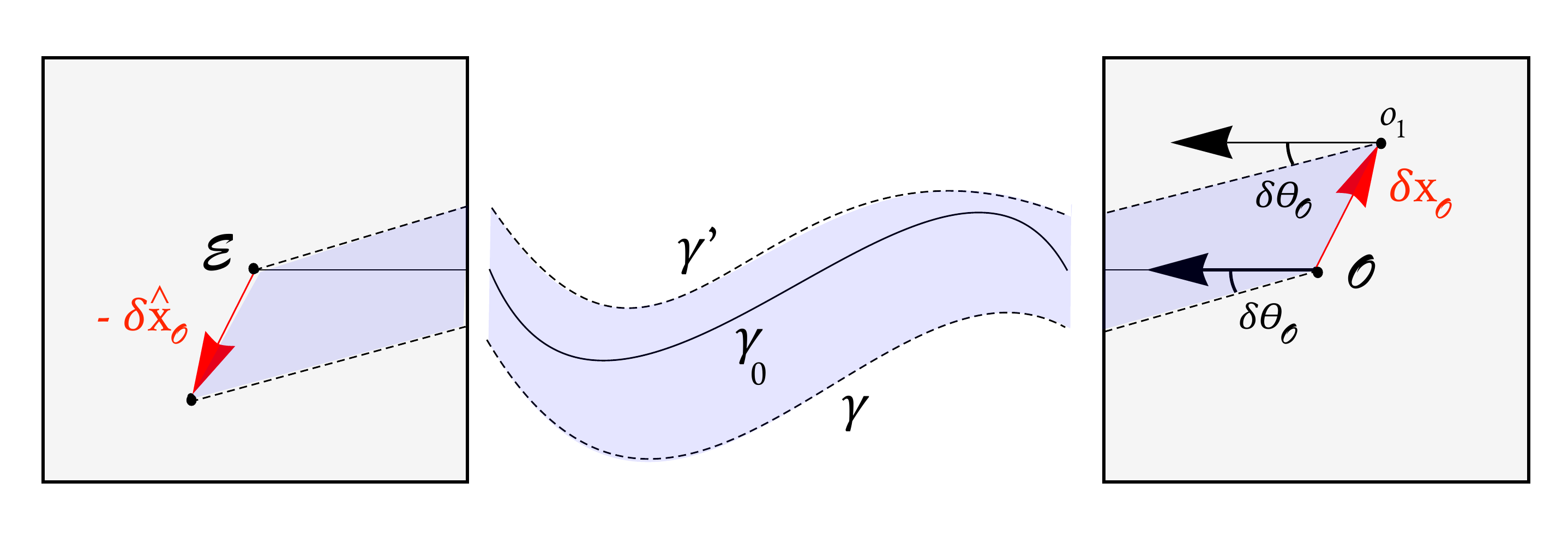}
\caption{If the perpendicular part of the emitter-observer asymmetry operator $m_\perp$ vanishes, it is possible to draw a thin and long parallelogram made of two, initially parallel
null geodesics $\gamma$ and $\gamma'$, the displacement vector $\delta x_\calO$ and its  parallel transport $-\delta \hat x_\calO$. $\gamma_0$ plays the role
of its diagonal. In this
 case determining the parallax angle for an observer displaced by $\delta x_\calO$ is equivalent to the problem of the angular size of an object extending from 
 $\calE$ to $-\delta \hat x_\calO$. }
\label{fig:parallelogram}
\efi

However, if the curvature is present along $\gamma_0$ then the property of exact equivalence between the
position displacements at the two ends of the null geodesic may be broken by
a term proportional to ${m_\perp}\UD{\bm A}{\bm B}$. From (\ref{eq:dirdevsnf}) it is easy to see that the value of the LHS cannot simply shifted from $\delta x_\calE$ to 
$-\delta \hat x_\calO$ since the operator $m_\perp$ spoils the symmetry. This observation justifies
the name emitter-observer asymmetry operator, proposed in Sec. \ref{sec:directionvariationformula}.

One possible geometric interpretation of the linear operator $m_\perp$ is thus that it is
 the obstruction for the existence of the thin and long parallelogram discussed above, 
made of the two initially parallel, long null geodesics, and two vectors $\delta x_\calO$ and $-\delta \hat x_\calO$.
If $m_\perp$ happens to vanish between $\calO$ and $\calE$ then the parallelogram exists. It follows then that the problem of parallax is effectively equivalent to the problem of image distortion discussed in the previous section despite all gravitational lensing which may happen along the way, see again Fig. \ref{fig:parallelogram}.
If $m_\perp$ does not vanish, then the parallelogram does not exist in general and the equivalence of opposite displacements on both sides of $\gamma_0$ is broken.
 In this case (\ref{eq:directionfromOdisplacement})
yields
 \bean
\Delta l_\calO^{\bm A} = -{\calD^{-1}}\UD{\bm A}{\bm C}\left(\delta\UD{\bm C}{\bm B} +
{m_\perp}\UD{\bm C}{\bm B}\right)\,\delta \hat x_\calO^{\bm B}.
 \eean
 Recall now that we have assumed that all observers are comoving with the observer $\calO$,
i.e. their 4-velocities are equal to $u_\calO^\mu$. Then
the position on the celestial sphere of what a displaced observer will measure is
 \bea
 \delta \theta_\calO^{\bm A} =
-\frac{1}{u_\calO^\sigma\,l_\sigma}\,{\calD^{-1}}\UD{\bm A}{\bm C}\left(\delta\UD{\bm C}{\bm B}
+ {m_\perp}\UD{\bm C}{\bm B}\right)\,\delta \hat x_\calO^{\bm B}. \label{eq:positionfromOdisplacement}
 \eea
This is a linear relation between the perpendicular displacement of the
observer and the change of apparent position as measured for a source at a
single instant
of the source's time, defined with respect to the local geometry, i.e. the classic parallax in the terminology of Sec. \ref{sec:parallaxnotions}.

Note that if the 4-velocities of the observers $o_1$, $o_2$, ...  are not exactly equal we need to take into account the aberration effects when comparing
the registered positions on the sky between the observer. These effects will add a 4-velocity-dependent term on top of the
linear term from (\ref{eq:positionfromOdisplacement}). Calculating this term is a standard special relativity problem which we leave to the reader. Here we prefer to assume comoving observers in order to isolate the dependence of the position on the sky on the observer's displacement.


\paragraph{Parallax matrix.} In analogy with the magnification matrix from (\ref{eq:maginficationmatrix})  we can introduce the observer-dependent \emph{parallax matrix}
\bea
 \Pi\UD{\bm A}{\bm B } =
\frac{1}{u_\calO^\sigma\,l_{\calO\,\sigma}}\,{\calD^{-1}}\UD{\bm A}{\bm C}\left(\delta\UD{\bm C}{\bm B}
+ {m_\perp}\UD{\bm C}{\bm B}\right). \label{eq:parallaxmatrix}
\eea
The parallax matrix relates perpendicular distances on the observer's side and  two-dimensional angles measuring the observed position on the sky in comparison with
the position observed by $o_1$ at $\calO$.
Namely, the Eq. (\ref{eq:positionfromOdisplacement}) for the classic parallax takes the form of
\bean
\delta \theta_\calO^{\bm A} = -\Pi\UD{\bm A}{\bm B}\,\delta \hat x_\calO^{\bm B}.
\eean
$\Pi\UD{\bm A}{\bm B}$ is independent of $u_\calE$ because its definition relies on the observations from various points of view of the light emitted by the source in a single
moment. Therefore what matters for the
observation is
only the exact position of the emitter at the moment of observation $\calE$, while the rest of its worldline, which for short times can be approximated by the first two terms in the Taylor expansion, given by the momentary 4-velocity $u_\calE^\mu$ and 4-acceleration $w_\calE^\mu$, is irrelevant. On the other hand, the parallax matrix depends on the spacetime geometry and on the observer's 4-velocity. The dependence on the geometry is via the curvature tensor along $\gamma_0$ because in (\ref{eq:parallaxmatrix}) $\Pi\UD{\bm A}{\bm B}$ is expressed as a function of the optical operators, themselves functionals of the Riemann tensor. The dependence on $u_\calO^\mu$ enters only via the aberration effects, just like in $M\UD{\bm A}{\bm B}$. Thus we have
\bean
\Pi\UD{\bm A}{\bm B} \equiv \Pi\UD{\bm A}{\bm B}\left(R\UD{\mu}{\nu\alpha\beta}, u_\calO^\mu\right).
\eean
Just like $M\UD{\bm A}{\bm B}$, the parallax matrix has the dimension of $1/L$.

\paragraph{Parallax distance.} In astronomy determining the parallax is one of the standard methods of
measuring the distances to objects up to few kiloparsecs \cite{2001A&A...369..339P,  Riess:2014uga}. The method
relies again on the
flat space formula for the parallax matrix: in a flat space we have $\Pi\UD{\bm A}{\bm B} =
d^{-1}\,\delta\UD{\bm A}{\bm B}$, where $d$ is the spatial distance between $\calO$ and $\calE$ measured in the observer's frame.
If we include the relativistic effects of light bending
$\Pi\UD{\bm A}{\bm B}$ is not guaranteed to be proportional to the
unit matrix any more. This leads to the  dependence of
the parallax effect on the orientation of the baseline $\delta
x_\calE^\mu$ \cite{rasanen}.
It is therefore reasonable to try to extract an angle-averaged quantity
out of $\Pi\UD{\bm A}{\bm B}$. In \cite{rosquist, rasanen}  the following definition has been proposed: the parallax distance is
defined to be proportional to the inverse of the expansion of the congruence of null geodesics originating from $\calE$ at $\calO$. In the
language of this paper it is simply inverse of arithmetic average of elements of the diagonal, see \cite{Korzynski:2018}:
\bean
 \widetilde{D}_{par} = \frac{2}{\mathrm {Tr} \left( \Pi\UD{\bm A}{\bm B} \right)}.
\eean
This is a possible generalization, but in this paper we would like to put forward another approach,  analogous to the one used in the
standard definition of the angular diameter distance $D_{ang}$. We define the parallax distance using the
determinant of the
parallax matrix
\bean
 D_{par} = \left|\det \Pi\UD{\bm A}{\bm B}\right|^{-1/2},
\eean
or equivalently
\bea
 D_{par} = (l_{\calO\,\sigma}\,u_\calO^\sigma)\left|\det
\calD\UD{\bm A}{\bm B}\right|^{1/2}\,\left|\det \left(\delta\UD{\bm A}{\bm B} +
{m_\perp}\UD{\bm A}{\bm B}\right)\right|^{-1/2}. \label{eq:Dpar}
\eea
The reader may check that in a flat space this definition yields again the right answer, but in a nonflat space it averages the results over directions in a different way than the
standard one. Let $\pi_1$ and $\pi_2$ be the two roots of the characteristic equation of $\Pi\UD{\bm A}{\bm B}$, possibly real or possibly complex and conjugate to each other. Then
 $\widetilde D_{par}^{-1} = \frac{1}{2}\,\left(\pi_1 + \pi_2\right)$ while
$D_{par}^{-1} = \sqrt{|\pi_1\,\pi_2|}$. Thus the latter definition is equivalent to the inverse of the \emph{geometric} average of the
moduli of the roots, while the former uses the inverse of their \emph{arithmetic} average. As a consequence we see that both values coincide if no shear is present between $\calO$
and $\calE$:  $\Pi\UD{\bm A}{\bm B}$ is proportional to the unit matrix in that case and both roots are equal.

\bfi
\centering
 \includegraphics[width=0.8\textwidth]{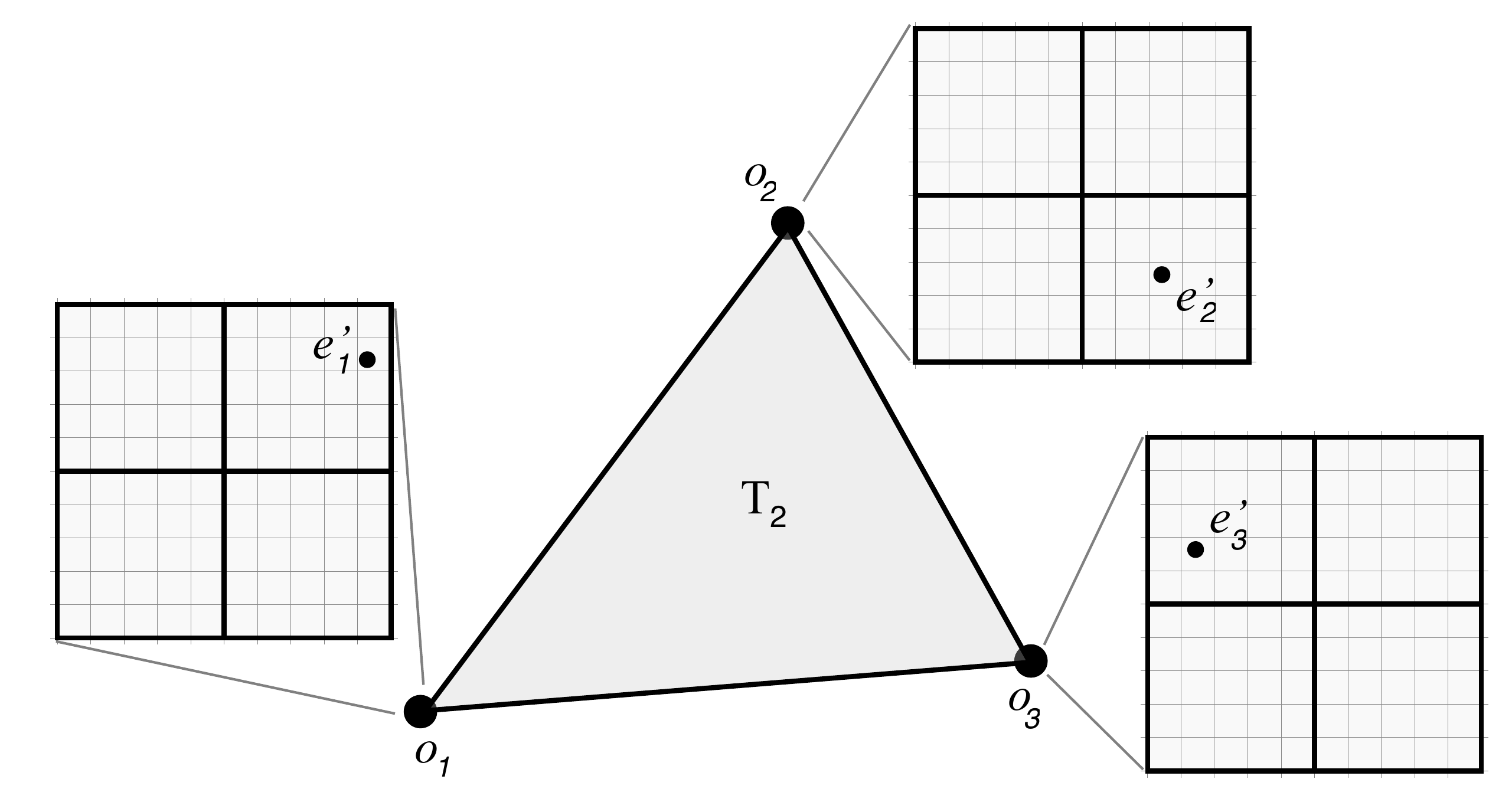}
\vspace{1cm}
\includegraphics[width=0.8\textwidth]{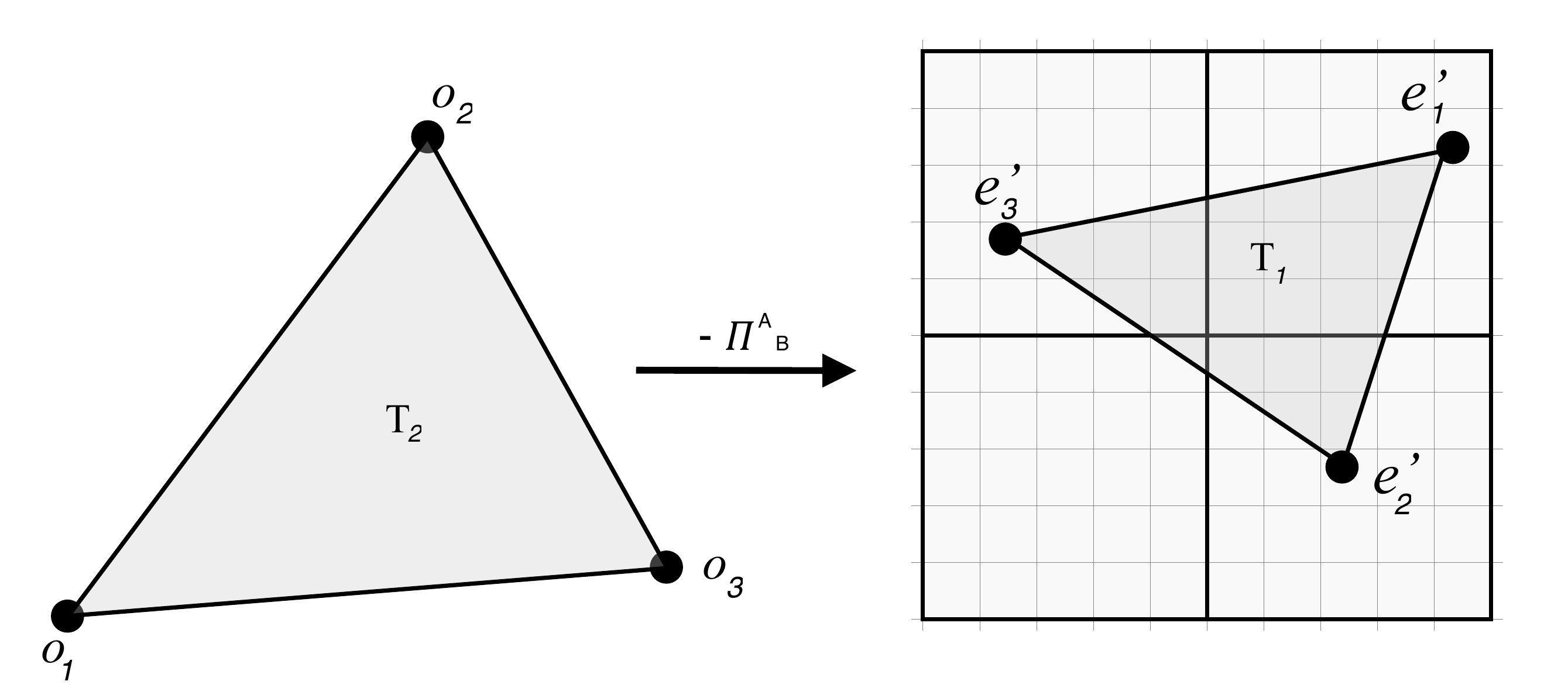}
\caption{A single, pointlike luminous source at $\calE$ observed by an observer $o_1$ at $\calO$ as well as two comoving observers $o_2$ and $o_3$ displaced with respect to $o_1$, performing their measurements on the same null hypersurface. The projection of their positions to the screen space perpendicular to $l_\calO$ yields a triangle
$T_2$ of area $A_\calO$. On the other hand, superimposing the registered positions of the sources by the three observers yields a triangle $T_1$ on the ``combined'' celestial sphere, whose
solid angle area is denoted by $\Omega_\calO$. }
\label{fig:triangles2}
\efi

We will now explain the physical interpretation of the new definition proposed above, highlighting this way the analogy to the notion of the angular diameter distance.  Consider an observer $o_1$ at $\calO$ and two additional  observers $o_2$
and $o_3$, displaced with respect to $\calO$ in two different directions (not necessary perpendicular to each other)
and comoving with $o_1$. They all measure the apparent position of the emitter when it passes through $\calE$ and combine their results of observation on a single celestial sphere $S^2$ by
identifying points on their celestial spheres corresponding to parallel directions in the sense of locally flat coordinates in $N_\calO$.
The result will be a triple of close points on $S^2$ corresponding the three observations and forming a solid triangle $T_1$. On the other hand, we may also consider a physical triangle
$T_2$ these observers form when we project their displacement vectors to the screen space perpendicular to $l_\calO$, see Figure \ref{fig:triangles2}. 
Let $A_\calO$ denote the area of the triangle $T_2$ (again it is independent of the observer's $o_1$ 4-velocity due to the Sachs theorem) and let $\Omega_\calO$ denote
the solid angle taken up by $T_1$. Then we have
\bean
D_{par} = \sqrt{\frac{A_\calO}{\Omega_\calO}}.
\eean
The analogy with the definition of the angular distance is now evident: the definition of $D_{par}$ is equivalent with $D_{ang}$, but with the displacements
considered on the observer's side of $\gamma_0$ instead of the emitter's side. As we will see in Sec. \ref{sec:peculiar}, the compatibility of both definitions opens up the possibility of defining a new observable measuring directly the spacetime curvature. 

Summarizing, the curvature along the line of sight produces two types of effect: the gravitational lensing, which modifies the Jacobi map, resulting in the (de)magnification
and distortion of images seen by observers, and introducing asymmetry between the displacements of the two end of the null geodesic. Both effects are independent and both affect the classic parallax.

\subsection{Position drift formula} \label{sec:positiondriftformula}

Before we consider more realistic models of measurements of the parallax we will rederive the general position drift formula for the momentary rate of change of the apparent position of a source at $\calE$ as observed by an observer in $\calO$ in any spacetime. The formula relates the position drift (or proper motion) to the optical operators between $\calO$ and $\calE$ and to the quantities describing the momentary motions
both the emitter and the observer, namely the emitter's 4-velocity $u_\calE^\mu$, the observer's 4-velocity $u_\calO^\mu$ and the observer's 4-acceleration  $w_\calO^\mu$.
It has already been presented in \cite{Korzynski:2018}, but the derivation there is rather involved. The derivation using the formalism developed here on the other hand is
conceptually simpler and computationally rather
straightforward.

Let $\tau_\calE$ and $\tau_\calO$ denote the proper time as measured by the emitter and observer respectively. Additionally, let $\tau_\calE = \tau_\calO = 0$
when both objects cross $\calE$ and $\calO$ respectively, i.e. at the moment of emission and  observation. After a short while we have $\delta x_\calO^\mu = u_\calO^\mu\,\delta \tau_\calO$ and
$\delta x_\calE^\mu = u_\calE^\mu\,\delta \tau_\calE$ in the leading, linear order in each of the proper times. The time lapse formula (\ref{eq:timelapse1}) yields
\bean
u_\calE^\mu\,l_{\calE\,\mu}\,\delta \tau_\calE =  u_\calO^\mu\,l_{\calO\,\mu}\,\delta \tau_\calO.
\eean
This relation can be turned into a formula for the emitter's time lapse as registered by the observer and compared with his/her proper time lapse:
\bean
\frac{\delta \tau_\calE}{\delta \tau_\calO} = \frac{u_\calO^\sigma\,l_{\calO\,\sigma}}{u_\calE^\rho\,l_{\calE\,\rho}}.
\eean
The ratio on the right-hand side is obviously related to the redshift defined as the relative difference between the photon energy as measured
by $\calE$ and $\calO$:
\bean
z = \frac{u_\calE^\rho\,l_{\calE\,\rho}}{u_\calO^\sigma\,l_{\calO\,\sigma}} - 1.
\eean
Therefore we see that (\ref{eq:timelapse1})  is equivalent to the following relation between the redshift $z$ defined by photon energy change and the rate of the emitter's time lapse the
 observer's time lapse:
\bean
\delta \tau_\calE = \frac{1}{1+z}\,\delta\tau_\calO
\eean
(derived earlier in \cite{perlick, kermack_mccrea_whittaker_1934, Korzynski:2018}).

We can now evaluate the position drift. After substituting the position deviation and simple manipulations the direction variation formula (\ref{eq:dirdevsnf}) yields
\bea
\frac{\Delta l_\calO^{\bm A}}{\delta\tau_\calO} =  {\calD^{-1}}\UD{\bm A}{\bm B}\,\left(\left(\frac{1}{1+z}\,u_\calE - 
\hat{u}_\calO\right)^{\bm B} - m\UD{\bm B}{\bm i}\,u_\calO^{\bm i}\right). \label{eq:driftformula0}
\eea
The ratio on the left-hand side is simply the covariant derivative of the null tangent vector along the observer's worldline evaluated  at $\calO$ and pulled back  to $\calP_\calO$:
\bean
\frac{\Delta l_\calO^{\bm A}}{\delta\tau_\calO} =   \left(u_\calO^\nu\,\nabla_{\nu} l_\calO \right) \Big |_\calO^{\bm A}
\eean
This can be related to the parallax with respect to the local inertial frame by simple rescaling according to (\ref{eq:positionapprox}) and (\ref{eq:position2}). 
For a nongeodesic observer, we may also obtain the expression for the position drift.   
By definition we need to evaluate the Fermi-Walker derivative of $r^\mu$ \cite{Hellaby:2017soj, Korzynski:2018}, given by
\bea
 \delta_\calO r^\mu =  \left(u_\calO^\nu\,\nabla_{\nu} r^\mu \right) + \left(-u_\calO^\mu\,w_{\calO\,\nu} +  w_\calO^\mu\,u_{\calO\,\nu}\right)\,r^\nu,  \label{eq:FWderivative}
\eea
with the transverse components of the term $u_\calO^\nu\,\nabla_{\nu} r^\mu $ given by (\ref{eq:position2}) and (\ref{eq:driftformula0}).
$\delta_\calO r^A$ corresponds to the position drift measured with respect to inertially dragged fixed directions. Combining (\ref{eq:FWderivative}) and (\ref{eq:driftformula0}) yields
\bea
\delta_\calO r^{\bm A} = \frac{1}{l_{\calO\,\sigma}\,u_\calO^\sigma}\,{\calD^{-1}}\UD{\bm A}{\bm B}\,\left(\left(\frac{1}{1+z}\,u_\calE - 
\hat{u}_\calO\right)^{\bm B} - m\UD{\bm B}{\bm i }\,u_\calO^{\bm i}\right) + w_\calO^{\bm A} \label{eq:Positiondrift}
\eea
for the only 2 nonvanishing components of $\delta_\calO r^{\bm \mu}$ in a SNF.
The last term is the perpendicular component of the observer's 4-acceleration. It corresponds to the special relativistic effect of the position drift due to the drift of the aberration
\cite{rasanen,Korzynski:2018,Marcori:2018cwn}.
Its influence on the drift of the positions of sources at cosmological distances has been recently discussed in \cite{Marcori:2018cwn}.
For a longer discussion of the position drift formula and its physical and astrophysical consequences see \cite{Korzynski:2018}, here we will just briefly look at the role of the emitter-observer asymmetry operator.

First, consider the situation in which $u_\calE^\mu$ and the parallel-transported $\hat{u}_\calO^\mu$ differ only by a component along the line of sight.
In this case, it is easy to see that the perpendicular component of the 4-velocity difference $\frac{1}{1+z}\,u_\calE - 
\hat{u}_\calO$ vanishes, so the first term in (\ref{eq:Positiondrift}) does not contribute to the drift. In the absence of $m\UD{\bm A}{\bm i}$, this means no drift seen by the observer. However, if the emitter-observer asymmetry operator is not 0 the observer can perceive a ``curvature-induced'' position drift even when both 4-velocities $u_\calE$, $\hat u_\calO$ and
the null vector $l_\calE$ lie on a single 2-plane and the first term in (\ref{eq:Positiondrift}) vanishes. This type of drift is proportional to the timelike component of $m\UD{\bm A}{\bm i}$, i.e. $m\UD{\bm A}{\bm i}\,u_\calO^{\bm i}$.
It is independent of the velocity of the emitter's radial motion with respect to the observer.

Second, we note that in (\ref{eq:Positiondrift})  operator $m$ appears once again in a term which breaks the symmetry between the emitter and observer.
This is not so easy to see at first inspection, because, unlike (\ref{eq:dirdevsnf}), the formula (\ref{eq:Positiondrift}) even without the
$m$ term does not seem antisymmetric with respect to the exchange of $u_\calE$ and $u_\calO$ at first glance because of the $\frac{1}{1+z}$ factor in front of $u_\calE$. This is because in the derivation above we have made a choice which ties the formula to the observer's frame: namely, we relate the rate of change of the apparent position to the lapse of the \emph{observer's} proper time. This choice introduces an asymmetry between $u_\calO$ and $u_\calE$ in (\ref{eq:Positiondrift}) at the level of pure special relativity, even without any GR effects present. Nevertheless, the symmetry breaking role of $m$ can again be seen in the following example.

Consider an emitter at $\calE$ and an observer at $\calO$ for whom not only the perpendicular components of $\frac{1}{1+z}\,u_\calE^\mu - \hat u_\calO^\mu$  vanish, but actually both 4-velocities coincide. This means that both objects are at rest with respect to each other with
the comparison made using the parallel transport along $\gamma_0$. Assume as before that the observer is geodesic ($w_\calO^\mu=0$). It follows then that  $z=0$ and the first term in (\ref{eq:Positiondrift}) vanishes. Just like in the previous example, in the absence of $m$ the curvature-induced drift vanishes, but additionally the whole situation is symmetric with respect to boosting the emitter and observer in the following sense:
consider another emitter $\tilde u_\calE^\mu$ passing through $\calE$, boosted with respect to $u^\mu_\calE$ in a direction orthogonal to the line of sight (orthogonality determined in the $u_\calE$ reference frame). This new emitter will exhibit drift
according to the observer, although in general not in the direction related to the parallel transported direction of its motion. On the other hand, the reader may check that exactly the \emph{same} position drift $\delta_\calO r^A$ can be induced by considering a free-falling, moving \emph{observer}, boosted with respect to $u_\calO^\mu$ with the same velocity but
in the opposite direction (again in the sense of parallel transport) and observing the unboosted emitter.  Just like the displacement equivalence on both ends of $\gamma_0$ noted in Sec. \ref{sec:EOasymmetry}, this boost equivalence property is lost whenever $m_\perp \neq 0$.

We would like to remark that even though we have used the flat light cones approximation and the parallel light rays approximation, the formula derived here is valid without any restrictions regarding the distance between $\calO$ and $\calE$ or the bilocal geodesic operators. This is because the position drift is by definition the first derivative of the position on the sky and the derivative is always obtained by linearization of all relations involved, including the null condition (\ref{eq:nullcond1}) and
the position on the sky formula (\ref{eq:position1}).

\subsection{Single worldline parallax in a general situation} \label{sec:generalparallax}

\bfi
\centering
 \includegraphics[width=0.8\textwidth]{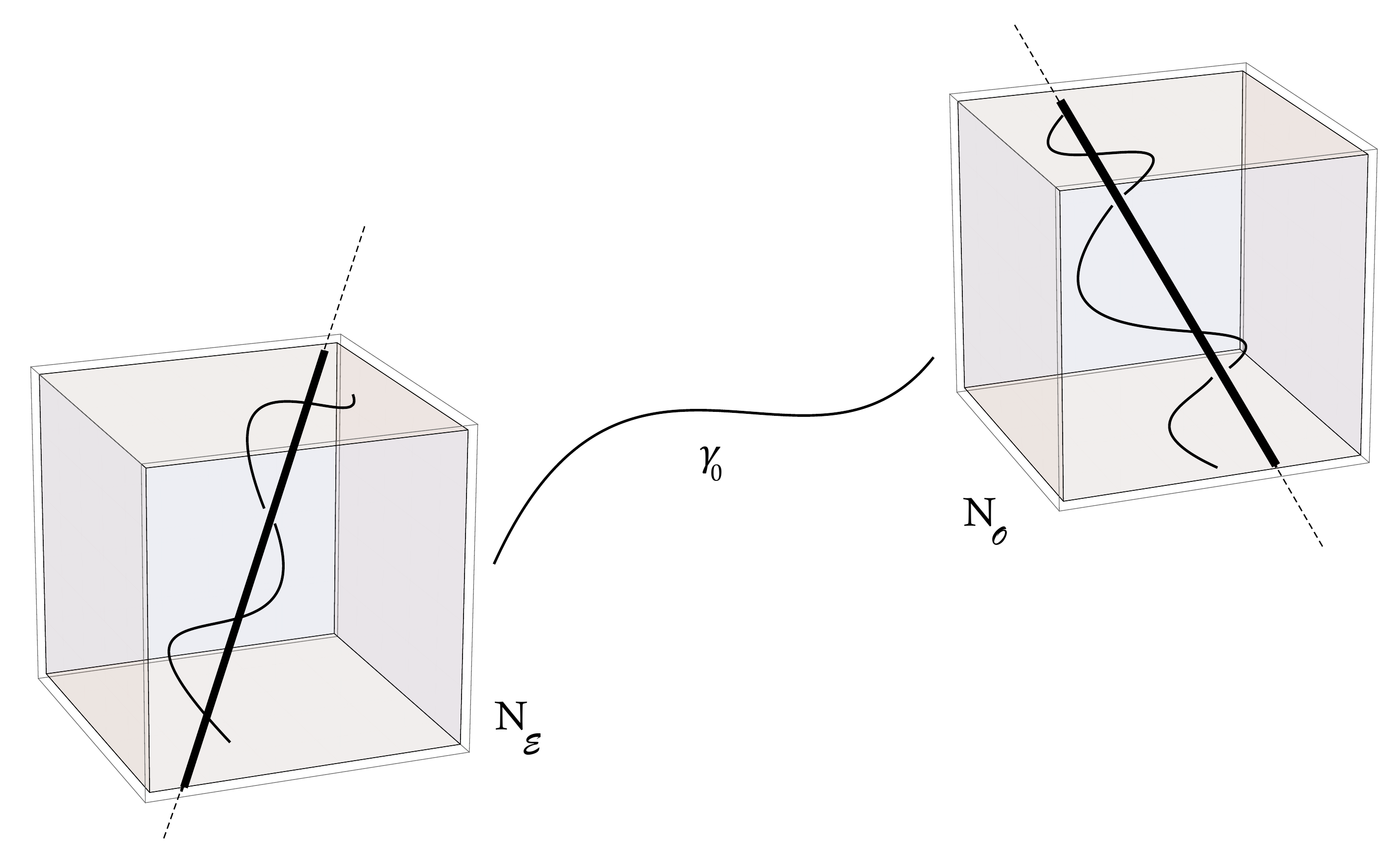}
\caption{ The observer and the emitter in gravitationally bound systems, undergoing short-time orbital motions around their free-falling barycenters. }
\label{fig:generalsituation}
\efi

Finally, we move on to discussing a more realistic model of parallax observations. Assume that both the observer and the emitter are located in gravitationally bound systems whose barycenters
undergo a geodesic motion (free fall) with a good approximation. The gravitational fields binding the systems are assumed to be so weak that the light bending they induce
is negligible -- if it is not then they may be introduced later as small, perturbative corrections, for example using the parametrized post-Newtonian formalism (PPN) \cite{klioner2003}. We assume here that the motion of the observer and the emitter in their respective local inertial frames can be well approximated by short-time, nonrelativistic orbital motions around the barycenters, superimposed on top of a secular motion
of the barycenters with constant 4-velocities $U_\calO$ and $U_\calE$ respectively, see Fig. \ref{fig:generalsituation}. The characteristic timescale of this short-time motion is assumed to be smaller that the size of $N_\calO$ and $N_\calE$, i.e. $L$. Assume that
points $\calO$ and $\calE$ lie on the worldlines of the respective barycenters and 
let $t_\calE$ and $t_\calO$ denote the proper time in the appropriate barycentric reference system (the barycentric coordinate times in the astronomical terminology), defined such that
$t_\calE = 0$ at $\calE$ and $t_\calO = 0$ at $\calO$. Then the momentary position
of the observer can be decomposed according to
\bean
  \delta x_\calO^\mu = U_\calO^\mu \,t_\calO + \sigma^\mu(t_\calO),
\eean
where the momentary position vector $\sigma^\mu(t_\calO)$ is assumed to be orthogonal to $l_\calO^\mu$ rather than $U_\calO^\mu$, i.e.  $\sigma_\mu\,l_\calO^\mu = 0$ (vector $\sigma^\mu$ can be spacelike or null). Similar decomposition
can be used for the emitter:
 \bean
  \delta x_\calE^\mu = U_\calE^\mu \,t_\calE + \rho^\mu(t_\calE)
\eean
with the condition $\rho_\mu\,l_\calE^\mu = 0$. In $N_\calO$ this decomposition is effectively equivalent to introducing a null time coordinate $v_\calO$, consistent with the barycentric coordinate time $t_\calO$ at the barycenter, but whose gradient is proportional to $l_{\calO\,\mu}$, and then splitting the momentary displacement vector into the timelike component and the other component lying on the null surface of constant $v_\calO$ (Fig. \ref{fig:secularmotion}).
\bfi
\centering
 \includegraphics[width=0.8\textwidth]{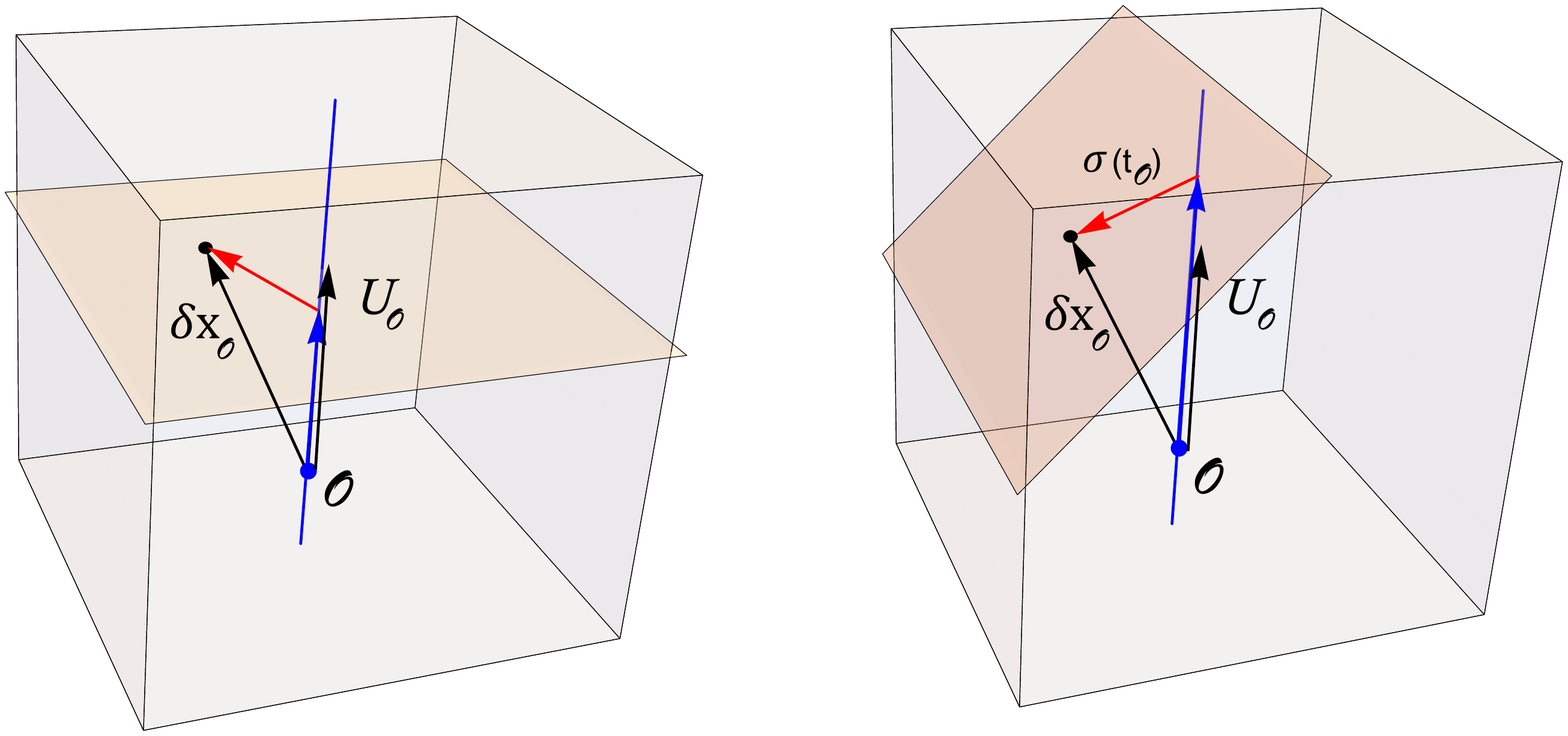}
\caption{ The standard way to describe the position of a point mass $\delta x_\calO$ is to decompose the position vector into the tangent and perpendicular components with respect to the barycenter 4-velocity $U_\calO$ (on the left). Here we use a modified decomposition: $\delta x_\calO$ decomposed into the part parallel to $U_\calO$ and $\sigma$ orthogonal to $l_\calO$. The decompositions differ by the tilt of the 3-plane containing the second component. }
\label{fig:secularmotion}
\efi
This is, in turn, equivalent to the decomposition of $\delta x_\calO$ in an SNF with $U_\calO^\mu$ as the first basis vector.
 The purpose of this operation in both $N_\calO$ and $N_\calE$ is to take into account the R\o mer delays on the observer's and the emitter's side.
Note that the observer's
and the emitter's proper times need to be related their respective barycentric coordinate times, but this is a fairly simple special relativity problem \cite{klioner2003}.

We now consider the apparent position of the emitter on the observer's sky. From the time lapse formula (\ref{eq:timelapse1}) we get the relation between the
barycentric time variables $t_\calO$ and $t_\calE$, calculated at the barycenters:
\bean
t_\calE = \frac{l_{\calO\,\mu}\,U_\calO^\mu}{l_{\calE\,\nu}\,U_\calE^\nu}\,t_\calO = \frac{1}{1+z}\,t_\calO,
\eean
where $z$ is the redshift between the two barycenter frames.
Then from the direction deviation formula in the SNF (\ref{eq:dirdevsnf}) we get 
\bea
\calD\UD{\bm A}{\bm B}\,\Delta l_\calO^{\bm B} = \left( \left( \frac{1}{1+z}\,U_\calE - \hat U_\calO  \right)^{\bm A} - m\UD{\bm A}{\bm i}\,U_\calO^{\bm i}\right)\,t_\calO
+ (\rho - \hat \sigma)^{\bm A} - {m_\perp}\UD{\bm A}{\bm B}\,\sigma^{\bm B}. \label{eq:equationabove1}
\eea
Assume now the observer measures the emitter's apparent position on the sky along his or her worldline, but subtracting the effects of aberration due to his or her motion with respect to the barycenter. In the terminology of Sec. \ref{sec:parallaxnotions} this amounts to the parallax with respect to the local inertial frame, connected with the free-falling barycenter, and given by $U_\calO$. The apparent position with respect to $l_\calO$
can be obtained from (\ref{eq:equationabove1}) combined with (\ref{eq:positionapprox}) and (\ref{eq:position2}). After rearranging the terms
and applying the definitions of the magnification matrix, the parallax matrix and the position drift formula (\ref{eq:Positiondrift}) we obtain
\bea
 \delta \theta^{\bm A} = \delta_\calO r^{\bm A}\,t_\calO + M\UD{\bm A}{\bm B}\,\rho^{\bm B}(t_\calO) - \Pi\UD{\bm A}{\bm B}\,\hat \sigma^{\bm B}(t_\calO). \label{eq:driftplusparallax}
\eea
In this formula $\delta_\calO r^{\bm A}$ in the first term is the position drift rate calculated for a fictitious emitter-observer pair in which both are located at the 
free-falling barycenters of their gravitationally bound systems:
\bean
\delta_\calO r^{\bm A} = \frac{1}{l_{\calO\,\mu}\,U_\calO^\mu}\,{\calD^{-1}}\UD{\bm A}{\bm B}\, \left( \left( \frac{1}{1+z}\,U_{\calE} - {\hat U}_{\calO} \right)^{\bm B} - 
m\UD{\bm B}{\bm i}\,U_\calO^{\bm i} \right).
\eean
$M\UD{\bm A}{\bm B}$ and $\Pi\UD{\bm A}{\bm B}$ are the magnification and the parallax matrix in the observer's barycenter reference frame.
We can see that the total effect is a sum of a secular drift caused by the motions  of the barycenters of both gravitationally bound systems (proper motion), an
oscillatory term due to the
emitter's orbital motions and another oscillatory term due to the observer's orbital motion. The latter two terms depend only on the perpendicular components of the deviation vectors $\sigma^\mu$ and $\rho^\mu$. This is a straightforward consequence of the PRA which neglects the perspective distortions. Note however that in the
presence of curvature they are multiplied by two different matrices: the magnification matrix and the parallax matrix, differing by a curvature correction
proportional to ${m_\perp}\UD{\bm A}{\bm B}$.

If the emitter's short-time motion term is negligible with respect to the third term
then the result is simply a sum of the proper motion and the classic parallax term \cite{gaia-astrometry, gaiadr2-parallaxes}. In the most important case
for modern astronomy, i.e. the question of the solar parallax $\sigma^{\bm A}(t_\calO)$ has a very precisely known functional form with an annual periodicity and can be easily disentangled from the first, secular term \cite{gaiadr2-parallaxes}. This way we may measure the parallax matrix in the
 Solar System's barycenter frame using the standard astrometric observations performed along a timelike worldline of the Earth-based observatory or a spacecraft. Note however that the applicability of this
 procedure relies on the assumptions that underlie the analysis above: that the emitter does not undergo short-scale motions of similar time scale,
 the gravitational field is such that the curvature is roughly constant across a connecting tube of the size of the Earth's orbit, the distortions due to the light bending from nearby masses can be disregarded or subtracted and we know sufficiently well the short-scale  motion of the observer around
 the Solar System's barycenter. The analysis above applies also to the case when the source is positioned at cosmological distances and
we need to take into account the nonflat geometry between the source and the observer, or when the image has undergone strong lensing.
 
 \subsection{Magnification, parallax and position drift near a caustic}

 We can now use formulas (\ref{eq:maginficationmatrix}), (\ref{eq:parallaxmatrix}) and (\ref{eq:Positiondrift}) to discuss the behavior
 of the magnification, classical parallax and the position drift near a typical caustic. Recall that on a caustic the Jacobi map becomes degenerate in at least one direction.
 Assume now $\calD\UD{\bm A}{\bm B}$ is degenerate along a single direction $n^{\bm A}$, i.e. $\calD\UD{\bm A}{\bm B}\, n^{\bm B} = 0$. 
This means that an infinitesimally small image undergoes a formally infinite distortion along $n^{\bm B}$ and an infinite magnification. 
 Consider now the parallax matrix: as long as the curvature along the line of sight 
 is bounded, the emitter-observer asymmetry operator $m_\perp$
 should stay finite. Moreover we may expect the combination $\delta\UD{\bm A}{\bm B} + {m_\perp}\UD{\bm A}{\bm B}$ to be
 an invertible operator in a generic case. Now,
  since (\ref{eq:parallaxmatrix}) contains the inverse of the Jacobi map, we see that unless the term $\delta\UD{\bm A}{\bm B} + {m_\perp}\UD{\bm A}{\bm B}$ happens
 by chance to be degenerate along the very same
 direction $n^{\bm A}$ the parallax matrix will blow up as well.
 In that case there obviously must be a direction $\bar n^{\bm A}$ in $\calP_\calO$ such that the 
 parallax for an observer displaced along it is formally infinite. It is given by $\bar n^{\bm A} = C\,(\delta\UD{\bm A}{\bm B} + {m_\perp}\UD{\bm A}{\bm B})^{-1}\,n^{\bm B}$,
 with $C$ being a positive normalization factor.

 We can also obtain a similar conclusion works for the position drift.
  Looking at (\ref{eq:Positiondrift}) we note that unless  the 4-velocities of the observer and the emitter happen to be aligned in a special way, such that
 the combination $[(1+z)^{-1}\,u_\calE - \hat u_\calO] - m([u_\calO])$ has a vanishing component along $n^{\bm A}$,
 the position drift becomes infinite as well. 
 
 Summarizing, we have just proved that near a caustic the magnification matrix, the parallax matrix and the value of the position drift, registered for a generic observer and emitter pair, will blow up simultaneously as measured on the observer's sky. Of course for real sources, of small but finite size, and for observational instruments of finite resolution, the measured values of the drift, parallax, and magnification will be large but finite. Nevertheless, in a generic situation, all of those effects
become amplified \emph{simultaneously} when the emitter approaches a caustic.

\section{Momentary motions-independent observables} \label{sec:peculiar}

As an example of an application of the formalism presented above, we will show that by combining the data about the classical parallax and the image distortion and magnification
it is possible to define quantities which are entirely insensitive to the momentary 4-velocities of both the observer and the emitter. In other words, we will introduce
momentary motions-independent observables, which measure the geometry of the spacetime between the emission point $\calE$ and the observation point $\calO$, encoded in the optical operators.

Recall that the magnification matrix $M\UD{\bm A}{\bm B}$ and the parallax matrix $\Pi\UD{\bm A}{\bm B}$ do not depend on the emitter's 4-velocity $u_\calE$, but they do
depend on the observer's 4-velocity $u_\calO$ because of the stellar aberration effect. Consider now the combination
\bea
{w_\perp}\UD{\bm A}{\bm B} = \left(M^{-1}\right)\UD{\bm A}{\bm C}\, \Pi\UD{\bm C}{\bm B} \label{eq:wperpdef}
\eea
calculated in a SNF. From (\ref{eq:maginficationmatrix}) and (\ref{eq:parallaxmatrix}) we see that the formula above defines an $u_\calO$-independent quantity, i.e. a combination of observables depending only on the curvature along the line of sight and not on
the kinematical quantities describing the motions:
\bean
  {w_\perp}\UD{\bm A}{\bm B} \equiv {w_\perp}\UD{\bm A}{\bm B}(R\UD{\mu}{\nu\alpha\beta}).
\eean
Geometrically this matrix defines a frame-independent operator $w_\perp: \calP_\calO \to \calP_\calO$. Unlike $M$ and $\Pi$ it is dimensionless.
From (\ref{eq:maginficationmatrix}) and we see that it has a simple expression in terms of the optical operators
\bea
 {w_\perp}\UD{\bm A}{\bm B} = \delta\UD{\bm A}{\bm B} + {m_\perp}\UD{\bm A}{\bm B}. \label{eq:wperpfromm}
\eea
Therefore its deviation from the unit matrix may serve as a measure of the spacetime curvature.

The value of the determinant of  $w_\perp$, calculated in a SNF, is of particular interest. We define the  dimensionless, scalar parameter $\mu$ by
\bea
 \mu = 1 - \det {w_\perp}\UD{\bm A}{\bm B}, \label{eq:mudef}
\eea
or equivalently
\bea
 \mu = 1 - \frac{\det {\Pi}\UD{\bm A}{\bm B}}{\det M\UD{\bm A}{\bm B}}. \label{eq:mudef4}
\eea
Just like $w_\perp$, it depends only on the curvature along the optical axis, but not on the motions of the observer and the emitter. Using (\ref{eq:Dang}), (\ref{eq:Dpar})  and
(\ref{eq:mudef4}) we show that $\mu$ can be
expressed via the parallax and the angular diameter distance:
\bea
\mu = 1 - \sigma\,\frac{D_{ang}^2}{D_{par}^2}, \label{eq:mudef2}
\eea
where $\sigma = \pm 1$ determines the sign of the second term. $\sigma$ depends on sign of the determinants of the magnification and the parallax matrices:
\bean
 \sigma = \textrm{sgn} \det M \UD{\bm A}{\bm B} \cdot \textrm{sgn} \det \Pi \UD{\bm A}{\bm B}.
\eean
In short, we take the minus sign in the second term of (\ref{eq:mudef2}) (i.e. $\sigma = 1$) if both matrices are orientation-preserving (i.e. the image the observer sees is not flipped and neither is the
dependence of the parallax on position deviation) or both are negative  (i.e. if the observer sees an inverted image and at the same time the
linear dependence of the
parallax  on the position has inverted parity with respect to the standard one) and the plus sign ($\sigma = -1$) if only one of them is flipped. The reader may check that for sufficiently small perturbation of
the null geodesics by curvature both  determinants should be positive, so
we have
\bea
\mu = 1 - \frac{D_{ang}^2}{D_{par}^2} \label{eq:mudef3}
\eea
for sufficiently short distances and/or sufficiently weak bending of light rays between $\calO$ and $\calE$. Note that there are no simple  relations analogous to (\ref{eq:mudef2}),  (\ref{eq:mudef}) and (\ref{eq:wperpfromm}) for the other parallax distance $\tilde D_{par}$.

We can see that $\mu$ measures the deviation of the metric from the flat one by comparing the parallax and the angular diameter distances measured to an object
positioned far away. Obviously, both definitions must give the same answer in a flat spacetime, i.e. $\mu = 0$ in the Minkowski space, but nonvanishing curvature along
$\gamma_0$ gives rise to the asymmetry between the observer and the emitter, as we have discussed in Sec. \ref{sec:EOasymmetry}. This, in turn, can make the two
optical methods of determining the distance inequivalent, giving rise to $\mu \neq 0$.
  
It is very instructive to consider $\mu$ in the case of fairly small curvature along the line of sight. We assume we may use the first order perturbation theory in the 
GDE, effectively treating the curvature tensor as a small perturbation. In that case we obtain from (\ref{eq:mODE1}) 
\bean
 { {\ddot m}_\perp}{}\UD{\bm A}{\bm B} \equiv {\ddot m}\UD{\bm A}{\bm B} = R\UD{\bm A}{\bm \mu\bm \nu \bm B}\,l^{\bm \mu}\,l^{\bm \nu},
\eean
in the leading order. After imposing the initial conditions (\ref{eq:mODE2})-(\ref{eq:mODE3}) we obtain the solution as an iterated integral:
\bean
 {m_\perp}\UD{\bm A}{\bm B} \approx \int_{\lambda_\calO}^{\lambda_\calE} \dd \lambda_1 \,\int_{\lambda_\calO}^{\lambda_1}\dd\lambda_2\,R\UD{\bm A}{\bm\mu\bm\nu \bm B}(\lambda_2)\,l^{\bm \mu}\,l^{\bm\nu}.
\eean
We also linearize (\ref{eq:mudef}) around $\mu = 0$, obtaining
\bean
 \mu \approx -{m_\perp}\UD{\bm A}{\bm A} = -{\rm tr}\, m_\perp .
\eean
Thus the leading order contribution to $\mu$ has also the form of an iterated integral
\bea
\mu \approx -\int_{\lambda_\calO}^{\lambda_\calE} \dd \lambda_1 \,\int_{\lambda_\calO}^{\lambda_1}\dd\lambda_2\,R\UD{\bm A}{\bm \mu\bm \nu \bm A}\,l^{\bm\mu}\,l^{\bm\nu}.
\label{eq:muintegral1}
\eea 
Finally we apply the standard decomposition of the Riemann tensor into the Weyl tensor $C\UD{\mu}{\nu\alpha\beta}$ and the Ricci tensor $R_{\mu\nu}$. We note first that the trace over the screen space of the Riemann tensor contracted twice with the null vector $l^\mu$, or the optical tidal matrix, is equal to the full trace of the Riemann tensor contracted in the same way, i.e.
\bean
R\UD{\bm A}{\bm \mu\bm \nu \bm A}\,l^{\bm\mu}\,l^{\bm\nu} = R\UD{\bm \alpha}{\bm\mu\bm \nu \bm \alpha}\,l^{\bm\mu}\,l^{\bm\nu} = -R_{\bm \mu\bm \nu}\,l^{\bm \mu}\,l^{\bm \nu}.
\eean
Thus the Weyl tensor does not contribute to the integral (\ref{eq:muintegral1}). We also notice that for a null vector $l^\mu$ we have
$R_{\mu\nu}\,l^\mu\,l^\nu = G_{\mu\nu}\,l^\mu\,l^\nu$, where $G_{\mu\nu}$ is the Einstein tensor.

In the final step we apply the Einstein field equations
\bean
G_{\bm\mu\bm\nu} + \Lambda \,g_{\bm\mu\bm\nu}= 8\pi G\,T_{\bm\mu\bm\nu}
\eean
contracted with $l^{\bm\mu}\,l^{\bm\nu}$. Again we see that because $l^{\bm\mu}$ is null the cosmological constant $\Lambda$ does not contribute to the integral, so
we obtain
\bea
\mu \approx {8\pi G}\,\int_{\lambda_\calO}^{\lambda_\calE} \dd \lambda_1 \,\int_{\lambda_\calO}^{\lambda_1}\dd\lambda_2\,
T_{\bm \mu\bm \nu}(\lambda_2)\,l^{\bm\mu}\,l^{\bm\nu}
\label{eq:muintegral2}
\eea 
in the leading order in the curvature. Note that although we have used a parallel-propagated SNF to derive it, this formula is valid in any basis, including every coordinate basis.
We see that for small curvature effects the Weyl tensor, carrying the information about tidal forces and gravitational waves along the line of sight, and the cosmological constant drop out from the integral, leaving only the dependence on the matter content along $\gamma_0$ and in its vicinity.
The iterated integral in (\ref{eq:muintegral2}) can be converted into a single integral of the same expression with a linear weight function:
\bea
\mu \approx {8\pi G}\,\int_{\lambda_\calO}^{\lambda_\calE} 
T_{\bm \mu\bm \nu}(\lambda)\,l^{\bm\mu}\,l^{\bm\nu}\,\left(\lambda_\calE - \lambda\right)\,\dd\lambda.
\label{eq:muintegral3}
\eea 
The proof of equivalence of (\ref{eq:muintegral2}) and (\ref{eq:muintegral3}) proceeds via the integration by parts of (\ref{eq:muintegral3}), with the linear function  $(\lambda_\calE - \lambda)$ undergoing
differentiation and the stress-energy tensor term being integrated.

Finally, the reader may verify that all expressions for $\mu$ are invariant with respect to affine reparametrizations of $\gamma_0$ given by (\ref{eq:affinereparametrization})-(\ref{eq:laffine}).

\paragraph{Applications. }We will sketch now a simple application of the result above. Since the formulas (\ref{eq:muintegral2}) or (\ref{eq:muintegral3})
relate $\mu$, a quantity potentially measurable using optical observations, with the amount of matter along the line of sight, we may use them to devise a purely optical method of determining the spacetime matter distribution. 
Consider a very precise, momentary measurement of the size of the image of a small object of known physical size at $\calE$, performed by a telescope at $\calO$ from very far away. We assume that the DOA holds in this configuration.
At the same moment, we need to perform equally precise measurements of its classic parallax from at least two other, noncollinear points nearby, displaced orthogonally to the direction of light propagation. The measurement is done
by comparing the two-dimensional position on the sky of the source, seen by the two displaced and comoving auxiliary observers, with the source's position recorded by the telescope $\calO$. Since both auxiliary observers are displaced strictly orthogonally to the direction of propagation of light at $\calO$, performing the
measurements simultaneously in the observer's frame will yield the measurement of parallax of a single event along the emitter's worldline.   Therefore what we measure this way is indeed the classic parallax in the terminology of Sec. \ref{sec:parallaxnotions}.

We assume that the physical size of the emitter (in its own frame) and the positions of the auxiliary observers with respect to the central one (in the observer's frame) are known very accurately.
From these data we can determine with high precision the matrices $\Pi\UD{\bm A}{\bm B}$ and $M\UD{\bm A}{\bm B}$ along a null geodesic using
directly the relations (\ref{eq:maginficationmatrix}) and (\ref{eq:parallaxmatrix}). Then from (\ref{eq:mudef4}) we obtain $\mu$.
This way we have effectively
weighed the whole matter content in the spacetime along the line of sight: from (\ref{eq:muintegral2}) we see that this method determines the amount of any kind of matter between the observer and the emitter.

We note that the method sketched above seems
to be well-suited for a space-based mission. Due to the insensitivity of $\mu$ to the momentary motions on both sides, there is absolutely no need to know the relative motions or the precise distance \emph{between} the observer and the emitter.
It is only the distances and velocities \emph{within} the group of observers measured \emph{in the observer's own frame}, as well as the emitter's size,  measured \emph{in its own frame}, which are used in this measurement and which need to be determined with high precision. The
exact shape of the emitter is also irrelevant since it
only provides a background image. The angular size of this image and its apparent shift as observed from 2 other points carry the information about the spacetime curvature along the line of sight. 

The measurement is also highly selective regarding the matter it takes into account. Masses located off the optical axis may introduce a measurable image displacement due to the gravitational light bending as well as image
distortions by their tidal fields. Nevertheless, their influence on the value of $\mu$ is negligible. This is because distant masses may only influence the results of the observations via the tidal effects encoded in the Weyl curvature
tensor along the line of sight. This influence, however, drops out of the trace in (\ref{eq:muintegral1}), leaving just the integral of the energy density \emph{on} the optical axis.
Thus, at least within the range of applicability of the approximations from this paper, the measurement of $\mu$ effectively cuts out a thin tube around
the fiducial geodesic through the matter distribution and neglects any influence
of the gravitational field sources lying outside it. On the other hand, note that repeated measurements along \emph{different} null geodesics may provide a tomography-like method to determine a map of the matter content of a given region.
A more detailed discussion of the parameter $\mu$ and its applications will be provided in a separate paper \cite{kgs2}. 

Comparing the parallax distance $D_{par}$ or $\tilde D_{par}$  and the angular diameter distance $D_{ang}$ (or the closely related luminosity distance $D_{lum}$) as a method of determining the spacetime geometry has a long history in relativistic cosmology,
beginning with McCrea \cite{mccrea}. Weinberg \cite{weinberg-letter} noticed that in a perfectly homogeneous FLRW model comparing the parallax distance and the luminosity distance as functions of the redshift allows one to obtain the spatial curvature, which is impossible to determine by the luminosity distance observations alone. Rosquist \cite{rosquist} derived a differential relation between the two distances valid under the assumption of no shear and noticed that the comparison of both distances yields information about an otherwise unobservable component of the spacetime metric in the observational coordinates introduced by Ellis, Nel, Maartens, Stoeger and Whitman \cite{ELLIS1985315}.  
Kasai \cite{kasai} considered the parallax distance $\tilde D_{par}$ in an FLRW model with first-order perturbations, comparing of the results to the expressions for the luminosity distance $D_{lum}$ in the same setting.
Finally, Räsänen \cite{rasanen} proposed an FLRW consistency condition based on the comparison of the two types of distance measures. None of these works, however, mentions the independence of $\mu$ from momentary motions or its
direct relation to the curvature and the matter content along the line of sight.

\section{Summary and remarks\label{eq:summary}}

We have presented a general, geometric approach to the problem of geometric optics in general relativity. It concerns the problems of observations of
luminous objects or outbursts of radiation, contained within a small stage region $N_\calE$, from a large distance by observers contained in another small
auditorium region $N_\calO$. Both regions are assumed to be small enough to be considered effectively flat and light propagation is treated within the geometric optics approximation. The approach works under rather general assumptions and should, therefore, apply to observations within the Solar System, the parallax measurements within the Galaxy, as well as cosmological observations. 

The problem of observation is divided into two separate problems: the question of propagation of light through the inhomogeneous spacetime between the regions
and the problem of the dependence of the results of observations from the motions in both regions. The second problem lies within the range of applicability
of special relativity and is fairly easy to formulate in a geometric way.
The first problem is considered using the first order geodesic deviation equation around a known, fiducial null geodesic, or the optical axis. In this formulation, the behavior of geodesics near the fiducial one is determined by a second-order linear ODE with the curvature playing the role of one of the coefficients.
The problem of light propagation, considered within the distant observer approximation, turns out to be
fairly simple to reformulate in a frame- and observer-independent way: the tangent vector the fiducial geodesic defines two corresponding foliations of the two regions $N_\calO$ and $N_\calE$ by null hypersurfaces.
Only points lying on the corresponding leaves of the foliations in $N_\calO$ and $N_\calE$ can be connected by null geodesics [Eq. (\ref{eq:timelapse1})]. The action of curvature on the light propagation, on the other hand, is completely encoded in two optical operators, the well-known Jacobi operator $\calD$ and the emitter-observer asymmetry operator $m$, defined in Sec. \ref{sec:directionvariationformula}. Both are most conveniently defined as bilocal operators acting from a quotient space on the observer's side to an appropriate quotient space at the emitter's end. Both can be expressed as functionals of the curvature
along the line of sight [Eqs. (\ref{eq:DODE1})-(\ref{eq:DODE3}) and (\ref{eq:mODE1})-(\ref{eq:mODE3})] and they
 do not depend on the coordinate systems, bases, observers or any other structures defined on the manifold. $m$ quantifies the difference between the direction variations measured by an observer due to perpendicular
displacements of the null geodesic endpoints in $N_\calO$ and $N_\calE$. Unlike $\calD$, $m$ vanishes in a flat space and therefore it measures directly the impact of the curvature on the optical observations. This makes $m$ and quantities derived from it excellent probes of the curvature along the line of sight.

We have then shown how one can state within this framework the problems of the parallax, the position drift (rate of change of the apparent position on the sky) and the gravitational lensing
in a frame- and observer-independent way. All observables (i.e. the positions on the observers' sky and their rates of change in the observers' proper time) can be obtained from the two optical operators and the data characterizing the motions of the emitter and observer: their momentary 4-velocities, their displacement with respect to the fiducial
null geodesic and the 4-acceleration of the observer with respect to his or her local inertial frame. In the resulting formulas, the effects of spacetime geometry on the light propagation and the effects of motions
on both ends of the null geodesics are clearly separated. The geometric machinery allowed also to compare and contrast various definitions of parallax appearing in the relativistic literature and discuss relations between them. Additionally, we managed to show that in a generic situation the parallax and drift effects blow up along with the magnification of the image of the source as it passes through a caustic. The underlying reason is that the expressions for those effects involve the inverse of the Jacobi map, which by definition becomes degenerate at a caustic.

We remind the reader that the kinematic quantities appearing in the formalism, i.e. the 4-velocities, 4-accelerations and displacements
of the observers and emitters, are defined \emph{with respect the local inertial frames} in the stage and auditorium regions. Those frames represent the local
nonrotating reference frames falling freely in the large-scale gravitational fields. Therefore in the formalism presented here, the results do not depend explicitly on the potentially complicated details of motions of the observers and the emitters with respect to any external masses generating the gravitational fields: all that matters is their motions
expressed in the locally flat coordinates defined within their respective neighborhoods.
This is, of course, a consequence of the GR equivalence principle applied to geometric optics: within our approximation, the dependence of observables on the momentary motions is a consequence
of purely special relativistic effects. The SR effects, of course, cannot depend on the local details of the large-scale gravitational field. Thus the only possible dependence
on the gravitational field generated by external objects enters via the curvature tensor along the line of sight appearing in the formulas for the optical operators.

The reformulation of the geometric optics in terms of the optical operators yielded finally an unexpected result. It turns out that by combining the full data about the parallax of a faraway object, contained in the parallax matrix (\ref{eq:parallaxmatrix}),
and the data about the magnification and distortion of its image, given by the magnification matrix (\ref{eq:maginficationmatrix}), we may define observables which are entirely independent of the motions of both the observer and the source, given by equation (\ref{eq:wperpdef}).
In other words, neither the relative motion of the observer and the emitter, difficult to determine for very long distances nor the motion of any of them with respect to a local inertial frame has any influence on their values.
This is in a stark contrast to the standard observables like the redshift or luminosity distance, which always depend additionally at least on the 4-velocity of the observer \cite{Korzynski:2018}.
The new observables probe exclusively the spacetime geometry between the regions $N_\calE$ and $N_\calO$. The simplest of them, i.e. the ratio between the
suitably defined parallax distance and the angular diameter distance [Eqs. (\ref{eq:mudef4})-(\ref{eq:mudef2})], contains information about the amount of matter along the line of sight for short distances, as seen in Eq. (\ref{eq:muintegral3}).

\section*{Acknowledgments}
 The work was supported by the National Science Centre, Poland (NCN) via the SONATA BIS programme, Grant No.~2016/22/E/ST9/00578 for the project
\emph{``Local relativistic perturbative framework in hydrodynamics and general relativity and its application to cosmology''}. 
The authors would also like to thank 
 Andrzej Krasi\'nski, Nezihe Uzun, Eleonora Villa, Thomas Buchert and  Edward Malec for useful discussions and comments.
 M.K. would also like to thank the Max Planck Institute for Gravitational Physics in Potsdam (Albert-Einstein-Institut) for hospitality and support.

\bibliography{main-Displacement}

\end{document}